\documentclass[twocolumn,traditabstract,longauth]{aa}
\usepackage{graphicx,amsmath}
\usepackage{epsf}
\usepackage{txfonts}
\usepackage[hyphens]{url}
\usepackage{longtable}
\usepackage{lscape}
\usepackage[hyperindex,breaklinks=true, colorlinks, citecolor=blue]{hyperref}
\usepackage{breakurl}
\usepackage{comment}
\usepackage{natbib}
\usepackage{upgreek}
\usepackage{float}
\usepackage[switch,pagewise]{lineno}
\usepackage{multirow}
\newcommand{\hpeter}[1]{}
\newcommand{\juan}[1]{{ #1}}%\newcommand{\juan}[1]{{\bf \color{magenta}#1}}
\newcommand{\hjuan}[1]{}
\newcommand{\langed}[1]{{#1}}%\newcommand{\juan}[1]{{\bf \color{magenta}#1}}
\newcommand{\commentproof}[1]{{\bf \color{green}#1}}
\renewcommand{\commentproof}[1]{} % turn off mark-up

\def\setsymbol#1#2{\expandafter\def\csname #1\endcsname{#2}}
\def\getsymbol#1{\csname #1\endcsname}

\def\Planck{\textit{Planck}}

\def\all2013resultspapers{\nocite{planck2013-p01, planck2013-p02, planck2013-p02a, planck2013-p02d, planck2013-p02b, planck2013-p03, planck2013-p03c, planck2013-p03f, planck2013-p03d, planck2013-p03e, planck2013-p01a, planck2013-p06, planck2013-p03a, planck2013-pip88, planck2013-p08, planck2013-p11, planck2013-p12, planck2013-p13, planck2013-p14, planck2013-p15, planck2013-p05b, planck2013-p17, planck2013-p09, planck2013-p09a, planck2013-p20, planck2013-p19, planck2013-pipaberration, planck2013-p05, planck2013-p05a, planck2013-pip56, planck2013-p06b, planck2013-p01a}}

\newbox\tablebox    \newdimen\tablewidth
\def\leaderfil{\leaders\hbox to 5pt{\hss.\hss}\hfil}
\def\endPlancktable{\tablewidth=\columnwidth 
    $$\hss\copy\tablebox\hss$$
    \vskip-\lastskip\vskip -2pt}
\def\endPlancktablewide{\tablewidth=\textwidth 
    $$\hss\copy\tablebox\hss$$
    \vskip-\lastskip\vskip -2pt}
\def\tablenote#1 #2\par{\begingroup \parindent=0.8em
    \abovedisplayshortskip=0pt\belowdisplayshortskip=0pt
    \noindent
    $$\hss\vbox{\hsize\tablewidth \hangindent=\parindent \hangafter=1 \noindent
    \hbox to \parindent{$^#1$\hss}\strut#2\strut\par}\hss$$
    \endgroup}
\def\doubleline{\vskip 3pt\hrule \vskip 1.5pt \hrule \vskip 5pt}

\def\L2{\ifmmode L_2\else $L_2$\fi}

\def\DeltaT{\ifmmode \Delta T\else $\Delta T$\fi}
\def\deltat{\ifmmode \Delta t\else $\Delta t$\fi}
\def\fknee{\ifmmode f_{\rm knee}\else $f_{\rm knee}$\fi}
\def\Fmax{\ifmmode F_{\rm max}\else $F_{\rm max}$\fi}
\def\solar{\ifmmode{\rm M}_{\mathord\odot}\else${\rm M}_{\mathord\odot}$\fi}
\def\Msolar{\ifmmode{\rm M}_{\mathord\odot}\else${\rm M}_{\mathord\odot}$\fi}
\def\Lsolar{\ifmmode{\rm L}_{\mathord\odot}\else${\rm L}_{\mathord\odot}$\fi}
\def\inv{\ifmmode^{-1}\else$^{-1}$\fi}
\def\mo{\ifmmode^{-1}\else$^{-1}$\fi}
\def\sup#1{\ifmmode ^{\rm #1}\else $^{\rm #1}$\fi}
\def\expo#1{\ifmmode \times 10^{#1}\else $\times 10^{#1}$\fi}
\def\,{\thinspace}
\def\lsim{\mathrel{\raise .4ex\hbox{\rlap{$<$}\lower 1.2ex\hbox{$\sim$}}}}
\def\gsim{\mathrel{\raise .4ex\hbox{\rlap{$>$}\lower 1.2ex\hbox{$\sim$}}}}

\def\simprop{\mathrel{\raise .4ex\hbox{\rlap{$\propto$}\lower 1.2ex\hbox{$\sim$}}}}
\def\deg{\ifmmode^\circ\else$^\circ$\fi}
\def\pdeg{\ifmmode $\setbox0=\hbox{$^{\circ}$}\rlap{\hskip.11\wd0 .}$^{\circ}
          \else \setbox0=\hbox{$^{\circ}$}\rlap{\hskip.11\wd0 .}$^{\circ}$\fi}
\def\arcs{\ifmmode {^{\scriptstyle\prime\prime}}
          \else $^{\scriptstyle\prime\prime}$\fi}
\def\arcm{\ifmmode {^{\scriptstyle\prime}}
          \else $^{\scriptstyle\prime}$\fi}
\newdimen\sa  \newdimen\sb
\def\parcs{\sa=.07em \sb=.03em
     \ifmmode \hbox{\rlap{.}}^{\scriptstyle\prime\kern -\sb\prime}\hbox{\kern -\sa}
     \else \rlap{.}$^{\scriptstyle\prime\kern -\sb\prime}$\kern -\sa\fi}
\def\parcm{\sa=.08em \sb=.03em
     \ifmmode \hbox{\rlap{.}\kern\sa}^{\scriptstyle\prime}\hbox{\kern-\sb}
     \else \rlap{.}\kern\sa$^{\scriptstyle\prime}$\kern-\sb\fi}
\def\ra[#1 #2 #3.#4]{#1\sup{h}#2\sup{m}#3\sup{s}\llap.#4}
\def\dec[#1 #2 #3.#4]{#1\deg#2\arcm#3\arcs\llap.#4}
\def\deco[#1 #2 #3]{#1\deg#2\arcm#3\arcs}
\def\rra[#1 #2]{#1\sup{h}#2\sup{m}}

\def\dots{\relax\ifmmode \ldots\else $\ldots$\fi}
\def\WHzsr{\ifmmode $W\,Hz\mo\,sr\mo$\else W\,Hz\mo\,sr\mo\fi}
\def\mHz{\ifmmode $\,mHz$\else \,mHz\fi}
\def\GHz{\ifmmode $\,GHz$\else \,GHz\fi}
\def\mKs{\ifmmode $\,mK\,s$^{1/2}\else \,mK\,s$^{1/2}$\fi}
\def\muKs{\ifmmode \,\mu$K\,s$^{1/2}\else \,$\mu$K\,s$^{1/2}$\fi}
\def\muKRJs{\ifmmode \,\mu$K$_{\rm RJ}$\,s$^{1/2}\else \,$\mu$K$_{\rm RJ}$\,s$^{1/2}$\fi}
\def\muKHz{\ifmmode \,\mu$K\,Hz$^{-1/2}\else \,$\mu$K\,Hz$^{-1/2}$\fi}
\def\MJysr{\ifmmode \,$MJy\,sr\mo$\else \,MJy\,sr\mo\fi}
\def\MJysrmK{\ifmmode \,$MJy\,sr\mo$\,mK$_{\rm CMB}\mo\else \,MJy\,sr\mo\,mK$_{\rm CMB}\mo$\fi}
\def\microns{\ifmmode \,\mu$m$\else \,$\mu$m\fi}

\def\muK{\ifmmode \,\mu$K$\else \,$\mu$\hbox{K}\fi}
\def\microK{\ifmmode \,\mu$K$\else \,$\mu$\hbox{K}\fi}
\def\muW{\ifmmode \,\mu$W$\else \,$\mu$\hbox{W}\fi}
\def\kms{\ifmmode $\,km\,s$^{-1}\else \,km\,s$^{-1}$\fi}
\def\kmsMpc{\ifmmode $\,\kms\,Mpc\mo$\else \,\kms\,Mpc\mo\fi}
\providecommand{\sorthelp}[1]{}

\newcommand{\mft}{{\rm mft}}
\newcommand{\bb}{B^{\mft}}% use in math mode
\newcommand{\nhb}{N_{{\rm H}_{2}}^{\mft}}% use in math mode

\renewcommand{\zeta}{\xi}

\newcommand{\LOS}{line of sight}
\newcommand{\LOSh}{line-of-sight}

\newcommand{\thresh}{21.7} % typical value of log column density where \xi changes sign 

\newcommand{\htot}{h_{\rm tot}} % total number of samples in the histogram

\newcommand{\planck}{\Planck}  % for \planck rather than \Planck
\def\Herschel{\textit{Herschel}}
\newcommand{\nh}{$N_{\textsc{H}}$}
\newcommand{\nhd}{N_{\textsc{H}}} % use in math mode

\newcommand{\lognh}{$\log_{10}(N_{\textsc{H}}/\mbox{cm}^{-2})$}
\newcommand{\microG}{$\mu$G}

\newcommand{\IRAS}{IRAS}

\newcommand{\healpix}{{\tt HEALPix}}

\newcommand{\preferentially}{\langed{mostly }}
\newcommand{\pseudovectors}{\langed{pseudo-vectors}}

\providecommand{\sorthelp}[1]{}

\newcommand{\hkd}{\cite{hildebrand2009}} 
\renewcommand{\hkd}{\juan{DCF+SF}}%\renewcommand{\hkd}{HKD}
\newcommand{\HIL}{\juan{DCF+SF}}%\newcommand{\HIL}{HKD}

\begin{document}
%%%%%%%%%%%%%%%%%%%%%%%%%%%%%%%%%%%%%%%%%%%%%%%%%%%%%%%%%%%%
%\linenumbers
%%%%%%%%%%%%%%%%%%%%%%%%%%%%%%%%%%%%%%%%%%%%%%%%%%%%%%%%%%%%

\title{\Planck\ intermediate results. XXXV. Probing the role of the magnetic field in the formation of structure in molecular clouds}
\titlerunning{Probing the role of the magnetic field in the formation of structure in molecular clouds}
\authorrunning{\Planck\ Collaboration}
%This author list corresponds to \title{Author list for PIP\_113\_Boulanger}
%Prepared by M. Lopez-Caniego (Marcos.Lopez.Caniego@sciops.esa.int), ESAC/ESA
%This version is from Thu Jun 25 11:45:47 2015 CET
%\subtitle{There are 197 co-authors in this list}
\author{\small
Planck Collaboration: P.~A.~R.~Ade\inst{83}
\and
N.~Aghanim\inst{56}
\and
M.~I.~R.~Alves\inst{56}
\and
M.~Arnaud\inst{70}
\and
D.~Arzoumanian\inst{56}
\and
M.~Ashdown\inst{66, 5}
\and
J.~Aumont\inst{56}
\and
C.~Baccigalupi\inst{81}
\and
A.~J.~Banday\inst{89, 8}
\and
R.~B.~Barreiro\inst{61}
% ...
}
\institute{\small
APC, AstroParticule et Cosmologie, Universit\'{e} Paris Diderot, CNRS/IN2P3, CEA/lrfu, Observatoire de Paris, Sorbonne Paris Cit\'{e}, 10, rue Alice Domon et L\'{e}onie Duquet, 75205 Paris Cedex 13, France\goodbreak
\and
African Institute for Mathematical Sciences, 6-8 Melrose Road, Muizenberg, Cape Town, South Africa\goodbreak
\and
Agenzia Spaziale Italiana Science Data Center, Via del Politecnico snc, 00133, Roma, Italy\goodbreak
\and
Aix Marseille Universit\'{e}, CNRS, LAM (Laboratoire d'Astrophysique de Marseille) UMR 7326, 13388, Marseille, France\goodbreak
\and
Astrophysics Group, Cavendish Laboratory, University of Cambridge, J J Thomson Avenue, Cambridge CB3 0HE, U.K.\goodbreak
\and
Astrophysics \& Cosmology Research Unit, School of Mathematics, Statistics \& Computer Science, University of KwaZulu-Natal, Westville Campus, Private Bag X54001, Durban 4000, South Africa\goodbreak
\and
CITA, University of Toronto, 60 St. George St., Toronto, ON M5S 3H8, Canada\goodbreak
\and
CNRS, IRAP, 9 Av. colonel Roche, BP 44346, F-31028 Toulouse cedex 4, France\goodbreak
\and
California Institute of Technology, Pasadena, California, U.S.A.\goodbreak
\and
Centro de Estudios de F\'{i}sica del Cosmos de Arag\'{o}n (CEFCA), Plaza San Juan, 1, planta 2, E-44001, Teruel, Spain\goodbreak
\and
Computational Cosmology Center, Lawrence Berkeley National Laboratory, Berkeley, California, U.S.A.\goodbreak
\and
DSM/Irfu/SPP, CEA-Saclay, F-91191 Gif-sur-Yvette Cedex, France\goodbreak
\and
DTU Space, National Space Institute, Technical University of Denmark, Elektrovej 327, DK-2800 Kgs. Lyngby, Denmark\goodbreak
\and
D\'{e}partement de Physique Th\'{e}orique, Universit\'{e} de Gen\`{e}ve, 24, Quai E. Ansermet,1211 Gen\`{e}ve 4, Switzerland\goodbreak
\and
Departamento de Astrof\'{i}sica, Universidad de La Laguna (ULL), E-38206 La Laguna, Tenerife, Spain\goodbreak
\and
Departamento de F\'{\i}sica, Universidad de Oviedo, Avda. Calvo Sotelo s/n, Oviedo, Spain\goodbreak
\and
Department of Astronomy and Astrophysics, University of Toronto, 50 Saint George Street, Toronto, Ontario, Canada\goodbreak
\and
Department of Astrophysics/IMAPP, Radboud University Nijmegen, P.O. Box 9010, 6500 GL Nijmegen, The Netherlands\goodbreak
\and
Department of Physics \& Astronomy, University of British Columbia, 6224 Agricultural Road, Vancouver, British Columbia, Canada\goodbreak
\and
Department of Physics and Astronomy, Dana and David Dornsife College of Letter, Arts and Sciences, University of Southern California, Los Angeles, CA 90089, U.S.A.\goodbreak
\and
Department of Physics and Astronomy, University College London, London WC1E 6BT, U.K.\goodbreak
\and
Department of Physics, Florida State University, Keen Physics Building, 77 Chieftan Way, Tallahassee, Florida, U.S.A.\goodbreak
\and
Department of Physics, Gustaf H\"{a}llstr\"{o}min katu 2a, University of Helsinki, Helsinki, Finland\goodbreak
\and
Department of Physics, Princeton University, Princeton, New Jersey, U.S.A.\goodbreak
\and
Department of Physics, University of California, Santa Barbara, California, U.S.A.\goodbreak
\and
Department of Physics, University of Illinois at Urbana-Champaign, 1110 West Green Street, Urbana, Illinois, U.S.A.\goodbreak
\and
Dipartimento di Fisica e Astronomia G. Galilei, Universit\`{a} degli Studi di Padova, via Marzolo 8, 35131 Padova, Italy\goodbreak
\and
Dipartimento di Fisica e Scienze della Terra, Universit\`{a} di Ferrara, Via Saragat 1, 44122 Ferrara, Italy\goodbreak
\and
Dipartimento di Fisica, Universit\`{a} La Sapienza, P. le A. Moro 2, Roma, Italy\goodbreak
\and
Dipartimento di Fisica, Universit\`{a} degli Studi di Milano, Via Celoria, 16, Milano, Italy\goodbreak
\and
Dipartimento di Fisica, Universit\`{a} degli Studi di Trieste, via A. Valerio 2, Trieste, Italy\goodbreak
\and
Dipartimento di Matematica, Universit\`{a} di Roma Tor Vergata, Via della Ricerca Scientifica, 1, Roma, Italy\goodbreak
\and
Discovery Center, Niels Bohr Institute, Blegdamsvej 17, Copenhagen, Denmark\goodbreak
\and
Escola de Artes, Ci\^encias e Humanidades, Universidade de S\~ao Paulo, Rua Arlindo Bettio 1000, CEP 03828-000, S\~ao Paulo, Brazil\goodbreak
\and
European Space Agency, ESAC, Planck Science Office, Camino bajo del Castillo, s/n, Urbanizaci\'{o}n Villafranca del Castillo, Villanueva de la Ca\~{n}ada, Madrid, Spain\goodbreak
\and
European Space Agency, ESTEC, Keplerlaan 1, 2201 AZ Noordwijk, The Netherlands\goodbreak
\and
Facolt\`{a} di Ingegneria, Universit\`{a} degli Studi e-Campus, Via Isimbardi 10, Novedrate (CO), 22060, Italy\goodbreak
\and
Gran Sasso Science Institute, INFN, viale F. Crispi 7, 67100 L'Aquila, Italy\goodbreak
\and
HGSFP and University of Heidelberg, Theoretical Physics Department, Philosophenweg 16, 69120, Heidelberg, Germany\goodbreak
\and
Helsinki Institute of Physics, Gustaf H\"{a}llstr\"{o}min katu 2, University of Helsinki, Helsinki, Finland\goodbreak
\and
INAF - Osservatorio Astrofisico di Catania, Via S. Sofia 78, Catania, Italy\goodbreak
\and
INAF - Osservatorio Astronomico di Padova, Vicolo dell'Osservatorio 5, Padova, Italy\goodbreak
\and
INAF - Osservatorio Astronomico di Roma, via di Frascati 33, Monte Porzio Catone, Italy\goodbreak
\and
INAF - Osservatorio Astronomico di Trieste, Via G.B. Tiepolo 11, Trieste, Italy\goodbreak
\and
INAF/IASF Bologna, Via Gobetti 101, Bologna, Italy\goodbreak
\and
INAF/IASF Milano, Via E. Bassini 15, Milano, Italy\goodbreak
\and
INFN, Sezione di Bologna, Via Irnerio 46, I-40126, Bologna, Italy\goodbreak
\and
INFN, Sezione di Roma 1, Universit\`{a} di Roma Sapienza, Piazzale Aldo Moro 2, 00185, Roma, Italy\goodbreak
\and
INFN, Sezione di Roma 2, Universit\`{a} di Roma Tor Vergata, Via della Ricerca Scientifica, 1, Roma, Italy\goodbreak
\and
INFN/National Institute for Nuclear Physics, Via Valerio 2, I-34127 Trieste, Italy\goodbreak
\and
IPAG: Institut de Plan\'{e}tologie et d'Astrophysique de Grenoble, Universit\'{e} Grenoble Alpes, IPAG, F-38000 Grenoble, France, CNRS, IPAG, F-38000 Grenoble, France\goodbreak
\and
Imperial College London, Astrophysics group, Blackett Laboratory, Prince Consort Road, London, SW7 2AZ, U.K.\goodbreak
\and
Infrared Processing and Analysis Center, California Institute of Technology, Pasadena, CA 91125, U.S.A.\goodbreak
\and
Institut N\'{e}el, CNRS, Universit\'{e} Joseph Fourier Grenoble I, 25 rue des Martyrs, Grenoble, France\goodbreak
\and
Institut Universitaire de France, 103, bd Saint-Michel, 75005, Paris, France\goodbreak
\and
Institut d'Astrophysique Spatiale, CNRS (UMR8617) Universit\'{e} Paris-Sud 11, B\^{a}timent 121, Orsay, France\goodbreak
\and
Institut d'Astrophysique de Paris, CNRS (UMR7095), 98 bis Boulevard Arago, F-75014, Paris, France\goodbreak
\and
Institute of Astronomy, University of Cambridge, Madingley Road, Cambridge CB3 0HA, U.K.\goodbreak
\and
Institute of Theoretical Astrophysics, University of Oslo, Blindern, Oslo, Norway\goodbreak
\and
Instituto de Astrof\'{\i}sica de Canarias, C/V\'{\i}a L\'{a}ctea s/n, La Laguna, Tenerife, Spain\goodbreak
\and
Instituto de F\'{\i}sica de Cantabria (CSIC-Universidad de Cantabria), Avda. de los Castros s/n, Santander, Spain\goodbreak
\and
Istituto Nazionale di Fisica Nucleare, Sezione di Padova, via Marzolo 8, I-35131 Padova, Italy\goodbreak
\and
Jet Propulsion Laboratory, California Institute of Technology, 4800 Oak Grove Drive, Pasadena, California, U.S.A.\goodbreak
\and
Jodrell Bank Centre for Astrophysics, Alan Turing Building, School of Physics and Astronomy, The University of Manchester, Oxford Road, Manchester, M13 9PL, U.K.\goodbreak
\and
Kavli Institute for Cosmological Physics, University of Chicago, Chicago, IL 60637, USA\goodbreak
\and
Kavli Institute for Cosmology Cambridge, Madingley Road, Cambridge, CB3 0HA, U.K.\goodbreak
\and
Kazan Federal University, 18 Kremlyovskaya St., Kazan, 420008, Russia\goodbreak
\and
LAL, Universit\'{e} Paris-Sud, CNRS/IN2P3, Orsay, France\goodbreak
\and
LERMA, CNRS, Observatoire de Paris, 61 Avenue de l'Observatoire, Paris, France\goodbreak
\and
Laboratoire AIM, IRFU/Service d'Astrophysique - CEA/DSM - CNRS - Universit\'{e} Paris Diderot, B\^{a}t. 709, CEA-Saclay, F-91191 Gif-sur-Yvette Cedex, France\goodbreak
\and
Laboratoire Traitement et Communication de l'Information, CNRS (UMR 5141) and T\'{e}l\'{e}com ParisTech, 46 rue Barrault F-75634 Paris Cedex 13, France\goodbreak
\and
Laboratoire de Physique Subatomique et Cosmologie, Universit\'{e} Grenoble-Alpes, CNRS/IN2P3, 53, rue des Martyrs, 38026 Grenoble Cedex, France\goodbreak
\and
Laboratoire de Physique Th\'{e}orique, Universit\'{e} Paris-Sud 11 \& CNRS, B\^{a}timent 210, 91405 Orsay, France\goodbreak
\and
Lawrence Berkeley National Laboratory, Berkeley, California, U.S.A.\goodbreak
\and
Lebedev Physical Institute of the Russian Academy of Sciences, Astro Space Centre, 84/32 Profsoyuznaya st., Moscow, GSP-7, 117997, Russia\goodbreak
\and
Max-Planck-Institut f\"{u}r Astrophysik, Karl-Schwarzschild-Str. 1, 85741 Garching, Germany\goodbreak
\and
National University of Ireland, Department of Experimental Physics, Maynooth, Co. Kildare, Ireland\goodbreak
\and
Nicolaus Copernicus Astronomical Center, Bartycka 18, 00-716 Warsaw, Poland\goodbreak
\and
Niels Bohr Institute, Blegdamsvej 17, Copenhagen, Denmark\goodbreak
\and
Optical Science Laboratory, University College London, Gower Street, London, U.K.\goodbreak
\and
SISSA, Astrophysics Sector, via Bonomea 265, 34136, Trieste, Italy\goodbreak
\and
SUPA, Institute for Astronomy, University of Edinburgh, Royal Observatory, Blackford Hill, Edinburgh EH9 3HJ, U.K.\goodbreak
\and
School of Physics and Astronomy, Cardiff University, Queens Buildings, The Parade, Cardiff, CF24 3AA, U.K.\goodbreak
\and
Sorbonne Universit\'{e}-UPMC, UMR7095, Institut d'Astrophysique de Paris, 98 bis Boulevard Arago, F-75014, Paris, France\goodbreak
\and
Space Sciences Laboratory, University of California, Berkeley, California, U.S.A.\goodbreak
\and
Special Astrophysical Observatory, Russian Academy of Sciences, Nizhnij Arkhyz, Zelenchukskiy region, Karachai-Cherkessian Republic, 369167, Russia\goodbreak
\and
Sub-Department of Astrophysics, University of Oxford, Keble Road, Oxford OX1 3RH, U.K.\goodbreak
\and
UPMC Univ Paris 06, UMR7095, 98 bis Boulevard Arago, F-75014, Paris, France\goodbreak
\and
Universit\'{e} de Toulouse, UPS-OMP, IRAP, F-31028 Toulouse cedex 4, France\goodbreak
% ...
}

\abstract{Within ten nearby ($d <$\,450\,pc) Gould Belt molecular clouds we evaluate statistically the relative orientation between the magnetic field projected on the plane of sky, inferred from the polarized thermal emission of Galactic dust observed by \Planck\ at 353\,GHz, and the gas column density structures, quantified by the gradient of the column density, \nh.
\juan{The selected regions, covering several degrees in size, are analysed at an effective angular resolution of 10\arcmin\ FWHM, thus sampling physical scales from 0.4 to 40\,pc in the nearest cloud. The column densities in the selected regions range from \nh\ $\approx 10^{21}$ to $10^{23}$\,cm$^{-2}$, and hence they correspond to the bulk of the molecular clouds.}
The relative orientation is evaluated pixel by pixel and analysed in bins of column density using the novel statistical tool called ``Histogram of Relative Orientations''.
\juan{Throughout this study, we assume that the polarized emission observed by Planck at 353\,GHz is representative of the projected morphology of the magnetic field in each region, i.e., we assume a constant dust grain alignment efficiency, independent of the local environment.}
%\juan{The presence of a correlation between the polarization orientation and the column density structure suggests that the dust polarized emission is sampling the magnetic field structure homogeneously at the considered scales.}
Within most clouds we find that the relative orientation changes progressively with increasing \nh, from \preferentially parallel or having no preferred orientation to \preferentially perpendicular. 
In simulations of magnetohydrodynamic turbulence in molecular clouds this trend in relative orientation is a signature of Alfv\'{e}nic or sub-Alfv\'{e}nic turbulence, implying that the magnetic field is significant for the gas dynamics at the scales probed by \Planck.
We compare the deduced magnetic field strength with estimates we obtain from other methods and discuss the implications of the \Planck\ observations for the general picture of molecular cloud formation and evolution.}
\date{Received 13 February 2015 / Accepted 25 May 2015}
\maketitle

%\tableofcontents
\clearpage

\section{Introduction}\label{section:introduction}

The formation and evolution of molecular clouds (MCs) and their substructures, from filaments to cores and eventually to stars, is the product of the interaction between turbulence, magnetic fields, and gravity \citep{bergin2007,mckee2007}. The study of the relative importance of these dynamical processes is limited by the observational techniques used to evaluate them. These limitations have been particularly critical when integrating magnetic fields into the general picture of MC dynamics \citep{elmegreen2004,crutcher2012,heiles2012,hennebelle2012}.

There are two primary methods of measuring magnetic fields in the dense interstellar medium (ISM). First, observation of the Zeeman effect in molecular lines provides the \LOSh\ component of the field $B_{\parallel}$ \citep{crutcher2005}. Second, polarization maps -- in extinction from background stars and emission from dust -- reveal the orientation of the field averaged along the line of sight and projected on the plane of the sky \citep{hiltner1949,davis1951,hildebrand1988,planck2014-XXI}.

Analysis of the Zeeman effect observations presented by~\cite{crutcher2010} shows that in the diffuse ISM sampled by H{\sc i} lines ($n_{\textsc{H}} < 300\,$cm$^{-3}$)\langed{,} the maximum magnetic field strength $B_{\rm max}$ does not scale with density. This is interpreted as the effect of diffuse clouds assembled by flows along magnetic field lines, which would increase the density but not the magnetic field strength. In \langed{the} denser regions ($n_{\textsc{H}} > 300\,$cm$^{-3}$)\langed{,} probed by OH and CN spectral lines\langed{,} the same study reports a scaling of the maximum magnetic field strength $B_{\rm max} \propto n^{0.65}_{\textsc{H}}$. The latter observation can be interpreted as the effect of isotropic contraction of gas too weakly magnetized for the magnetic field to affect the morphology of the collapse. However, given that the observations are restricted to pencil-like lines of sight and the molecular tracers are not homogeneously distributed, the Zeeman effect measurements alone are not sufficient to determine the relative importance of the magnetic field at \langed{the} multiple scales within MCs.

The observation of starlight polarization provides an estimate of the projected magnetic field orientation in particular lines of sight. Starlight polarization observations show coherent magnetic fields around density structures in MCs \citep{pereyra2004,franco2010,sugitani2011,chapman2011,santos2014}. The coherent polarization morphology can be interpreted as the result of dynamically important magnetic fields. However, these observations alone are not sufficient to map even the projected magnetic field morphology fully and in particular do not tightly constrain the role of magnetic fields in the formation of structure inside MCs.

The study of magnetic field orientation within the MCs is possible through the observation of polarized thermal emission from dust. Far-infrared and submillimetre polarimetric observations have been limited to small regions up to hundreds of square arcminutes within clouds \citep{li2006,matthews2014} or to large sections of the Galactic plane at a resolution of several degrees \citep{benoit2004,bierman2011}. 
%At
\langed{On} the scale of prestellar cores and cloud segments, these observations reveal both significant levels of polarized emission and coherent field morphologies \citep{wardthompson2000,dotson2000, matthews2009}.

The strength of the magnetic field projected on the plane of the sky ($B_{\perp}$) can be estimated from polarization maps using the Davis-Chandrasekhar-Fermi (DCF) method \citep{davis1951a,chandrasekhar1953}. 
As discussed in Appendix~\ref{section:bestimates}, it is assumed that the dispersion in polarization angle $\varsigma_{\psi}$\footnote{We use $\varsigma_{\psi}$ to avoid confusion with $\sigma_{\psi}$, which was introduced in \cite{planck2014-XIX} as the uncertainty in the polarization angle $\psi$.} is \langed{entirely due} to incompressible and isotropic turbulence.
Turbulence also affects the motion of the gas and so broadens profiles of emission and absorption lines, as quantified by dispersion $\sigma_{v_{\parallel}}$.
In the DCF interpretation $B_{\perp}$ is proportional to the ratio $\sigma_{v_{\parallel}}/\varsigma_{\psi}$.
Application of the DCF method to subregions of the Taurus MC gives estimates of $B_{\perp}\approx10$\,\microG\ in low\langed{-}density regions and $\approx 25$ to $\approx 42$\,\microG\ inside filamentary structures \citep{chapman2011}. Values of $B_{\perp} \approx760$\,\microG\ have been found in dense parts of the Orion MC region \citep{houde2009}. Because of the experimental difficulties involved in producing large polarization maps, a complete statistical study of the magnetic field variation across multiple scales is not yet available.

Additional information on the effects of the magnetic field on the cloud structure is found by studying the magnetic field orientation inferred from polarization observations relative to the orientation of the 
column density structures. Patterns of relative orientation have been described qualitatively in simulations of magnetohydrodynamic (MHD) turbulence with different degrees of magnetization.  This is quantified as half the ratio of the gas pressure to the mean-field magnetic pressure
\citep{ostriker2001,heitsch2001}, with the resulting turbulence ranging from sub-Alfv\'{e}nic to super-Alfv\'{e}nic.
Quantitative analysis of simulation cubes, where the orientation of $\vec{B}$ is available directly, 
%reveal a preferential 
\langed{reveals a preferred}
orientation relative to density structures 
%which 
\langed{that}
 depends on the initial magnetization of the cloud \citep{hennebelle2013a,soler2013}. 
Using simple models of dust grain alignment and polarization efficiency to produce synthetic observations of the simulations, \citet{soler2013} showed that the 
%preferential
\langed{preferred}
 relative orientation and its systematic dependence on the degree of magnetization are preserved.

%Until recently, 
 \juan{Observational} studies of relative orientation have mostly relied on visual inspection of polarization maps \citep[\juan{e.g.,}][]{myers1991,dotson1996}. This is adequate for evaluating general trends in the orientation of the field.  However, it is limited ultimately by the need to represent the field orientation with \pseudovectors, because when a large polarization map is to be overlaid on a scalar-field map, such as intensity or column density, only a selection of \pseudovectors\ can be plotted. 
On the one hand, if the plotted \pseudovectors\ are the result of averaging the Stokes parameters over a region, then the combined visualization illustrates different scales in the polarization and in the scalar field. On the other hand, if the plotted \pseudovectors\ correspond to the polarization in a particular pixel\langed{,} then the illustrated pattern is influenced by small-scale fluctuations that might not be significant in evaluating any trend in relative orientation.

\begin{figure*}[ht!]
\centerline{
\includegraphics[width=0.75\textwidth,angle=0,origin=c]{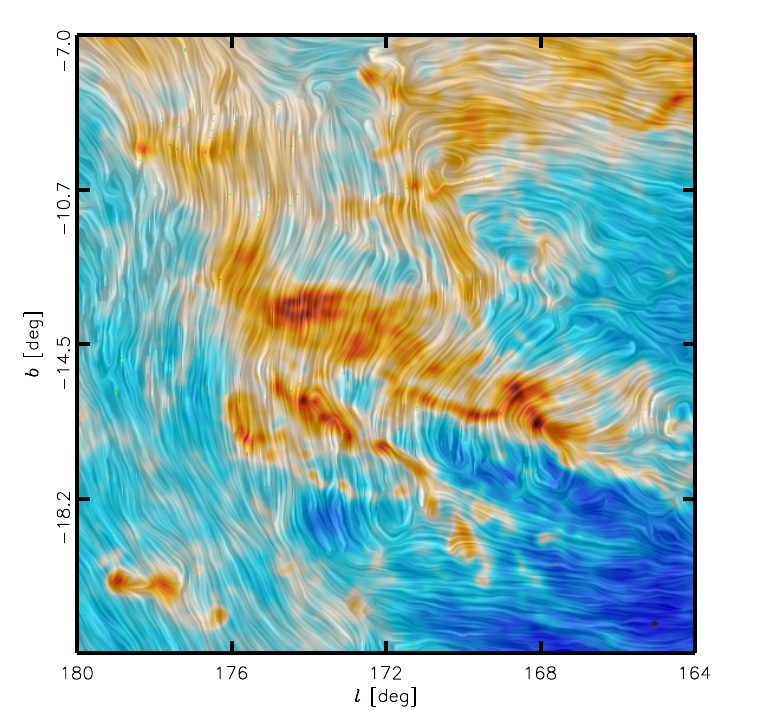}
}
\vspace{-0.5cm}
\caption{Magnetic field and column density measured by \Planck\ towards the Taurus MC. The colours represent column density.
The ``drapery" pattern,
produced using the line integral convolution method \citep[LIC,][]{cabral1993}, indicates the orientation of magnetic field lines, orthogonal to the orientation of the submillimetre polarization.
}
\label{fig:LICmap}
\end{figure*}

\cite{tassis2009} 
%presented
\langed{present} a statistical study of relative orientation between structures in the intensity and the inferred magnetic field from polarization measured at 350$\,\mu$m towards 32 Galactic clouds in maps of a few arcminutes in size. Comparing the mean direction of the field to the semi-major axis of each cloud, they 
%found
\langed{find} that the field is \preferentially perpendicular to that axis. Similarly\langed{,} \cite{li2013} compared the relative orientation in 13 clouds in the Gould Belt, calculating the main cloud orientation from the extinction map and the mean orientation of the intercloud magnetic field from starlight polarization. That study reported a bimodal distribution of relative cloud and field orientations; 
%, i.e.,
\langed{that is,} some MCs are oriented perpendicular and some parallel to the mean orientation of the intercloud field. In both studies each cloud constitutes one independent observation of relative orientation, so that the statistical significance of each study 
%is dependent
\langed{depends} on the total number of clouds observed. In a few regions of smaller 
%scale
\langed{scales}, roughly a few tenths of a parsec,
\cite{koch2013} 
%reported a preferential
\langed{report a preferred} orientation of the magnetic field, inferred from polarized dust emission, parallel to the gradient of the emission intensity.

%\setcounter{subsection}{-1}
%\subsection{New studies using \planck}

By measuring the intensity and polarization of thermal emission from Galactic dust over the whole sky and down to scales that probe the interiors of nearby MCs, \Planck\footnote{
\Planck\ (\url{http://www.esa.int/Planck}) is a project of the European Space Agency (ESA) with instruments provided by two scientific consortia funded by ESA member states (in particular the lead countries France and Italy), with contributions from NASA (USA) and telescope reflectors provided by a collaboration between ESA and a scientific consortium led and funded by Denmark.}
provides an unprecedented data set from a single instrument and with a common calibration scheme, for 
%the study of 
\langed{studying} the morphology of the magnetic field in MCs and the surrounding ISM, as illustrated for the Taurus region in Fig.~\ref{fig:LICmap}. 
We present a quantitative analysis of the relative orientation in a set of nearby ($d<450\,$pc) well-known MCs to quantify the role of the magnetic field in the formation of density structures on physical scales ranging from tens of parsecs 
to approximately one parsec in the nearest clouds. 

The present work is an extension of previous findings, \langed{as} reported by the \Planck\ collaboration, on 
%the
\langed{their} study of the polarized thermal emission from Galactic dust. Previous studies include an overview of 
%the polarized thermal emission from Galactic dust
\langed{this emission} \citep{planck2014-XIX}, which reported dust polarization 
%fractions
\langed{percentages} up to 20\,\% at low \nh, decreasing systematically with increasing \nh\ to a low plateau for regions with $\nhd > 10^{22}$\,cm$^{-2}$.
\cite{planck2014-XX} presented a comparison of the polarized thermal emission from Galactic dust with results from simulations of MHD turbulence, focusing on the statistics of the polarization fractions and angles.   Synthetic observations were made of the simulations under the simple assumption of homogeneous dust grain alignment efficiency. 
Both studies reported that the largest polarization fractions are reached in the most diffuse regions. Additionally, there is an anti-correlation between the polarization 
%fraction
\langed{percentage} and the dispersion of the polarization angle.  This anti-correlation is 
%well reproduced
\langed{reproduced well} by the synthetic observations, indicating that it is essentially 
%due to
\langed{caused by} the turbulent structure of the magnetic field.

Over most of the sky \cite{planck2014-XXXII} analysed the relative orientation between density structures,
\langed{which is} characterized by the Hessian matrix, and 
polarization, revealing that most of the elongated structures (filaments or ridges) have counterparts in the Stokes $Q$ and $U$ maps. This implies that in these structures\langed{,} the magnetic field has a 
%well defined mean direction at 
\langed{well-defined mean direction on} the scales probed by \Planck. Furthermore, the ridges are 
%preferentially
\langed{predominantly} aligned with the magnetic field measured on the structures. This statistical trend becomes more striking for  decreasing column density
%, and
\langed{and,} as expected from the potential effects of projection, for increasing polarization fraction. There is no alignment for the highest column density ridges in the $\nhd \gtrsim10^{22}$\,cm$^{-2}$ sample. \cite{planck2014-XXXIII} studied the polarization properties of three nearby filaments, showing by geometrical modelling that the magnetic field in those representative regions has a 
%well defined
\langed{well-defined} mean direction that is different from the field orientation in the surroundings.

In the present work, we quantitatively evaluate the relative orientation of the magnetic field inferred from the \Planck\ polarization observations with respect to the gas column density structures, using the 
%Histogram of Relative Orientations
\langed{histogram of relative orientations} \citep[HRO,][]{soler2013}. The HRO is a novel statistical tool that quantifies the relative orientation of each polarization measurement with respect to the column density gradient, making use of the unprecedented statistics provided by the \planck\ polarization observations. The HRO can also be evaluated in both 3D simulation data cubes and synthetic observations, \langed{thereby} providing a direct comparison between observations and the physical conditions included in MHD simulations. We compare the results of the HRO applied to the \Planck\ observations with the results of the same analysis applied to synthetic observations of MHD simulations of super-Alfv\'{e}nic, Alfv\'{e}nic, and sub-Alfv\'{e}nic turbulence. 

Thus by comparison with numerical simulations of MHD turbulence, the HRO provides estimates of the magnetic field strength without any of the assumptions involved in the DCF method.  For comparison, we estimate $B_{\perp}$ using the DCF method and the related method described by \citet{hildebrand2009} (\hkd\juan{, for DCF plus structure function}) and provide a critical assessment of their applicability.

This paper is organized as follows. 
Section~\ref{section:data} introduces the \Planck\ $353\,$GHz polarization observations, the gas column density maps, and the CO line observations used to derive the velocity information. 
The particular regions where we evaluate the relative orientation between the magnetic field and the column density structures are presented in Sect.~\ref{section:regions}.
Section~\ref{section:hro} describes the statistical tools used for the study of these relative orientations. 
In Sect.~\ref{section:discussion} we discuss our results and their implications in the general picture of cloud formation. 
Finally, Sect.~\ref{section:conclusions} summarizes the main results. 
Additional information on the selection of the polarization data, the estimation of uncertainties affecting the statistical method, and the statistical significance of the relative orientation studies can be found in Appendices~\ref{appendix:selection}, \ref{appendix:hro}, and \ref{appendix:statistics}, respectively.
Appendix~\ref{section:bestimates} presents alternative estimates of the magnetic field strength in each region.

% DATA SET ========================================================================================================

\section{Data}\label{section:data}

\subsection{Thermal dust polarization}

Over the whole sky \Planck\ observed the linear polarization (Stokes $Q$ and $U$) in seven frequency bands from 30 to 353$\,$GHz \citep{planck2013-p01}. In this study, we used data from the High Frequency Instrument \citep[HFI,][]{lamarre2010} at 353\,GHz, the highest frequency band that is sensitive to polarization.
Towards MCs the contribution of the cosmic microwave background (CMB) polarized emission is negligible at 353\,GHz, making this the \Planck\ map \langed{that is} best suited 
%for 
\langed{to} studying the spatial structure of the dust polarization \citep{planck2014-XIX,planck2014-XX}.

We 
%use
\langed{used} the Stokes $Q$ and $U$ maps and the associated noise maps made from five independent consecutive sky surveys of the \Planck\ cryogenic mission, 
%corresponding
\langed{which together correspond} to the DR3 (delta-DX11d) internal data release. We refer to previous \Planck\ publications for the data processing, map making, photometric calibration, and photometric uncertainties \citep{planck2013-p02,planck2013-p02b,planck2013-p03,planck2013-p03f}. As in the first \Planck\ polarization papers, we 
%use
\langed{used} the International Astronomical Union (IAU) conventions for the polarization angle, measured from the local direction to the north Galactic pole with positive values increasing towards the east. 

The maps of $Q$, $U$, their respective variances $\sigma^{2}_{\textsc{Q}}$, $\sigma^{2}_{\textsc{U}}$, and their covariance $\sigma_{\textsc{QU}}$ are initially at 4\parcm8 resolution in \healpix\ format\footnote{\citet{gorski2005}, \url{http://healpix.sf.net}} with a pixelization at $N_{\rm side} = 2048$, which corresponds to an effective pixel size of 1\parcm7. To increase the signal-to-noise ratio (S/N) of extended emission, we 
%smooth
\langed{smoothed} all the  maps to 10\arcmin\ resolution using a Gaussian approximation to the \Planck\ beam and the covariance smoothing procedures described in \cite{planck2014-XIX}. 

The maps of the individual regions are projected and resampled onto a Cartesian grid by using the gnomonic projection procedure described in \cite{paradis2012}. The HRO analysis is performed on these projected maps. 

\subsection{Column density}\label{columndensity}

We 
%use
\langed{used} the dust optical depth at 353\,GHz ($\tau_{353}$) as a proxy for the gas column density (\nh). The $\tau_{353}$ map \citep{planck2013-p06b} 
%is
\langed{was} derived from the all-sky \Planck\ intensity observations at 353, 545, and 857$\,$GHz, and the \IRAS\ observations at 100$\,\mu$m, which 
%are
\langed{were} fitted using a modified black body spectrum. Other parameters obtained from this fit are the temperature and the spectral index of the dust opacity.
The $\tau_{353}$ map, computed initially at 5\arcmin\ resolution, 
%is
\langed{was} smoothed to 10\arcmin\ to match the polarization maps. The errors resulting from smoothing the product $\tau_{353}$ map, rather than the underlying data, are negligible compared to the uncertainties in the dust opacity and do not significantly affect the results of this study. 

To scale from $\tau_{353}$ to \nh, following \cite{planck2013-p06b}, we 
%adopt
\langed{adopted} the dust opacity found using Galactic extinction measurements of quasars, 
\begin{equation}\label{eq:nhmap}
\tau_{353}/\nhd = 1.2 \times 10^{-26}\,\mbox{cm}^{2}\, .
\end{equation}
Variations in dust opacity are present even in the diffuse ISM and the opacity increases systematically by a factor of 2 from the diffuse to the denser ISM \citep{planck2011-7.12,martin2012,planck2013-p06b}, but our results do not \langed{critically} depend on this calibration 
%critically
.  

% STUDIED REGIONS 

\begin{figure*}[ht!]
\vspace{-1.0cm}
%\centerline{\includegraphics[height=0.98\textwidth,angle=90,origin=c]{DX11/tau_DX11d_5arcmin.eps}}
\centerline{\includegraphics[height=0.98\textwidth,angle=90,origin=c]{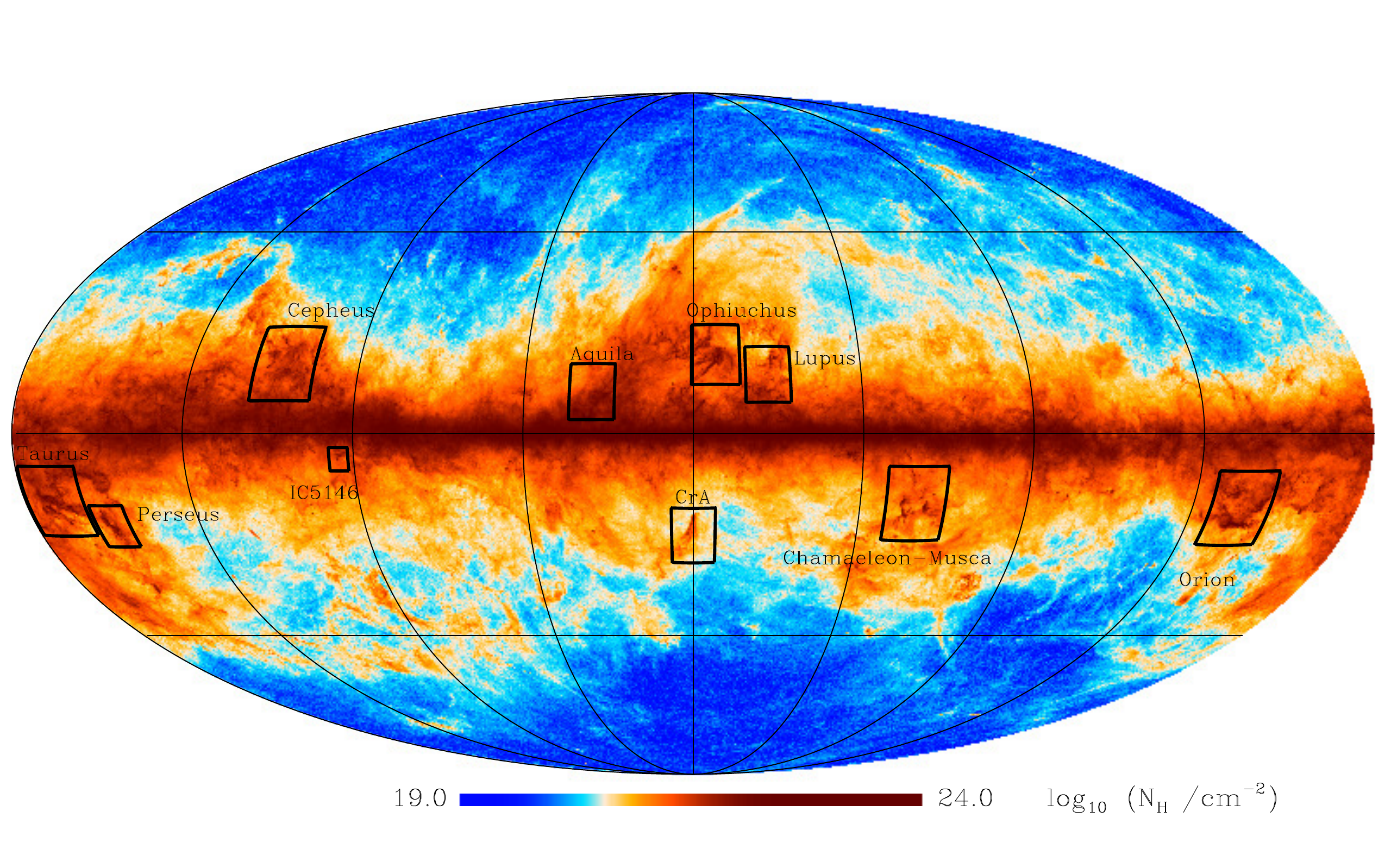}}
\vspace{-4.0cm}
\caption{Locations and sizes of the regions selected for analysis.  The background map is the gas column density, \nh, derived from the dust optical depth at $353\,$GHz \citep{planck2013-p06b}. 
}
\label{fig:Regions}
\end{figure*}

\begin{table*}[tmb]  % table* is a two-column table.  Drop the * for one column.
\begingroup
\newdimen\tblskip \tblskip=5pt
\caption{Locations and properties of the selected regions}
\label{table-fields}                            % Label goes here.
\nointerlineskip
\vskip -3mm
\footnotesize
\setbox\tablebox=\vbox{
   \newdimen\digitwidth 
   \setbox0=\hbox{\rm 0} 
   \digitwidth=\wd0 
   \catcode`*=\active 
   \def*{\kern\digitwidth}
   \newdimen\signwidth 
   \setbox0=\hbox{+} 
   \signwidth=\wd0 
   \catcode`!=\active 
   \def!{\kern\signwidth}
\halign{\hbox to 1.15in{#\leaderfil}\tabskip 2.2em&
\hfil#&
\hfil#&
\hfil#&
\hfil#&
\hfil#&
\hfil#&
\hfil#&
%\hfil#&
\hfil#\tabskip 0pt\cr
\noalign{\doubleline}
\omit\hfil Region\hfil& \hfil$l$\hfil & \hfil$b$\hfil & \hfil$\Delta l$\hfil & \hfil$\Delta b$\hfil & \hfil{Distance}$^a$\hfil & \hfil $\left< \nhd \right>$$^b$\hfil & \hfil$\mathrm{Max}\left( \nhd \right)$$^b$\hfil & \hfil$\left<N_{{\rm H}_{2}}\right>$$^c$\hfil \cr
\omit&\hfil[\deg]\hfil&\hfil[\deg]\hfil&\hfil[\deg]\hfil&\hfil[\deg]\hfil &[pc]\hfil & \hfil[$10^{21}$\,cm$^{-2}$]\hfil & \hfil[$10^{21}$\,cm$^{-2}$]\hfil & \hfil[$10^{21}$cm$^{-2}$]\hfil \cr
\noalign{\vskip 4pt\hrule\vskip 6pt}
%----------------------------------------------------------------------------------------------------------------
Taurus & 172.5 & $-$14.5 & 15.0 & 15.0 & 140\hfil & 5.4\hfil &   \phantom{0}51.9\hfil &    1.6\hfil \cr
%----------------------------------------------------------------------------------------------------------------
Ophiuchus & 354.0 & 17.0 & 13.0 & 13.0 & 140\hfil & 4.4\hfil &  103.3\hfil &  1.1\hfil \cr
%----------------------------------------------------------------------------------------------------------------
Lupus & 340.0 & 12.7 & 12.0 & 12.0 & 140\hfil & 3.8\hfil &   \phantom{0}30.8\hfil &   1.2\hfil \cr
%----------------------------------------------------------------------------------------------------------------
Chamaeleon-Musca & 300.0 & $-$15.0 & 16.0 & 16.0 & 160\hfil & 2.3\hfil &    \phantom{0}29.7\hfil &    1.3\hfil \cr
%----------------------------------------------------------------------------------------------------------------
Corona Australis (CrA) & 0.0 & $-$22.0 & 12.0 & 12.0 & 170\hfil &    1.1\hfil &    \phantom{0}40.5\hfil & 1.2\hfil \cr
%----------------------------------------------------------------------------------------------------------------
\noalign{\vskip 4pt\hrule\vskip 4pt}
%----------------------------------------------------------------------------------------------------------------
Aquila Rift & 27.0 & 8.0 & 12.0 & 12.0 & 260\hfil &    9.3\hfil &    \phantom{0}58.7\hfil &    1.9\hfil \cr
%----------------------------------------------------------------------------------------------------------------
Perseus & 159.0 & $-$20.0 & 9.0 & 9.0 & 300\hfil &    3.9\hfil &    \phantom{0}94.8\hfil &    2.6\hfil \cr
%----------------------------------------------------------------------------------------------------------------
\noalign{\vskip 4pt\hrule\vskip 4pt}
%----------------------------------------------------------------------------------------------------------------
IC\,5146 & 94.0 & $-$5.5 & 5.0 & 5.0 & 400\hfil &    3.7\hfil &    \phantom{0}22.6\hfil & 1.0\hfil \cr
%---------------------------------------------------------------------------------------------------------------
Cepheus & 110.0 & 15.0 & 16.0 & 16.0 & 440\hfil &    4.2\hfil &    \phantom{0}21.3\hfil &    1.2\hfil \cr
%----------------------------------------------------------------------------------------------------------------
Orion & 212.0 & $-$16.0 & 16.0 & 16.0 & 450\hfil &    5.0\hfil &    \phantom{0}93.6\hfil &    2.2\hfil  \cr
%----------------------------------------------------------------------------------------------------------------
\noalign{\vskip 3pt\hrule\vskip 4pt}}}
%\endPlancktable                    % ends one-column \halign
\endPlancktablewide                 % ends two-column \halign
% DISTANCES DISTANCES DISTANCES 
\tablenote a The estimates of distances are from: \cite{schlafly2014} for Taurus, Ophiuchus, Perseus, IC\,5146, Cepheus, and Orion; \cite{knude1998} for Lupus and CrA; \cite{whittet1997} for Chamaeleon-Musca; and \cite{straizys2003} for Aquila Rift.\par
\tablenote b Estimated from $\tau_{353}$ using Eq.~\eqref{eq:nhmap} for the selected pixels defined in Appendix~\ref{appendix:selection}.\par
\tablenote c Using the line integral $W_{\textsc{CO}}$ over $-10<v_{\parallel}/\mbox{(km\,s$^{-1}$)}<10$ from the \cite{dame2001} survey and $X_{\textsc{CO}}=1.8\times10^{20}$\,cm$^{-2}\,$K$^{-1}$\,km$^{-1}$\,s.\par
\endgroup
\end{table*}  

%========================================================================================================

\section{Analysed regions}\label{section:regions}

The selected regions, shown in Fig.~\ref{fig:Regions}, correspond to nearby ($d < 450\,$pc) MCs, whose characteristics are well studied and can be used for cloud-to-cloud comparison \citep{poppel1997,reipurth2008}. 
Their properties are summarized in Table \ref{table-fields}\langed{,} which includes: Galactic longitude $l$ and latitude $b$ at the centre of the field; field size $\Delta l\times\Delta b$; estimate of distance; mean and maximum total column densities from dust, $\left< \nhd \right>$ and $\mathrm{max}\left( \nhd \right)$, respectively; and mean ${\rm H}_{2}$ column density from CO.

In the table the regions are organized from the nearest to the farthest in three groups: 
(a) regions located at $d\approx 150$\,pc, namely Taurus, Ophiuchus, Lupus, Chamaeleon-Musca, and Corona Australis (CrA); 
(b) regions located at $d\approx 300$\,pc, Aquila Rift and Perseus; and 
(c) regions located at $d\approx 450$\,pc, IC\,5146, Cepheus, and Orion.    

%%%%%%%%%%%%%%%%%%%%%%%%%%%%%%%%%%%%%%%%%%%%%%%%%%%%%%%%%
%%%%%%%%%%%%%%%%%%%%%%%%%%%%%%%%%%%%%%%%%%%%%%%%%%%%%%%%%
\begin{figure*}
\centerline{
\includegraphics[height=0.31\textheight,angle=0,origin=c]{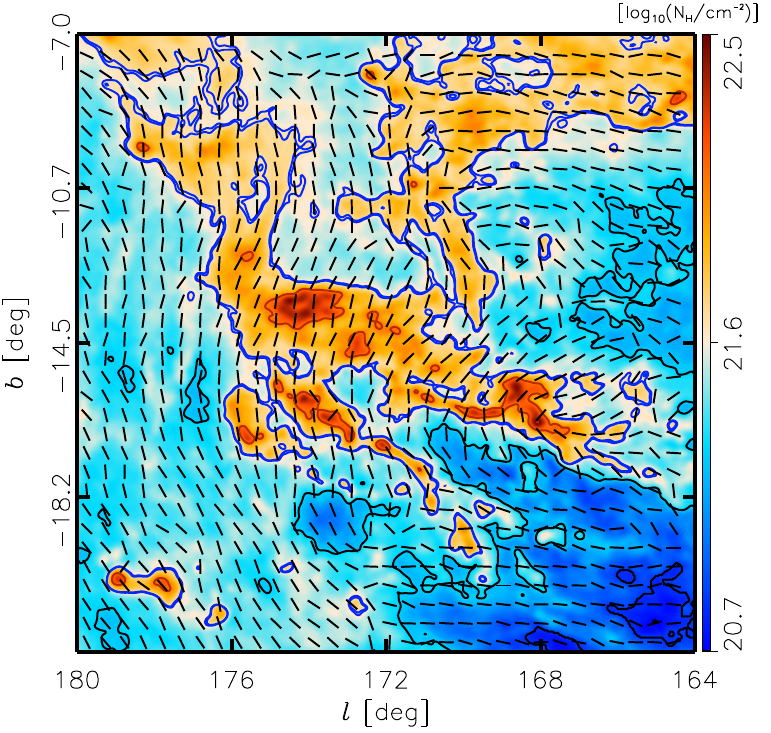}
\includegraphics[height=0.31\textheight,angle=0,origin=c]{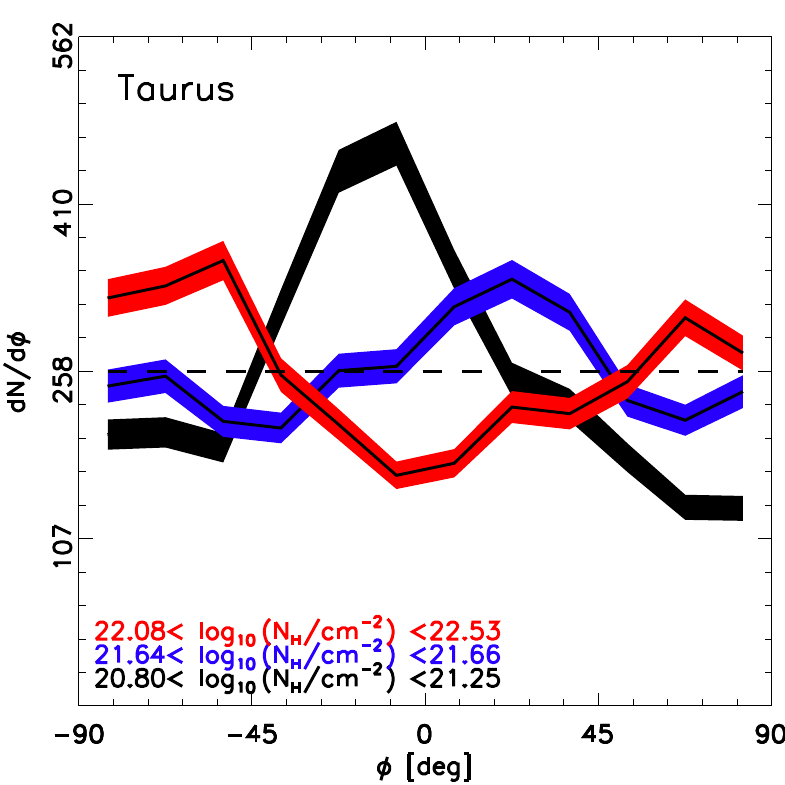}
}
\vspace{-0.1cm}
\centerline{
\includegraphics[height=0.31\textheight,angle=0,origin=c]{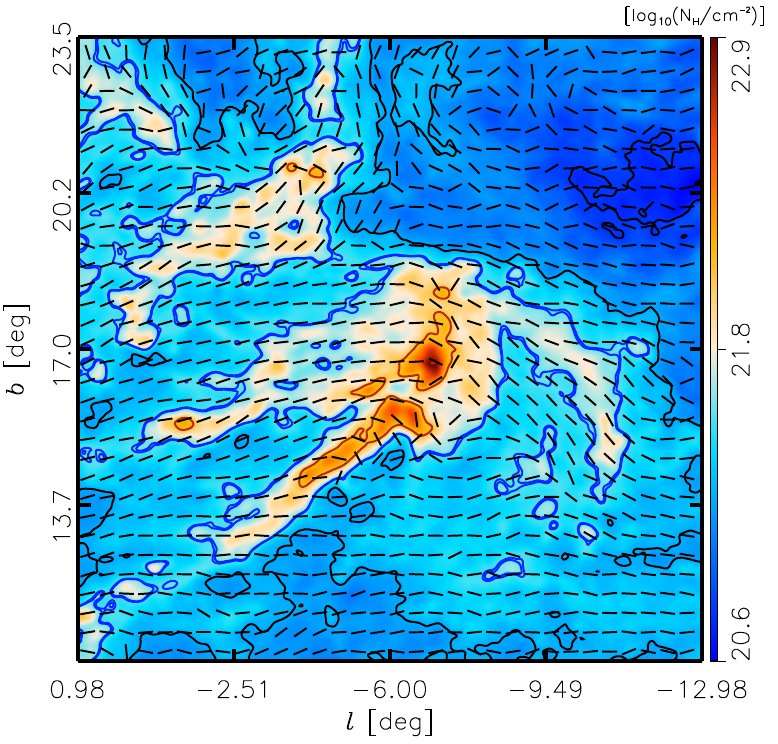}
\includegraphics[height=0.31\textheight,angle=0,origin=c]{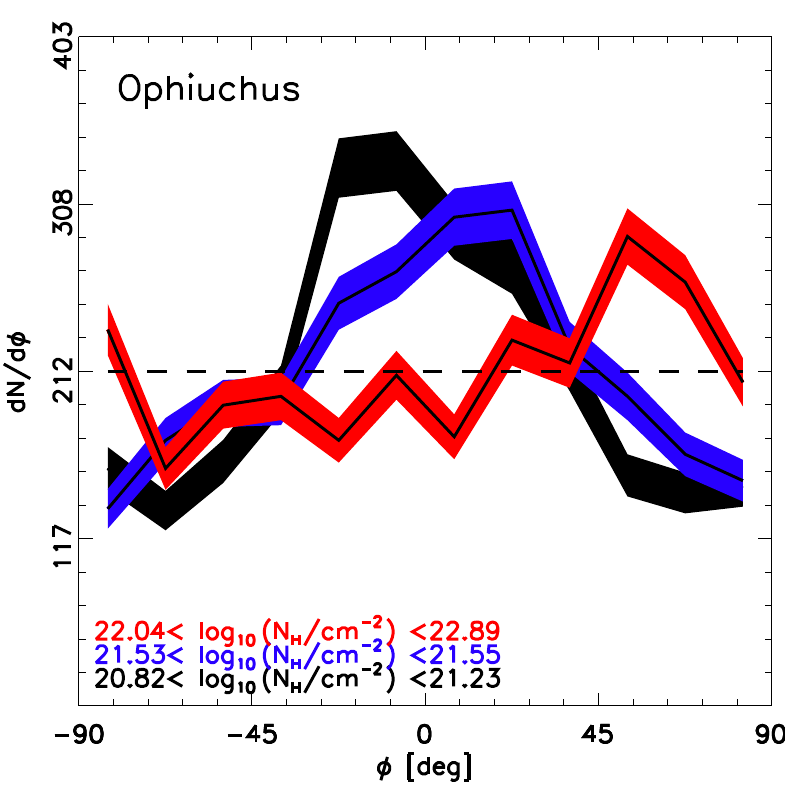}
}
\vspace{-0.1cm}
\caption{\emph{Left}: Columm density map, \lognh, overlaid with magnetic field \pseudovectors\ whose orientations are inferred from the \Planck\ 353$\,$GHz polarization observations. The length of the \pseudovectors\ is normalized so does not reflect the polarization fraction. 
In this first group, the regions analysed are, from top to bottom, Taurus, Ophiuchus, Lupus, Chamaeleon-Musca, and CrA.
\emph{Right}: HROs for the lowest, an intermediate, and the highest \nh\ bin (black, blue, and red, respectively).  
For a given region, bins have equal numbers of selected pixels (see Sect.~\ref{introhro} and Appendix~\ref{appendix:selection}) within the \nh\ ranges labelled.
The intermediate bin corresponds to selected pixels near the blue contours in the column density images.
The horizontal dashed line corresponds to the average per angle bin of 15\,\deg.
The widths of the shaded areas for each histogram correspond to the $\pm 1\,\sigma$ uncertainties related to the histogram binning operation.
Histograms peaking at 0\deg\ correspond to $\vec{B}_{\perp}$ predominantly aligned with iso-\nh\ contours. Histograms peaking at $90$\deg\ and/or $-90$\deg\ correspond to $\vec{B}_{\perp}$ predominantly perpendicular to iso-\nh\ contours.
}
\label{fig:HRO1}
\end{figure*}
\setcounter{figure}{2}
\begin{figure*}[ht!]
%\ContinuedFloat
\centerline{
\includegraphics[height=0.31\textheight,angle=0,origin=c]{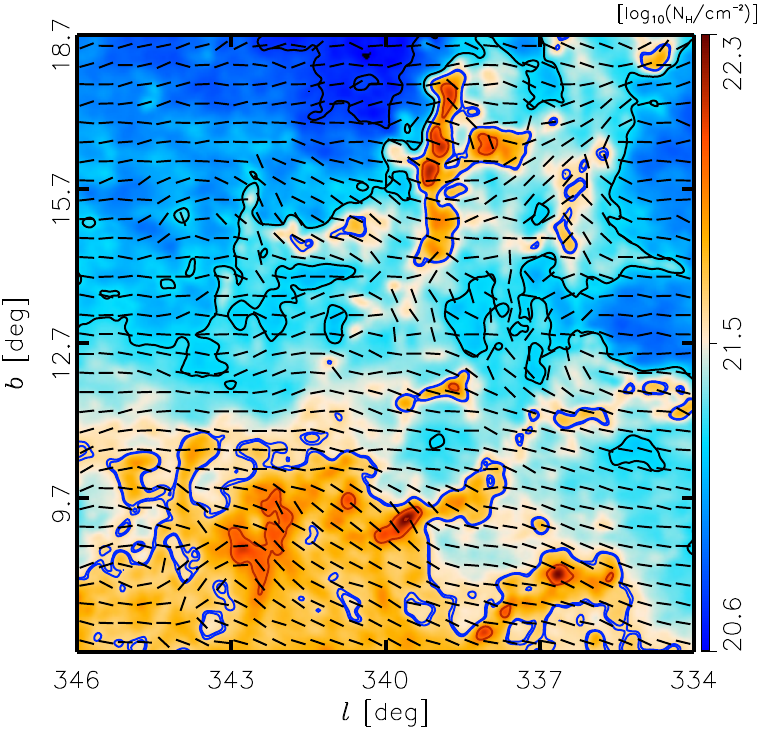}
\includegraphics[height=0.31\textheight,angle=0,origin=c]{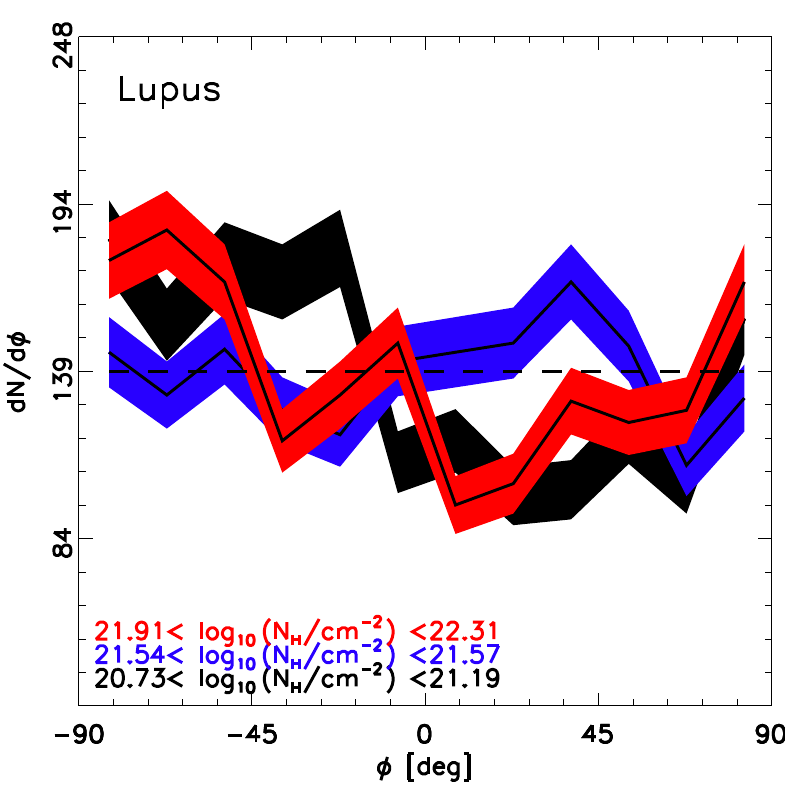}
}
%\vspace{-0.1cm}
\centerline{
\includegraphics[height=0.31\textheight,angle=0,origin=c]{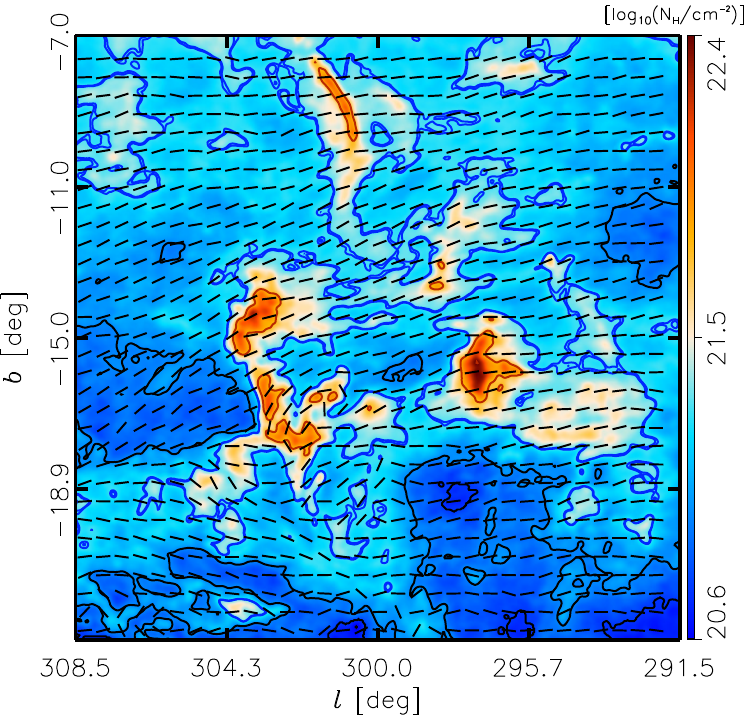}
\includegraphics[height=0.31\textheight,angle=0,origin=c]{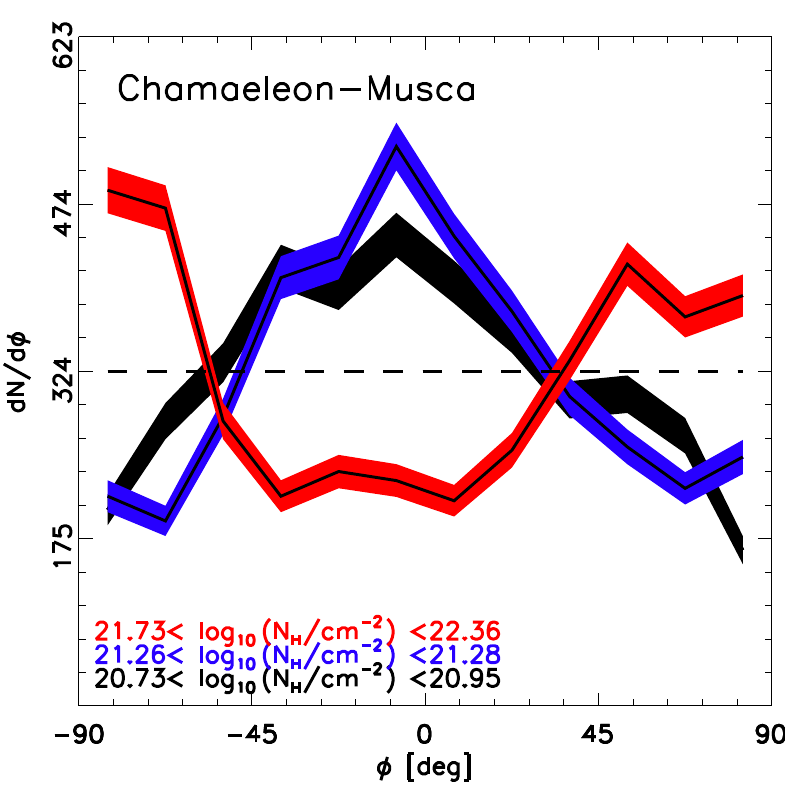}
}
%\vspace{-0.1cm}
\centerline{
\includegraphics[height=0.31\textheight,angle=0,origin=c]{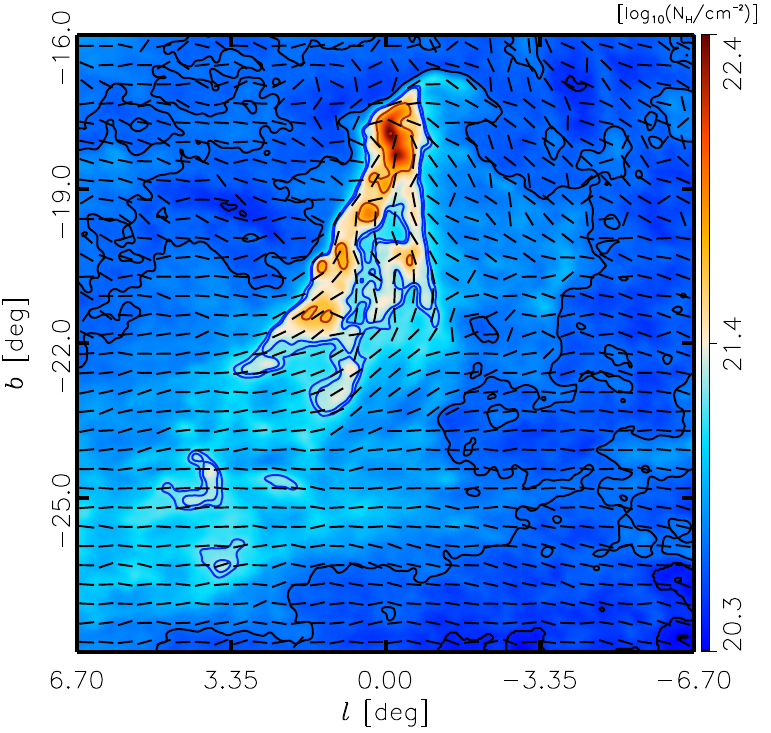}
\includegraphics[height=0.31\textheight,angle=0,origin=c]{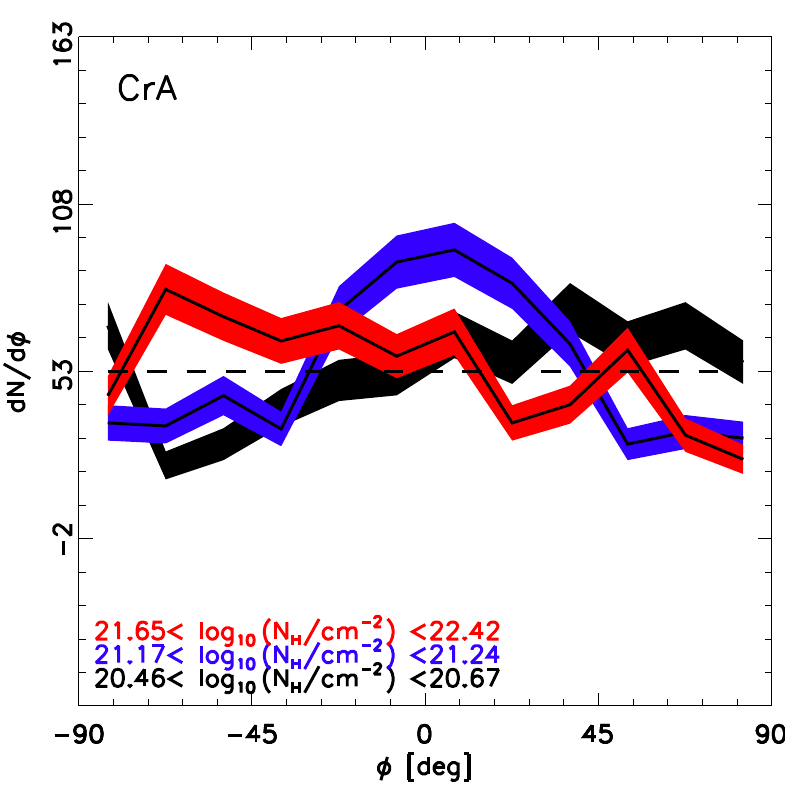}
}
%\vspace{-0.2cm}
\caption[]{(continued).}%\label{fig:HRO2}
\end{figure*}

Among the clouds in the first group (all shown in Fig.~\ref{fig:HRO1}, left column) are Ophiuchus and Lupus, \langed{which are} two regions with different star-forming 
%activity but
\langed{activities but are} close neighbours within an environment disturbed by the Sco-Cen OB association \citep{wilking2008,comeron2008}. 
Chamaeleon-Musca is a region evolving in isolation,
%and
\langed{ and it} is relatively unperturbed \citep{luhman2008}.  
Taurus (see also Fig.~\ref{fig:LICmap}), a cloud with low-mass star formation, appears to be formed by the material swept up by an ancient superbubble centred on the Cas-Tau group \citep{kenyon2008}. 
Finally, CrA is one of the nearest regions with recent intermediate- and low-mass star formation, possibly formed by a high-velocity cloud impact 
%onto
\langed{on} the Galactic plane \citep{neuhauser2008}. 

%%%%%%%%%%%%%%%%%%%%%%%%%%%%%%%%%%%%%%%%%%%%%%%%%%%%%%%%%
%%%%%%%%%%%%%%%%%%%%%%%%%%%%%%%%%%%%%%%%%%%%%%%%%%%%%%%%%
\begin{figure*}[ht!]
\centerline{
\includegraphics[height=0.31\textheight,angle=0,origin=c]{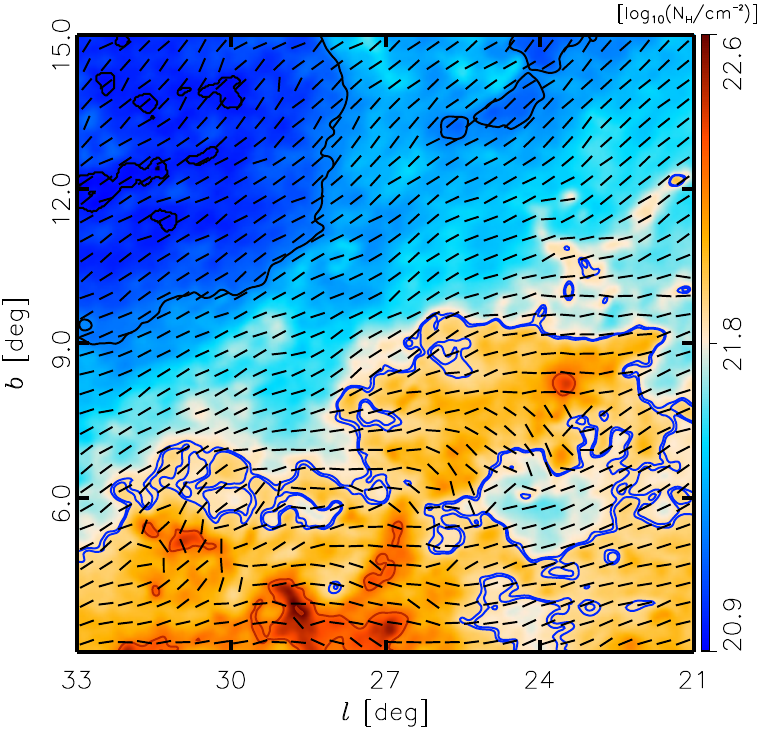}
\includegraphics[height=0.31\textheight,angle=0,origin=c]{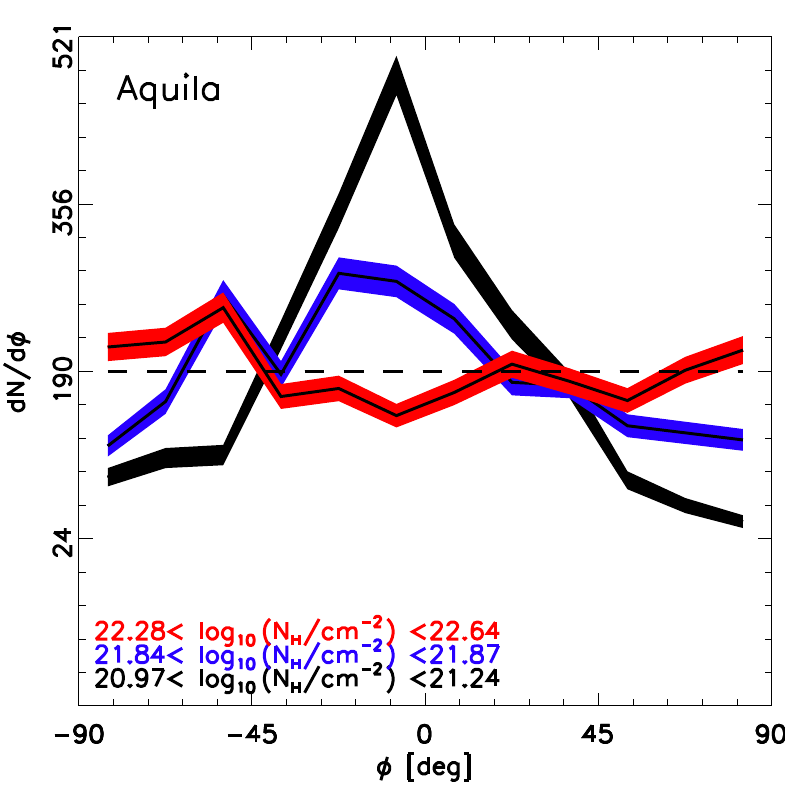}
}
%\vspace{-0.1cm}
\centerline{
\includegraphics[height=0.31\textheight,angle=0,origin=c]{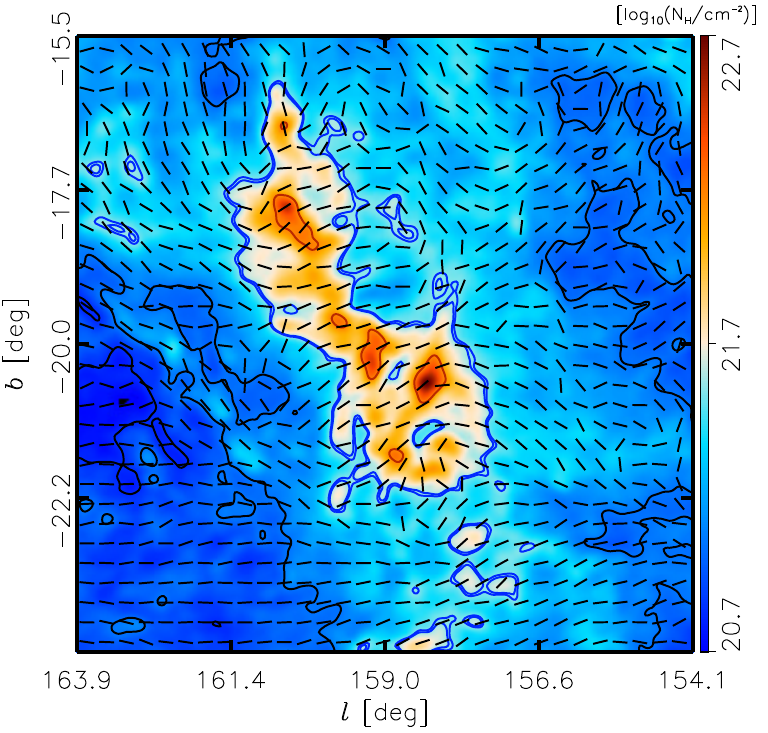}
\includegraphics[height=0.31\textheight,angle=0,origin=c]{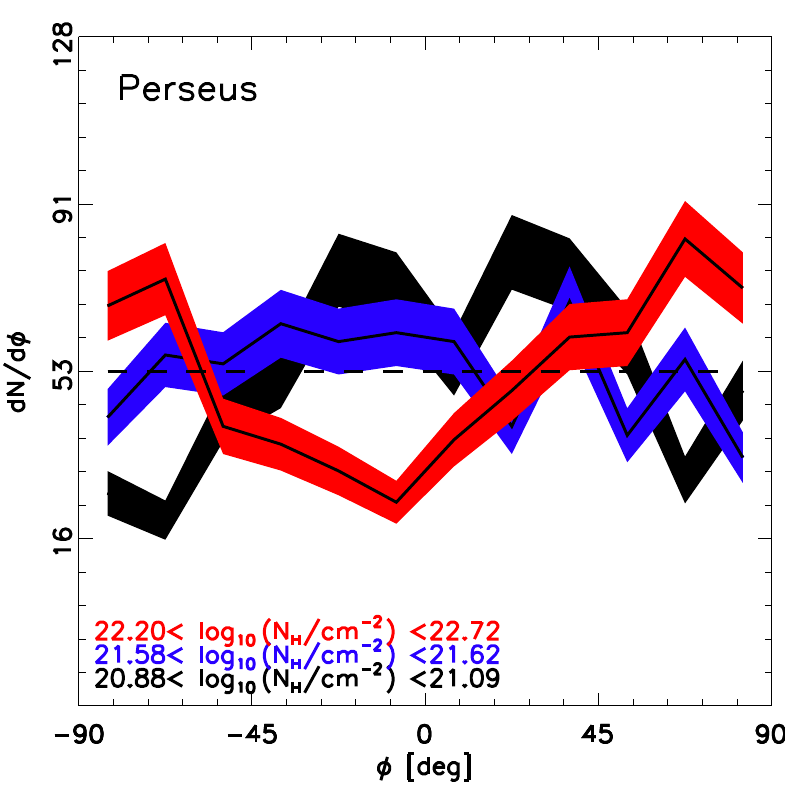}
}
%\vspace{-0.2cm}
\caption{Same as Fig.~\ref{fig:HRO1} for the second group, Aquila Rift and Perseus.}
\label{fig:HRO3}
\end{figure*}

In the second group we consider Aquila Rift and Perseus, shown in Fig.~\ref{fig:HRO3} (left column).
Aquila Rift is a large complex of dark clouds where star formation proceeds in isolated pockets \citep{eiroa2008,prato2008}. 
The Perseus MC is the most active site of on-going star formation within $300\,$pc of the Sun. It features a large velocity gradient and is located close to hot stars that might have impacted its structure \citep{bally2008a}. 

%%%%%%%%%%%%%%%%%%%%%%%%%%%%%%%%%%%%%%%%%%%%%%%%%%%%%%%%%
%%%%%%%%%%%%%%%%%%%%%%%%%%%%%%%%%%%%%%%%%%%%%%%%%%%%%%%%%
\begin{figure*}[ht!]
\centerline{
\includegraphics[height=0.31\textheight,angle=0,origin=c]{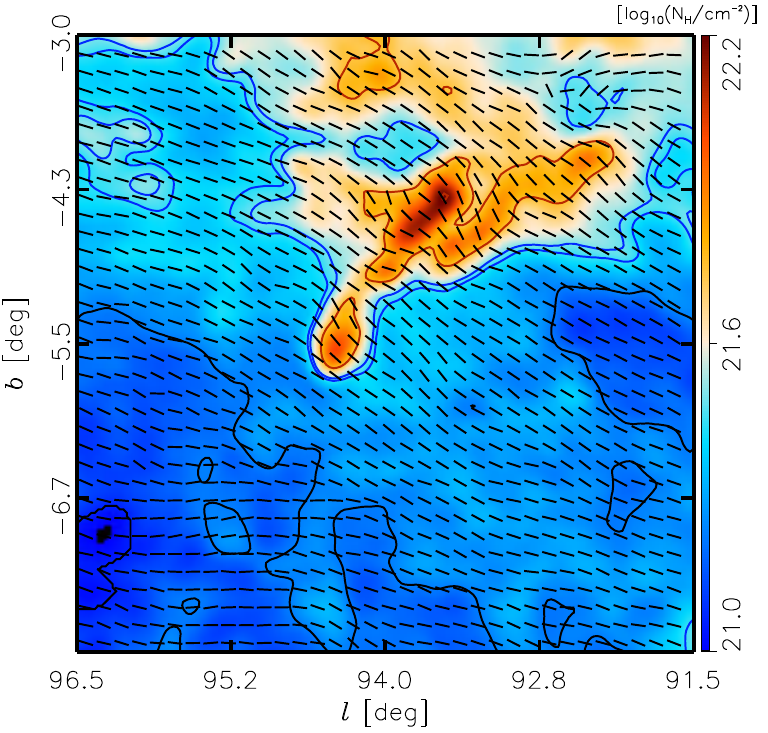}
\includegraphics[height=0.31\textheight,angle=0,origin=c]{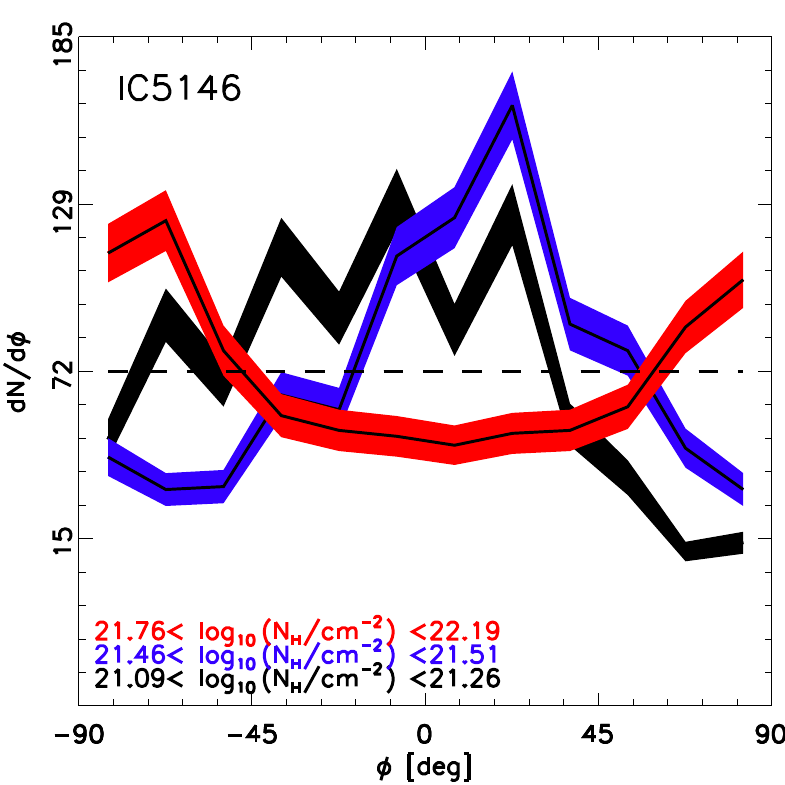}
}
%\vspace{-0.1cm}
\centerline{
\includegraphics[height=0.31\textheight,angle=0,origin=c]{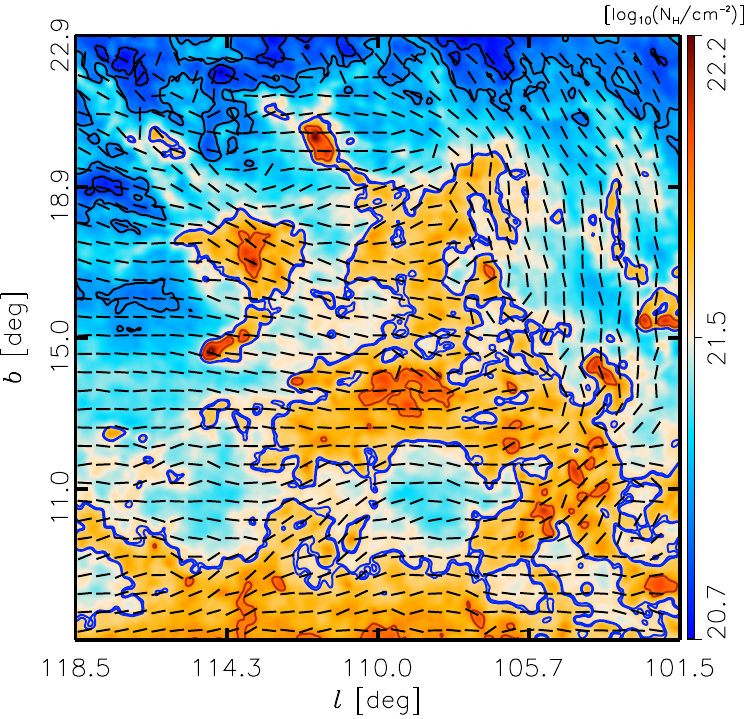}
\includegraphics[height=0.31\textheight,angle=0,origin=c]{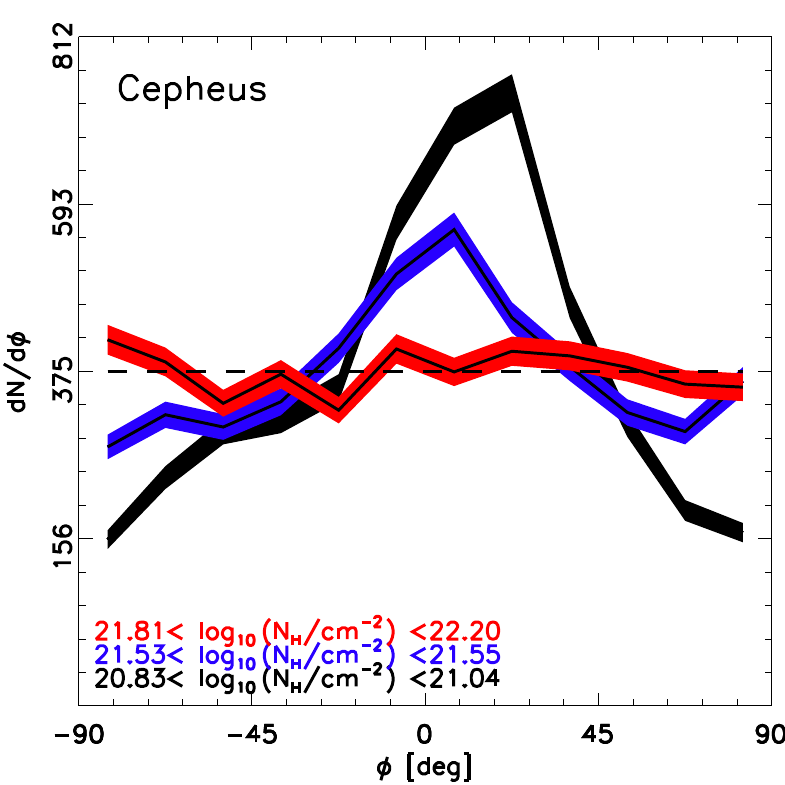}
}
%\vspace{-0.1cm}
\centerline{
\includegraphics[height=0.31\textheight,angle=0,origin=c]{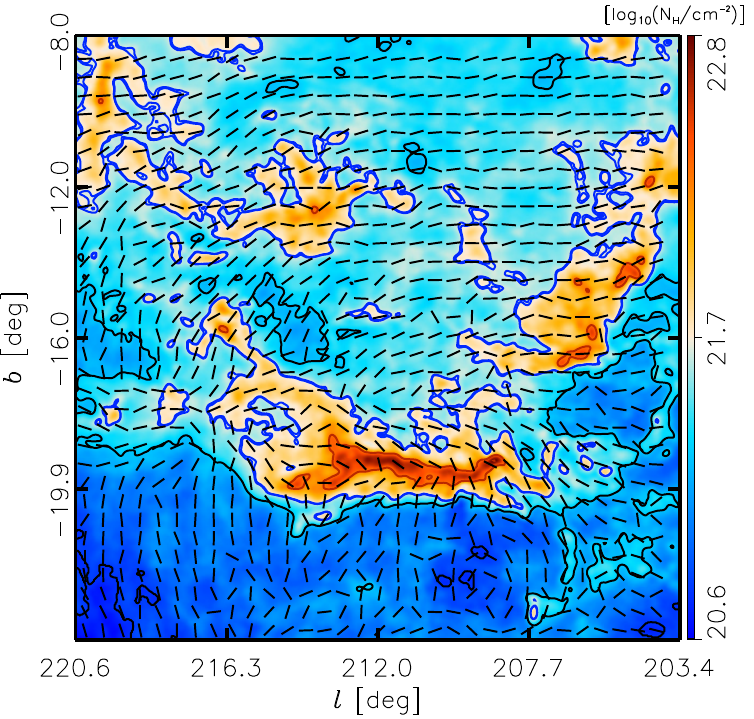}
\includegraphics[height=0.31\textheight,angle=0,origin=c]{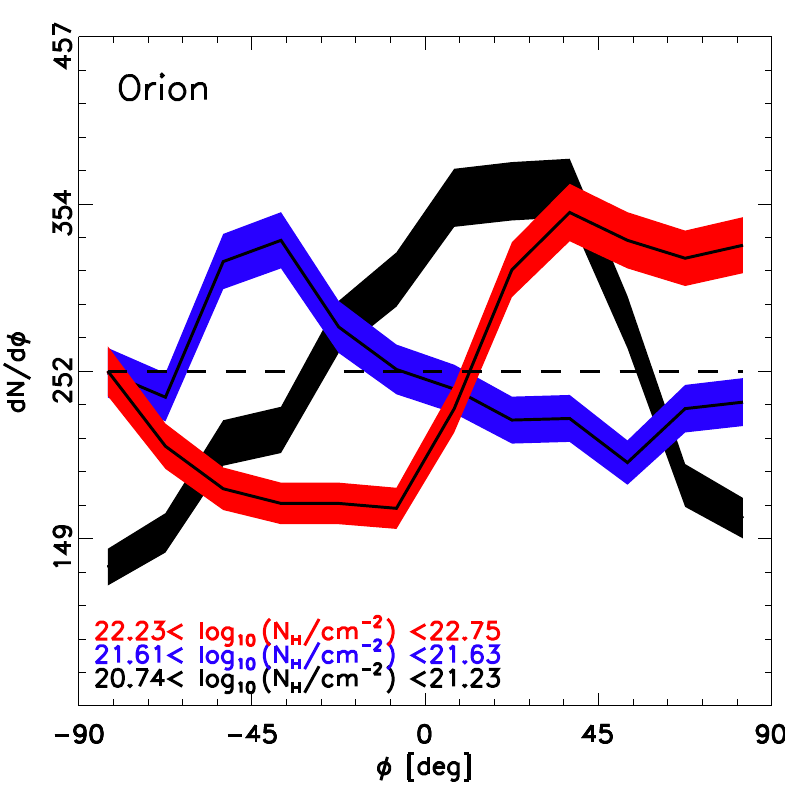}
}
%\vspace{-0.2cm}
\caption{Same as Fig.~\ref{fig:HRO1} for the third group, IC\,5146, Cepheus, and Orion.}
\label{fig:HRO4}
\end{figure*}

In the third group are IC\,5146, Cepheus, and Orion, shown in Fig.~\ref{fig:HRO4} (left column). 
IC\,5146 is an MC complex in Cygnus. It includes an open cluster surrounded by a bright optical nebulosity called the Cocoon nebula, and a region of embedded lower-mass star formation known as the IC\,5146 Northern\footnote{In equatorial coordinates.} Streamer \citep{harvey2008}. 
The Cepheus Flare, called simply Cepheus in this study, is a large complex of dark clouds that seems to belong to an even larger expanding shell from an old supernova remnant \citep{kun2008}. 
Orion is a dark cloud complex with %ongoing
\langed{on-going} high and low mass star formation, whose structure appears to be 
%impacted
\langed{affected} by multiple nearby hot stars \citep{bally2008}.

%Taking into account
\langed{When taking} background/foreground emission and noise within these regions
\langed{into account}, pixels are selected for analysis according to criteria for  the gradient of the column density (Appendix~\ref{appendix:selectionp}) and the polarization (Appendix~\ref{appendix:selectionp}).

% -------------------------------------------------------------------------------------------------------------------------------------------------------
% HRO HRO HRO HRO HRO HRO HRO HRO HRO HRO HRO HRO HRO HRO HRO HRO HRO 
% -------------------------------------------------------------------------------------------------------------------------------------------------------

\section{Statistical study of the relative orientation of the magnetic field and column density structure}\label{section:hro}

\subsection{Methodology}

\subsubsection{Histogram of relative orientations}\label{introhro}

We quantify the relative orientation of the magnetic field with respect to the column density structures using the HRO \citep{soler2013}. The column density structures are characterized by their gradients, which are 
by definition perpendicular to the iso-column density curves (see calculation in Appendix~\ref{appendix:gradtauunc}). The gradient constitutes a vector field that we compare pixel by pixel to the magnetic field orientation inferred from the polarization maps. 

\begin{figure}[ht!]
\vspace{-0.1cm}
\centerline{
\includegraphics[width=0.35\textwidth,angle=0,origin=c]{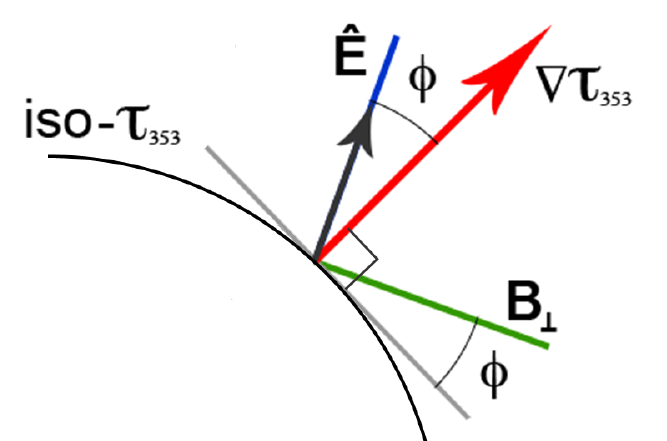}
}
\caption{Schematic of the vectors involved in the calculation of the relative orientation angle $\phi$.}
\label{fig:HROschema}
\end{figure}

In practice we use $\tau_{353}$ as a proxy for \nh\ (Sect.~\ref{columndensity}). The angle $\phi$ 
between $\vec{B}_{\perp}$ and the tangent to the $\tau_{353}$ contours is evaluated using\footnote{In this paper we use the version of $\arctan$ with two signed arguments to resolve the $\pi$ ambiguity in the orientation of \pseudovectors\ \citep{planck2014-XIX}.} 
\begin{equation}\label{eq:hroangle}
\phi = \arctan\left(\,|\nabla\,\tau_{353}\times \vec{\hat{E}}\,| \, , \, \nabla \tau_{353}\cdot \vec{\hat{E}}\,\right)\, ,
\end{equation}
where, as illustrated in Fig.~\ref{fig:HROschema}, $\nabla\,\tau_{353}$ is perpendicular to the tangent of the iso-$\tau_{353}$ curves, the orientation of the unit polarization \langed{pseudo-vector} $\vec{\hat{E}}$, perpendicular to $\vec{B}_{\perp}$, is characterized by the polarization angle
\begin{equation}\label{eq:polangle}
\psi = \frac{1}{2}\arctan(-U,Q)\, ,
\end{equation}
and in Eq.~\eqref{eq:hroangle}, as implemented, the norm actually carries a sign when the range used for $\phi$ is 
%[$-90$\deg, $90$\deg]
\langed{between $-90$\deg\ and $90$\deg}.

The uncertainties in $\phi$ due to the variance of the $\tau_{353}$ map and the noise properties of Stokes $Q$ and $U$ at each pixel are characterized in Appendix~\ref{appendix:hro}.

The gradient technique is one of multiple methods 
%to characterize
\langed{for characterizing} the orientation of structures in a scalar field. Other methods, which include the Hessian matrix analysis \citep{molinari2011,planck2014-XXXII} and the inertia matrix \citep{hennebelle2013a}, are appropriate for measuring the orientation of ridges, i.e., the central regions of filamentary structures. The gradient technique is sensitive to contours and in that sense it is better suited to 
%characterize changes of
\langed{characterizing changes in the} relative orientation in extended regions, not just on the crests of structures \citep{soler2013,planck2014-XXXII}. Additionally, the gradient technique can sample multiple scales by increasing the size of the vicinity of pixels used for its calculation (derivative kernel; see Appendix~\ref{appendix:gradtauunc}). Previous studies that assign 
%a global
\langed{an average} orientation of the cloud \citep{tassis2009,li2013} are equivalent to studying the relative orientation using a derivative kernel close to the cloud size.

The selected pixels belong to the regions of each map where the magnitude of the gradient $|\nabla \tau_{353}|$ is 
%larger
\langed{greater} than in a diffuse reference field (Appendix~\ref{appendix:selection}). This selection criterion aims at separating the structure of the cloud from the structure of the background using the reference field as a proxy. For each region the selected reference field is the region 
%of
\langed{with} the same size and Galactic latitude that has the lowest average \nh. See Appendix~\ref{appendix:selectiong}.

In addition to the selection on $|\nabla \tau_{353}|$, we only consider pixels where the norms of the Stokes $Q$ and $U$ are larger than in the diffuse reference field, therefore minimizing the effect of background/foreground polarization external to the cloud.
The relative orientation angle, $\phi$, is computed by using polarization measurements with a high 
%signal to noise
\langed{S/N} in Stokes $Q$ and $U$, i.e., only considering pixels with $|Q|/\sigma_{Q}$ or $|U|/\sigma_{U} > 3$. This selection allows the unambiguous definition of $\vec{\hat{E}}$ by constraining the uncertainty in the polarization angle (see Appendix~\ref{appendix:selectionp}).

Once we have produced a map of relative orientations for selected pixels following Eq.~\eqref{eq:hroangle}, we divide the map into bins of \nh\ containing an equal number of pixels and generate a histogram of $\phi$ for each bin. The shape of the histogram is used to evaluate 
%directly the preferential 
\langed{the preferred} relative orientation in each bin \langed{directly}. A concave histogram, peaking at 0\deg, corresponds to 
%preferential
\langed{the preferred} alignment of $\vec{B}_{\perp}$ with the \nh\ contours. A convex histogram, peaking at $90$\deg\ and/or $-90$\deg, corresponds to 
%preferential
\langed{the preferred} orientation of $\vec{B}_{\perp}$ perpendicular to the \nh\ contours. 
 
The HROs in each region are computed in 25 \nh\ bins having equal numbers of selected pixels (10 bins in two regions with fewer pixels, CrA and IC\,5146).
The number of \nh\ bins is determined by requiring enough bins to resolve the highest \nh\ regions and at the same time maintaining enough pixels per \nh\ bin to obtain significant statistics from each histogram. The typical number of pixels per bin of \nh\ ranges from approximately 600 in CrA to around 4\,000 in Chamaeleon-Musca. We use 12 angle bins of width 15\,\deg.

The HROs of the first group of regions, the nearest at $d\approx150\,$pc, are shown in the 
%right
\langed{right-hand} column of Fig.~\ref{fig:HRO1}. For the sake of clarity, we 
%present only the histograms corresponding
\langed{only present the histograms that correspond} to three bins, namely the lowest and highest \nh\ and an intermediate \nh value. The intermediate bin is the 12th (sixth in two regions with fewer pixels, CrA and IC\,5146), and \langed{it} corresponds to pixels near the blue contour in the image in the 
%left
\langed{left-hand} column of Fig.~\ref{fig:HRO1}. The widths of the shaded areas for each histogram correspond to the 1\,$\sigma$ uncertainties related to the histogram binning operation, which are greater than the uncertainties produced by the variances of $Q$, $U$, and $\tau_{353}$ (Appendix~\ref{appendix:hro}). 
The sharp and narrow features (``jitter") in the HROs are independent of these variances. They are the product of sampling the spatial correlations in the magnetic field over a finite region of the sky together with the histogram binning; these features average out when evaluating the relative orientation over larger portions of the sky \citep{planck2014-XXXII}. 

Although often asymmetric, most histograms reveal a change in the 
%preferential
\langed{preferred} relative orientation across \nh\ bins. 
The most significant feature in the HROs of Taurus, Ophiuchus, and Chamaeleon-Musca is the drastic change in relative orientation from parallel in the lowest \nh\ bin to perpendicular in the highest \nh\ bin. In Lupus the behaviour at low \nh\ is not clear, but at high \nh\ \langed{it} is clearly perpendicular. In contrast, CrA 
%shows
\langed{tends to show} $\vec{B}_{\perp}$ 
%preferentially
\langed{as} parallel in the intermediate \nh\ bin, but no 
%preferential
\langed{preferred} orientation in the other \nh\ bins. 

The HROs of the clouds located at $d\approx300\,$pc, Aquila Rift and Perseus, are shown in the 
%right
\langed{right-hand} column of Fig.~\ref{fig:HRO3}. They indicate that the relative orientation is 
%preferentially
\langed{usually} parallel in the lowest \nh\ bins and 
%preferentially 
perpendicular in the highest \nh\ bins.

The HROs of the third group, located at $d\approx400-450\,$pc, IC\,5146, Cepheus, and Orion, are presented in the 
%right
\langed{right-hand} column of Fig.~\ref{fig:HRO4}. In both IC\,5146 and Orion the HROs for the highest \nh\ bins reveal a 
%preferential
\langed{preferred} orientation of the field perpendicular to the \nh\ contours (Orion is quite asymmetric), whereas the HROs corresponding to the low and intermediate \nh\ bins reveal %preferential
\langed{a preferred} alignment of the field with \nh\ structures. This trend is also present, but less pronounced, in the Cepheus region.

%%%%% figure on shape parameter, test regions, table

\begin{figure*}[ht!]
\centerline{
\includegraphics[width=0.5\textwidth,angle=0,origin=c]{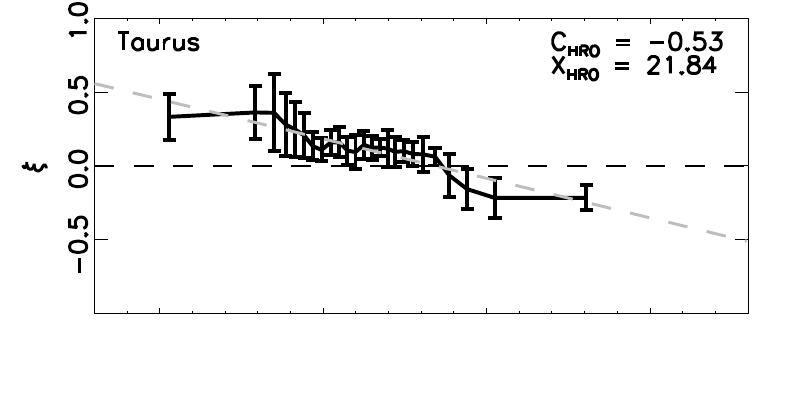}
\hspace{-0.4cm}
\includegraphics[width=0.5\textwidth,angle=0,origin=c]{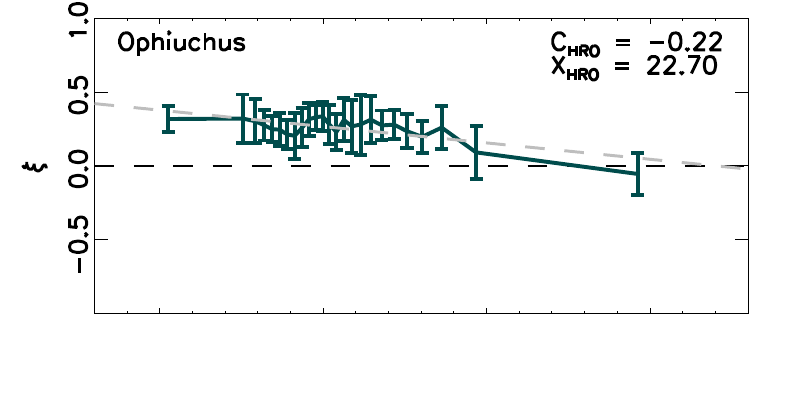}
}
\vspace{-1.1cm}
\centerline{
\includegraphics[width=0.5\textwidth,angle=0,origin=c]{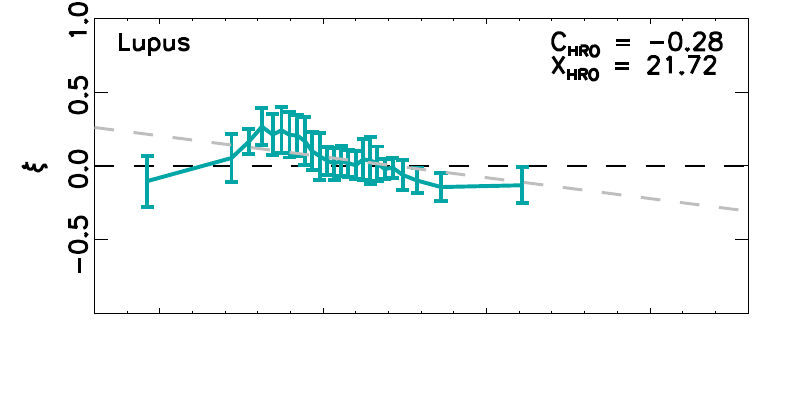}
\hspace{-0.4cm}
\includegraphics[width=0.5\textwidth,angle=0,origin=c]{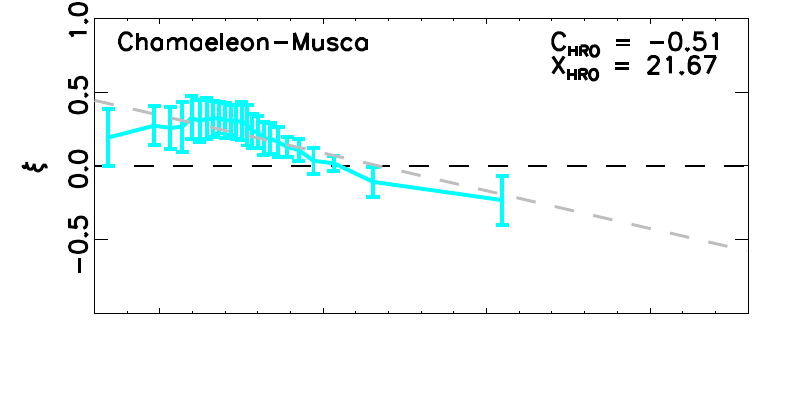}
}
\vspace{-1.1cm}
\centerline{
\includegraphics[width=0.5\textwidth,angle=0,origin=c]{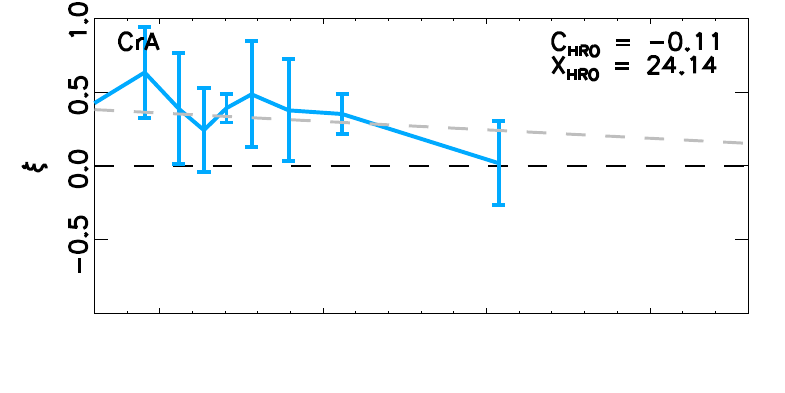}
\hspace{-0.4cm}
\includegraphics[width=0.5\textwidth,angle=0,origin=c]{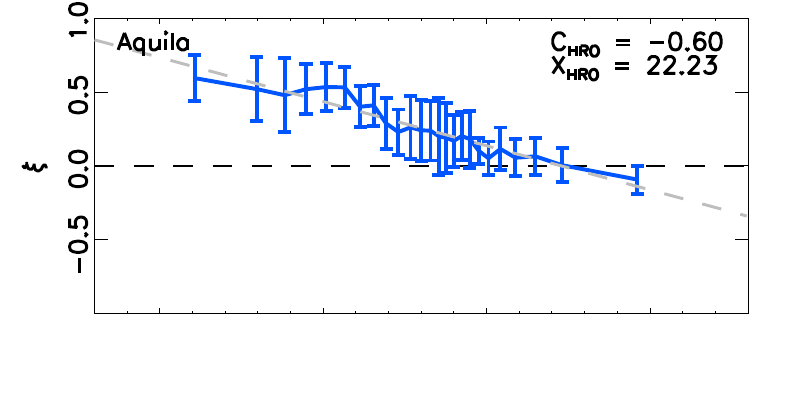}
}
\vspace{-1.1cm}
\centerline{
\includegraphics[width=0.5\textwidth,angle=0,origin=c]{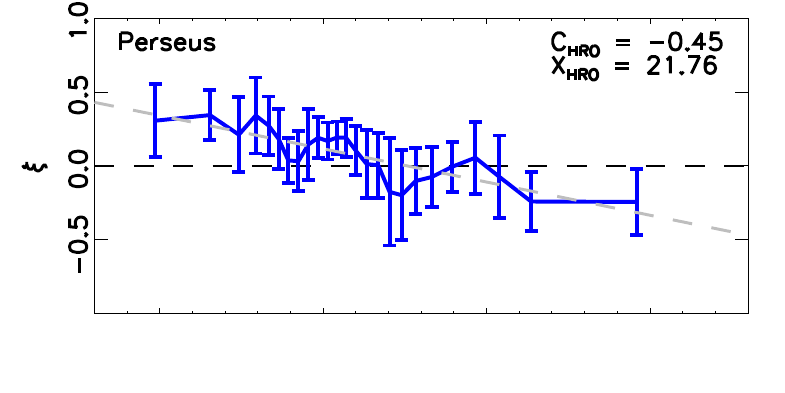}
\hspace{-0.4cm}
\includegraphics[width=0.5\textwidth,angle=0,origin=c]{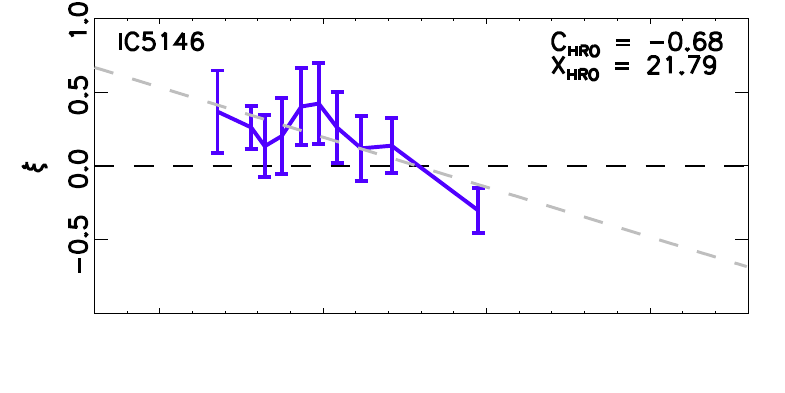}
}
\vspace{-1.1cm}
\centerline{
\includegraphics[width=0.5\textwidth,angle=0,origin=c]{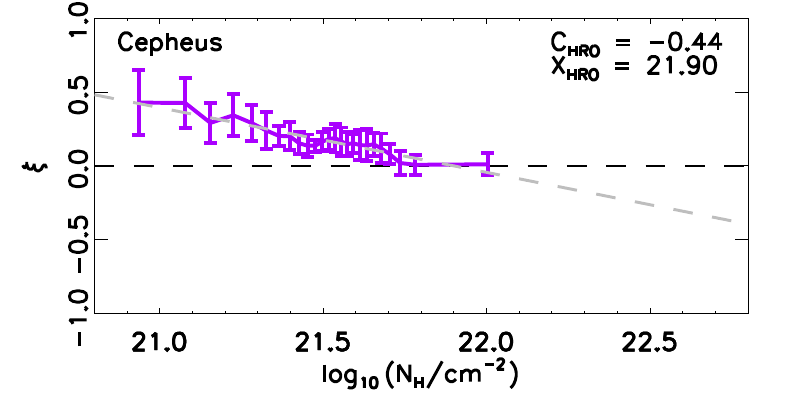}
\hspace{-0.4cm}
\includegraphics[width=0.5\textwidth,angle=0,origin=c]{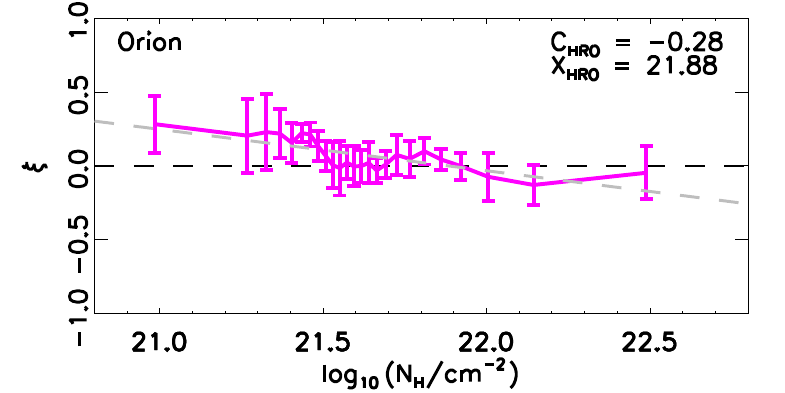}
}
\vspace{-0cm}
\caption{Histogram shape parameter $\zeta$ (Eqs.~\ref{eq:zeta} and \ref{eq:errzeta}) calculated for the different \nh\ bins in each region. 
The cases $\zeta > 0$ and $\zeta < 0$ correspond to the magnetic field oriented 
%preferentially
\langed{mostly} parallel and perpendicular to the structure contours, respectively.
For $\zeta \approx 0$ there is no preferred orientation.
The displayed values of $C_{\textsc{HRO}}$ and $X_{\textsc{HRO}}$ were calculated from Eq.~\eqref{eq:hrofit} and correspond to the grey dashed line in each plot.}
\label{fig:HROzeta}
\end{figure*}

\begin{figure*}[ht!]
\centerline{
\includegraphics[height=0.31\textheight,angle=0,origin=c]{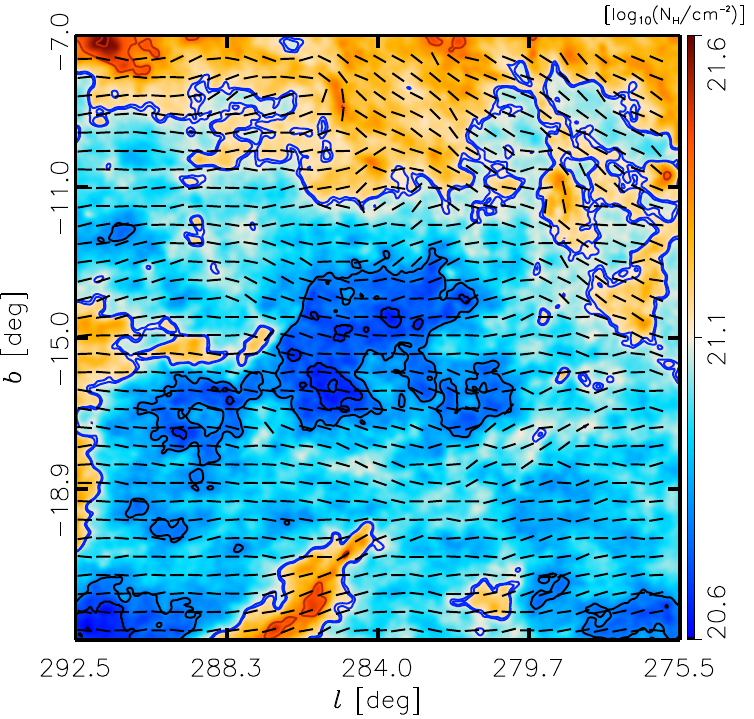}
\includegraphics[height=0.31\textheight,angle=0,origin=c]{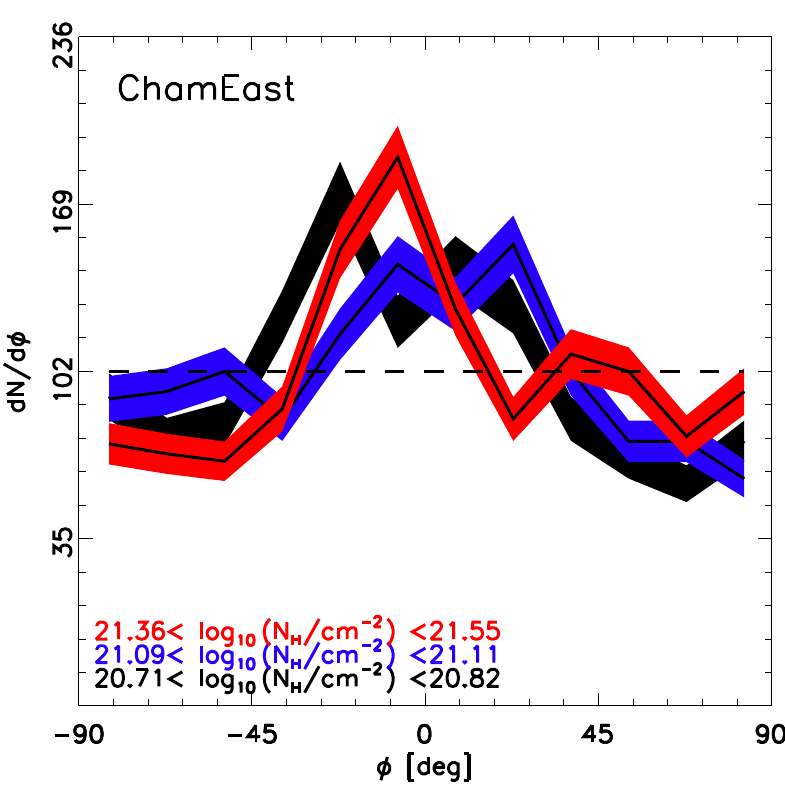}
}
\vspace{-0.1cm}
\centerline{
\includegraphics[height=0.31\textheight,angle=0,origin=c]{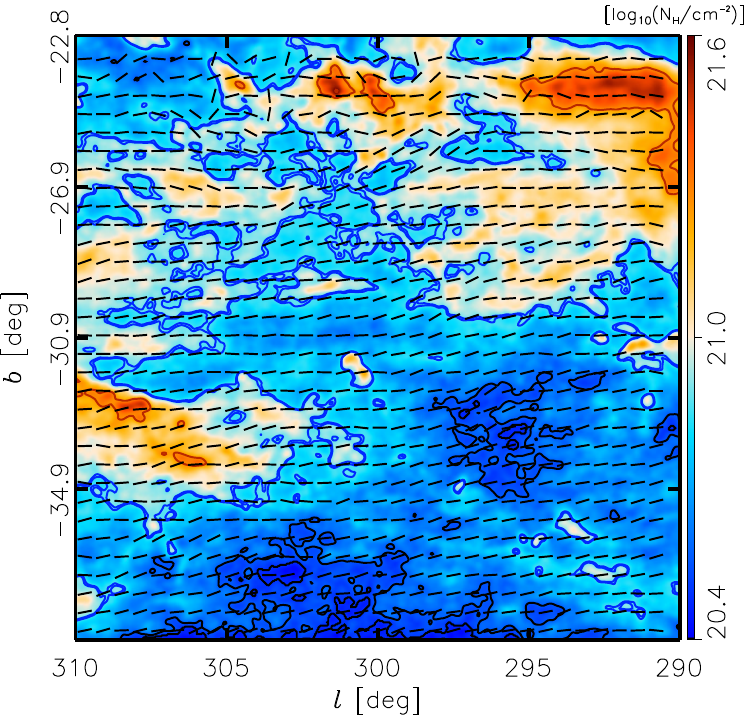}
\includegraphics[height=0.31\textheight,angle=0,origin=c]{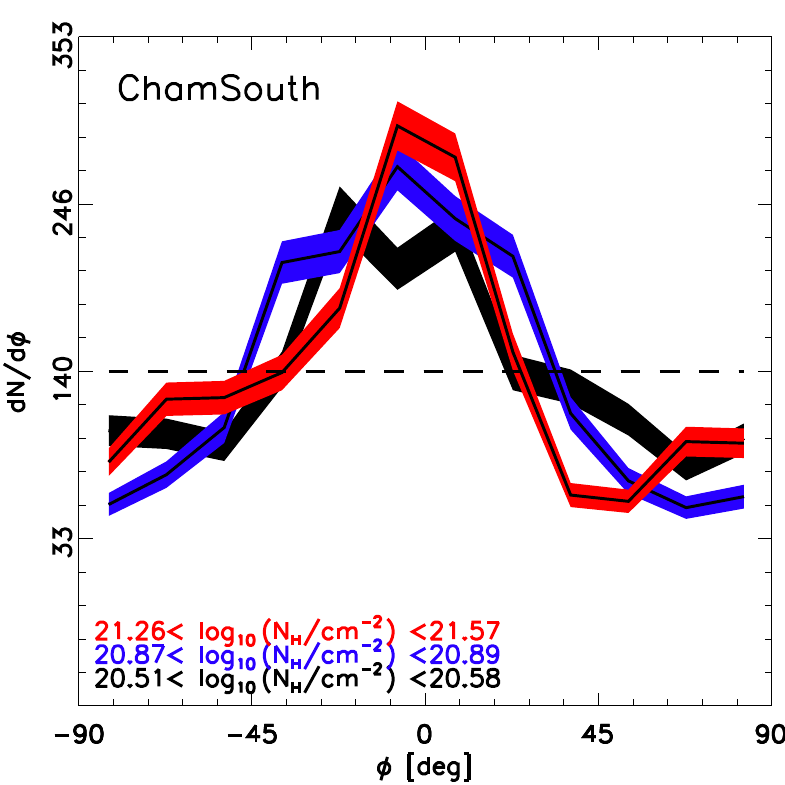}
}
\vspace{-0.2cm}
\caption{Same as Fig.~\ref{fig:HRO1} for the two test regions located directly east (ChamEast) and south of Chamaeleon-Musca (ChamSouth).}
\label{fig:HROtest}
\end{figure*}

\begin{table}[tmb]  % table* is a two-column table.  Drop the * for one column.
\begingroup
\newdimen\tblskip \tblskip=5pt
\caption{Fit of $\zeta$ vs.\ \lognh.$^{a}$}
\label{table-zeta}                            % Label goes here.
\nointerlineskip
\vskip -3mm
\footnotesize
\setbox\tablebox=\vbox{
   \newdimen\digitwidth 
   \setbox0=\hbox{\rm 0} 
   \digitwidth=\wd0 
   \catcode`*=\active 
   \def*{\kern\digitwidth}
   \newdimen\signwidth 
   \setbox0=\hbox{+} 
   \signwidth=\wd0 
   \catcode`!=\active 
   \def!{\kern\signwidth}
\halign{\hbox to 1.15in{#\leaderfil}\tabskip 2.2em&
\hfil#&\hfil#\tabskip 0pt\cr
\noalign{\doubleline}
\omit\hfil Region\hfil & \hfil$C_{\textsc{HRO}}$\hfil &  \hfil$X_{\textsc{HRO}}$ \hfil \cr
%\omit & \hfil[10$^{20}$\,cm$^{-2}$]\hfil & \hfil[\%]\hfil \cr
\noalign{\vskip 4pt\hrule\vskip 6pt}
%----------------------------------------------------------------------------------------------------------------
Taurus & \hfil$-0.53$\hfil & \hfil21.84\hfil \cr
%----------------------------------------------------------------------------------------------------------------
Ophiuchus & \hfil$-0.22$\hfil & \hfil22.70\hfil \cr
%----------------------------------------------------------------------------------------------------------------
Lupus & \hfil$-0.28$\hfil & \hfil21.72\hfil \cr
%----------------------------------------------------------------------------------------------------------------
Chamaeleon-Musca & \hfil$-0.51$\hfil & \hfil21.67\hfil \cr
%----------------------------------------------------------------------------------------------------------------
Corona Australia (CrA) & \hfil$-0.11$\hfil & \hfil24.14\hfil \cr
%----------------------------------------------------------------------------------------------------------------
\noalign{\vskip 4pt\hrule\vskip 6pt}
%----------------------------------------------------------------------------------------------------------------
Aquila Rift & \hfil$-0.60$\hfil & \hfil22.23\hfil \cr
%----------------------------------------------------------------------------------------------------------------
Perseus & \hfil$-0.45$\hfil & \hfil21.76\hfil \cr
%----------------------------------------------------------------------------------------------------------------
\noalign{\vskip 4pt\hrule\vskip 6pt}
IC\,5146 & \hfil$-0.68$\hfil & \hfil21.79\hfil  \cr
%---------------------------------------------------------------------------------------------------------------
Cepheus &  \hfil$-0.44$\hfil & \hfil21.90\hfill \cr
%----------------------------------------------------------------------------------------------------------------
Orion &  \hfil$-0.28$\hfil & \hfil21.88\hfil \cr
%----------------------------------------------------------------------------------------------------------------
\noalign{\vskip 3pt\hrule\vskip 4pt}}}
\endPlancktable                    % ends one-column \halign
%\endPlancktablewide                 % ends two-column \halign
\tablenote a See Eq.~\ref{eq:hrofit} and Fig.~\ref{fig:HROzeta}.\par
\endgroup
\end{table}  

\subsubsection{Histogram shape parameter $\zeta$}\label{section:zeta}

The changes in the HROs are quantified using the histogram shape parameter $\zeta$, defined as
\begin{equation}\label{eq:zeta}
\zeta = \frac{A_{\rm c}-A_{\rm e}}{A_{\rm c}+A_{\rm e}}\, ,
\end{equation}
where $A_{\rm c}$ is the area in the centre of the histogram ($-22\pdeg5< \phi < 22\pdeg5$) and $A_{\rm e}$ 
%is 
the area in the extremes of the histogram ($-90\pdeg0 < \phi < -67\pdeg5$ and $67\pdeg5 < \phi < 90\pdeg0$).   
The value of $\zeta$, the result of the integration of the histogram over 45\deg\ ranges, is independent of the number of bins selected to represent the histogram if the bin widths are smaller than the integration range.

A concave histogram corresponding to $\vec{B}_{\perp}$ 
%preferentially
\langed{mostly} aligned with \nh\ contours would have $\zeta > 0$.
A convex histogram corresponding to $\vec{B}_{\perp}$ 
%preferentially
\langed{mostly} perpendicular to \nh\ contours would have $\zeta < 0$.
A flat histogram corresponding to no 
%preferential
\langed{preferred} relative orientation would have $\zeta \approx 0$ .

The uncertainty in $\zeta$,  $\sigma_{\zeta}$, is obtained from
\begin{equation}\label{eq:errzeta}
\sigma^{2}_{\zeta} = \frac{4\,(A^{2}_{\rm e}\sigma^{2}_{A_{\rm c}}+A^{2}_{\rm c}\sigma^{2}_{A_{\rm e}})}{(A_{\rm c}+A_{\rm e})^{4}}\, .
\end{equation}
The variances of the areas, $\sigma^{2}_{A_{\rm e}}$ and $\sigma^{2}_{A_{\rm c}}$, characterize the jitter of the histograms. If the jitter is large, $\sigma_{\zeta}$ is large compared to $| \zeta |$ and the relative orientation is indeterminate. 
The jitter depends on the number of bins in the histogram, but $\zeta$ does not. 

Figure~\ref{fig:HROzeta} illustrates the change in $\zeta$ as a function of \lognh\ of each bin.  For most of the clouds, $\zeta$ is positive in the lowest and intermediate \nh\ bins and negative or close to zero in the highest bins.
The most pronounced changes in $\zeta$ from positive to negative are seen in Taurus, Chamaeleon-Musca, Aquila Rift, Perseus, and IC\,5146.

The trend in $\zeta$ vs.\ \lognh\ can be fit roughly by a linear relation
\begin{equation}\label{eq:hrofit}
\zeta =  C_{\textsc{HRO}}\,[\log_{10}(\nhd/{\rm cm}^{-2}) - X_{\textsc{HRO}}]\, .
\end{equation}
The values of $C_{\textsc{HRO}}$ and $X_{\textsc{HRO}}$ in the regions analysed are summarized in Table~\ref{table-zeta}. 
For the clouds with a pronounced change in relative orientation the slope $C_{\textsc{HRO}}$ is steeper than about $-0.5$\langed{,} 
and the value $X_{\textsc{HRO}}$ for the \lognh\ at which the relative orientation changes from parallel to perpendicular is greater than about 21.7.
Ophiuchus, Lupus, Cepheus, and Orion are intermediate cases, where $\zeta$ 
%does not definitely 
\langed{definitely does not} go negative in the data, but seems to \langed{do so} by extrapolation; these tend to have a shallower $C_{\textsc{HRO}}$ and/or a higher $X_{\textsc{HRO}}$. 

The least pronounced change in $\zeta$ is seen in CrA, where $\zeta$ is consistently positive in all bins\langed{,} and the slope is very flat.
We applied the HRO analysis to a pair of test regions (Fig.~\ref{fig:HROtest}) with even lower \nh\ values ($\log_{10}(\nhd/\mbox{cm}^{-2}) < 21.6$; see also 
%blue points in 
Fig.~\ref{fig:HROsims} below) located directly south and directly east of the Chamaeleon-Musca region. 
%Like
\langed{As} in CrA, we find 
%a 
%preferential
%\langed{preferred} parallel alignment of $\vec{B}_{\perp}$ 
\langed{that $\vec{B}_{\perp}$ is mostly} parallel to the \nh\ contours, a fairly constant $\zeta$, and no indication of 
%preferentially 
\langed{predominantly} perpendicular relative orientation. 

% -------------------------------------------------------------------------------------------------------------------------------------------------------------
\subsection{Comparisons with previous studies}

The above trends in relative orientation between $\vec{B}_{\perp}$ and the \nh\ contours in targeted MCs, 
%with
\langed{where} $\vec{B}_{\perp}$ \langed{tend to become}
%becoming
% preferentially
perpendicular to the \nh\ contours at high \nh, 
%are in agreement
\langed{agree} with the results of the Hessian matrix analysis applied to \Planck\ observations over the whole sky, as reported in Fig.~15 of \cite{planck2014-XXXII}.

Evidence for preferential orientations in sections of some regions included in this study has been reported previously.
% TAURUS
The Taurus region has been the target of many studies.  \cite{moneti1984} and \cite{chapman2011} 
%found
\langed{find} evidence using infrared polarization of background stars for a homogeneous magnetic field perpendicular to the embedded dense filamentary structure. 
High-resolution submillimetre observations of intensity with \Herschel\ find evidence 
%for
\langed{of} faint filamentary structures (``striations"), \langed{which are} well correlated with the magnetic field orientation inferred from starlight polarization \citep{palmeirim2013} and perpendicular to the filament B211.  \cite{heyer2008} 
%reported
\langed{report} a velocity anisotropy aligned with the magnetic field, which can be interpreted as evidence of the channeling effect of the magnetic fields.
But  %note that 
the magnetic field in B211 and the dense filamentary structures are not measured directly.  
As described above, using \Planck\ polarization we find that $\vec{B}_{\perp}$ is \preferentially aligned with the lowest \nh\ contours ($20.8 < \log_{10}(\nhd/\mbox{cm}^{-2}) < 21.3$) (see also \citealp{planck2014-XXXIII}), although the aligned structures do not correspond to the striations, which are not resolved at 10\arcmin\ resolution.  However, we also find that at higher \nh\ the relative orientation becomes perpendicular.

Similar studies in other regions have found evidence of striations correlated with the starlight-inferred magnetic field orientation and perpendicular to the densest filamentary structures. The regions studied include Serpens South \citep{sugitani2011}, which is part of Aquila Rift in this study, Musca \citep{pereyra2004}, and the Northern Lupus cloud \citep{matthews2014}.  By studying \Planck\ polarization in larger regions around the targets of these previous observations, we show that a systematic change in relative orientation is the prevailing statistical trend in clouds that reach \lognh\,$\gtrsim 21.5$.

% ----------------------------------------------------------------------------------------------------------------------------------------------------------------
% DISCUSSION DISCUSSION DISCUSSION DISCUSSION DISCUSSION DISCUSSION DISCUSSION DISCUSSION  
% ----------------------------------------------------------------------------------------------------------------------------------------------------------------

\section{Discussion}\label{section:discussion}

\begin{figure*}[ht!]
\vspace{-0.2cm}
\centerline{
\includegraphics[height=0.31\textheight,angle=0,origin=c]{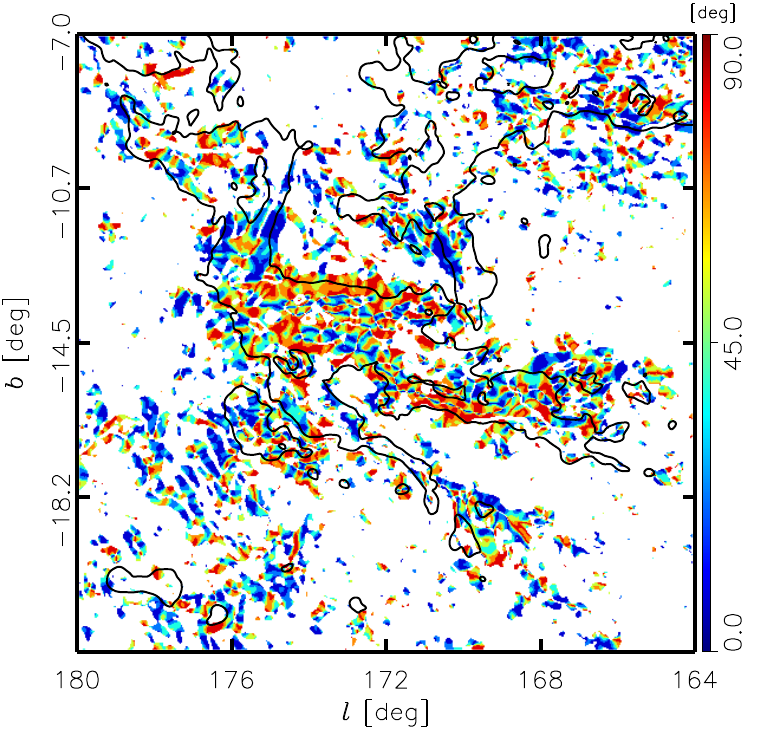}
\includegraphics[height=0.31\textheight,angle=0,origin=c]{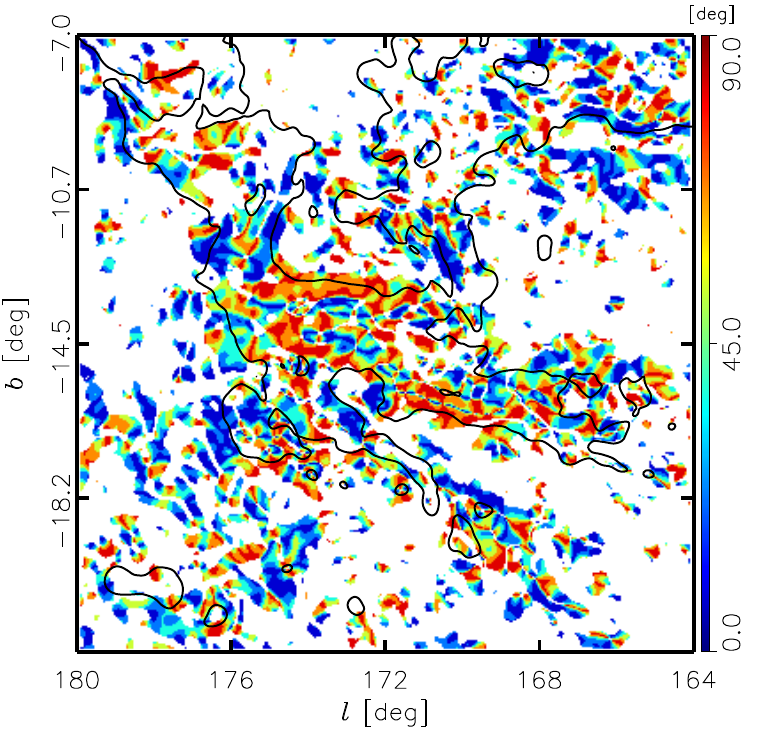}
}
\hspace{-0.2cm}
\centerline{
\includegraphics[height=0.31\textheight,angle=0,origin=c]{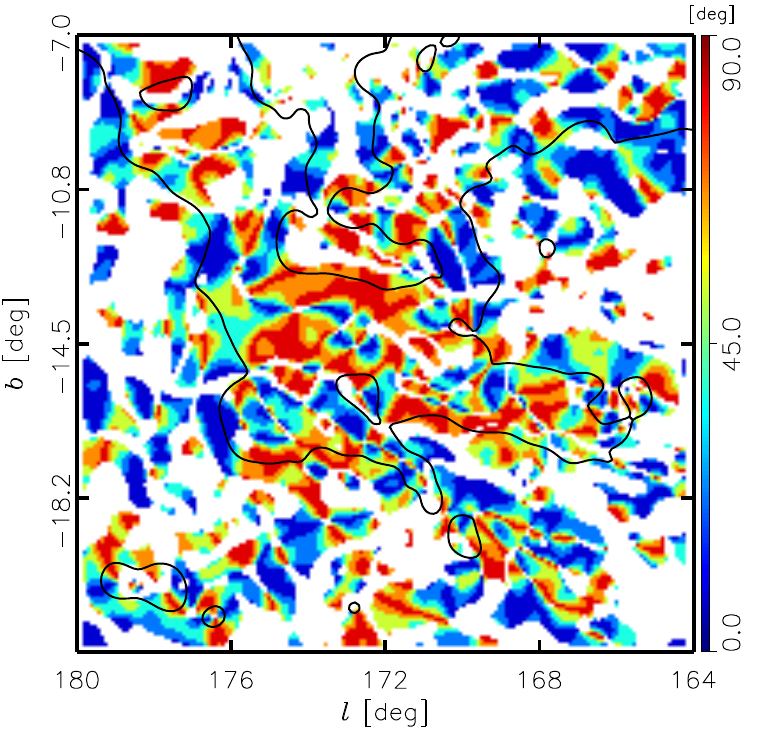}
\includegraphics[height=0.31\textheight,angle=0,origin=c]{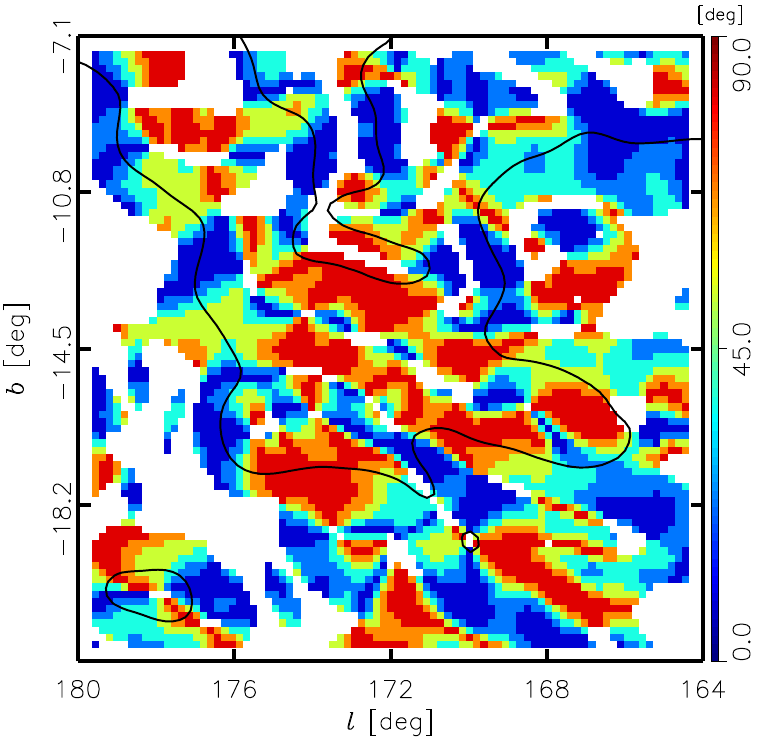}
}
\vspace{-0.1cm}
\caption{Maps of the absolute value of the relative orientation angle, $|\phi|$, in the Taurus region. These maps are produced after smoothing the input maps to beam FWHMs of 10\arcm, 15\arcm, 30\arcm, and 60\arcm\ and then resampling the grid to sample each beam FWHM with the same number of pixels. The regions in red correspond to $\vec{B}_{\perp}$ close to perpendicular to \nh\ structures. The regions in blue correspond to $\vec{B}_{\perp}$ close to parallel to \nh\ structures. The black contour, corresponding to the \nh\ value of the intermediate contour introduced in Fig.~\ref{fig:HRO1}, provides a visual reference to the cloud structure.}
\label{fig:anglemap}
\end{figure*}

\begin{figure*}[ht!]
\centerline{
\includegraphics[width=0.33\textwidth,angle=0,origin=c]{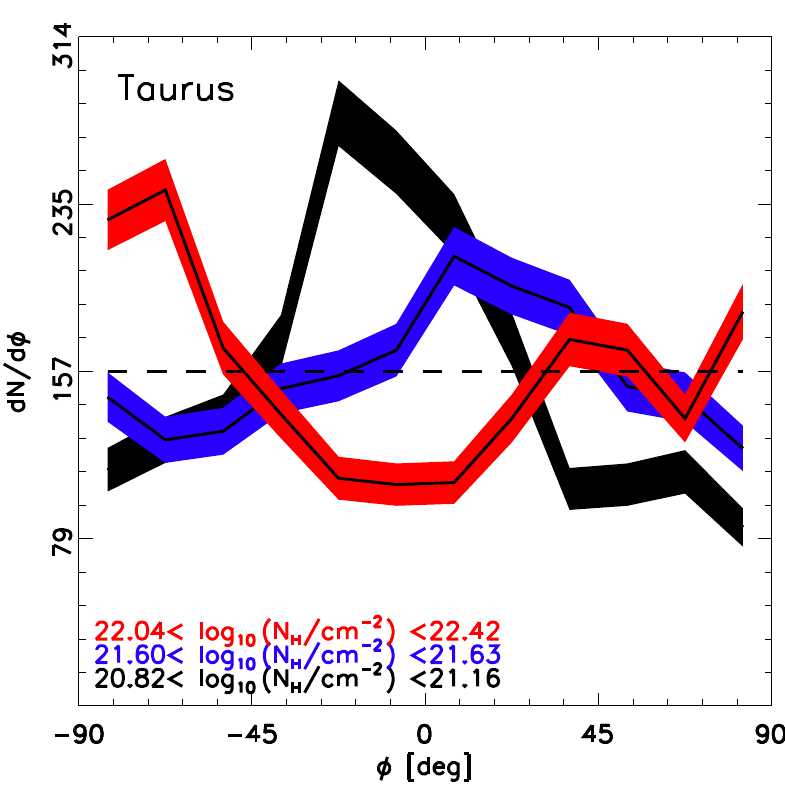}
\hspace{-0.2cm}
\includegraphics[width=0.33\textwidth,angle=0,origin=c]{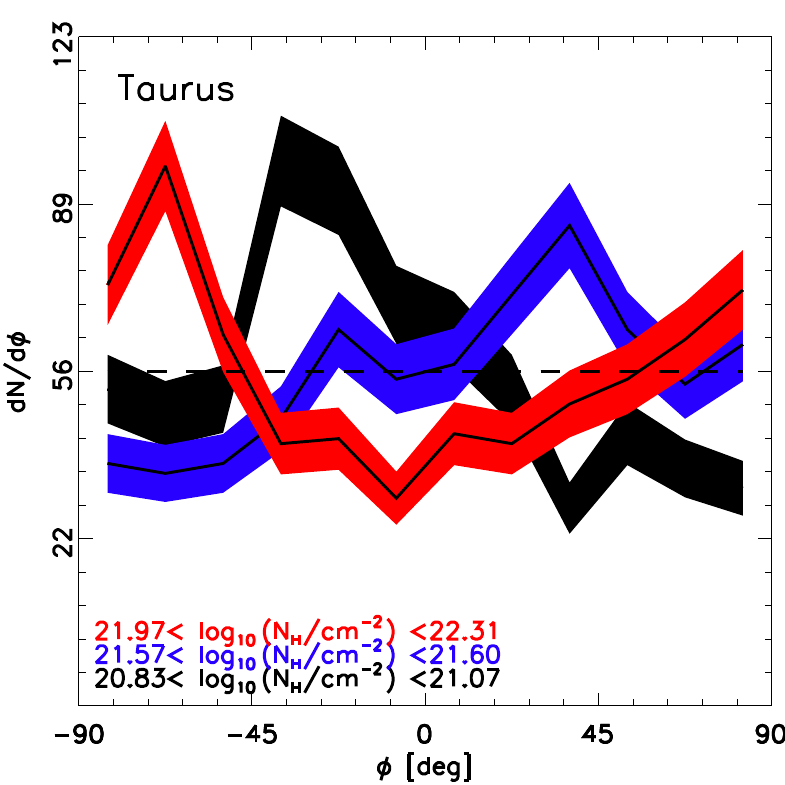}
\hspace{-0.2cm}
\includegraphics[width=0.33\textwidth,angle=0,origin=c]{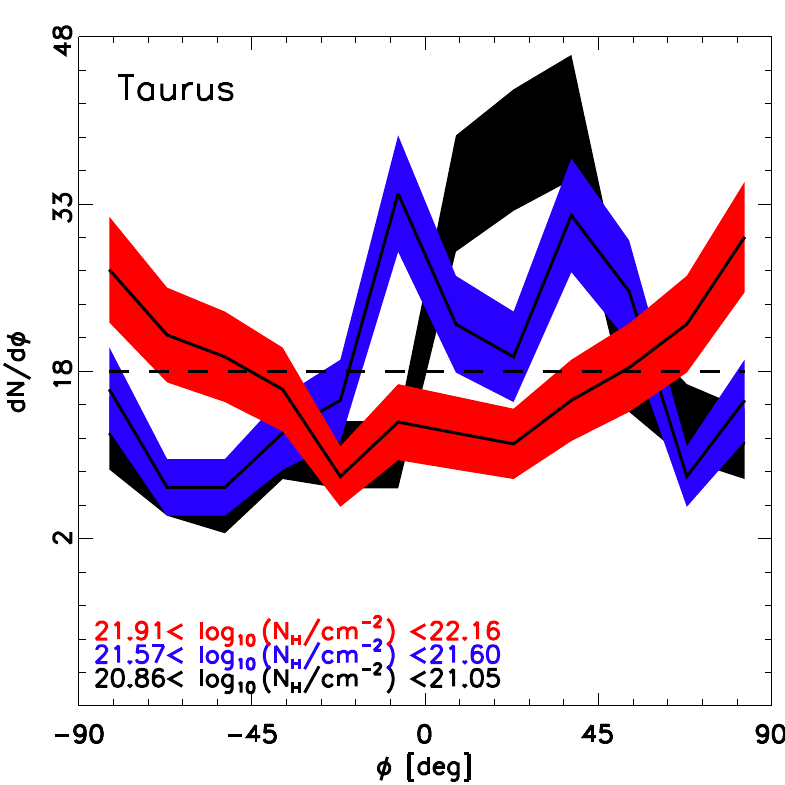}
}
\caption{HROs of the Taurus region after smoothing the input maps to beam FWHMs of 15\arcm, 30\arcm, and 60\arcm, shown from left to right, respectively.}
\label{fig:TaurusMultiScale}
\end{figure*}

\subsection{The relative orientation between $\vec{B}_{\perp}$ and \nh\ structures}

\subsubsection{Spatial distribution of the HRO signal}

The maps obtained using Eq.~\eqref{eq:hroangle} characterize the relative orientation in each region, without assuming an organization of the \nh\ structures in ridges or filaments; HROs basically just sample the orientation of \nh\ contours. However, the resulting maps of the relative orientation angle, shown for the Taurus region in Fig.~\ref{fig:anglemap}, reveal that the regions \langed{that are} \preferentially oriented parallel or perpendicular to the field form continuous patches, indicating that the HRO signal is not 
%coming only from variations of 
\langed{only coming from variations in} the field or the \nh\ contours at the smallest scales in the map.

HRO analysis 
%at
\langed{on} larger scales in a map can be achieved by considering a larger vicinity of pixels for 
%the calculation of
\langed{calculating} the gradient $\nabla\tau_{353}$. This operation is equivalent to calculating the 
%next-neighbours
\langed{next-neighbour} gradient on a map first smoothed to the scale of interest. The results 
%on
\langed{for} relative orientation after smoothing to resolutions of 15\arcm, 30\arcm, and 60\arcm\ are illustrated for the Taurus region in Fig.~\ref{fig:anglemap}.   Figure~\ref{fig:TaurusMultiScale}  shows that the corresponding HROs have a similar behaviour for the three representative \nh\ values.  These results confirm that  the 
%preferential
\langed{preferred} relative orientation is not particular to the smallest scales in the map, but corresponds to coherent structures in \nh.

A study of \langed{the} 
%preferential
\langed{preferred} orientation for the whole cloud would be possible by smoothing the column density and polarization maps to a scale comparable to the cloud size. The statistical significance of such a study would be limited to the number of clouds in the sample and would not be directly comparable to previous studies of relative orientation of clouds, where elongated structures were selected to characterize the mean orientation of each cloud \citep{li2013}.

\subsubsection{Statistical significance of the HRO signal}

Our results reveal a systemic change of  $\zeta$ with \nh, suggesting a systematic transition from magnetic field 
%preferentially
\langed{mostly} parallel to \nh\ contours in the lowest \nh\ bins to \preferentially perpendicular in the highest \nh\ bins of the clouds studied.
The statistical significance of this change can in principle be evaluated by considering the geometrical effects that influence this distribution. 
In Appendix~\ref{appendix:statisticsc}, using simulations of $Q$ and $U$ maps, we eliminate the possibility that this arises from random magnetic fields,
random spatial correlations in the field, or \langed{the} large scale structure of the field.  In Appendix~\ref{appendix:statisticsr} we simply displace the $Q$ and $U$ maps spatially and repeat the analysis, showing that the systemic trend of $\zeta$ vs.\ \nh\ disappears for displacements greater than 1\deg. 

Using a set of Gaussian models, \cite{planck2014-XXXII} estimated the statistical significance of this transition in terms of the relative orientation between two vectors in 3D and their projection in 2D. 
As these authors emphasized, two vectors that are close to parallel in 3D would be projected as parallel in 2D for almost all viewing angles for which the projections of both vectors have a non-negligible length, but on the other hand, 
%for two vectors that are perpendicular in 3D 
the situation is more ambiguous seen in projection \langed{for two vectors that are perpendicular in 3D}, because they can be projected as parallel in 2D depending on the angle of viewing.
The quantitative effects are illustrated by the simulations in Appendix~\ref{appendix:statisticsp}, where we consider distributions of vectors in 3D that are mostly parallel, mostly perpendicular, or have no 
%preferential
\langed{preferred} orientation. 
The projection tends to make vector pairs look more parallel in 2D, but the distribution of relative orientations
in 2D is quite similar though not identical to the distribution in 3D. 
In particular, the signal of perpendicular orientation is not erased and we can conclude that two projected vectors with non-negligible lengths that are close to perpendicular in 2D must also be perpendicular in 3D.  This would apply to the \preferentially perpendicular orientation for the highest bin of \nh\ in the Taurus region.

% this is called first as is correct, but it appears later because it is a long column %%%%%%%%%%%%%%%%%%%%%%
\begin{figure}[ht!]
\centerline{
\includegraphics[width=0.5\textwidth,angle=0,origin=c]{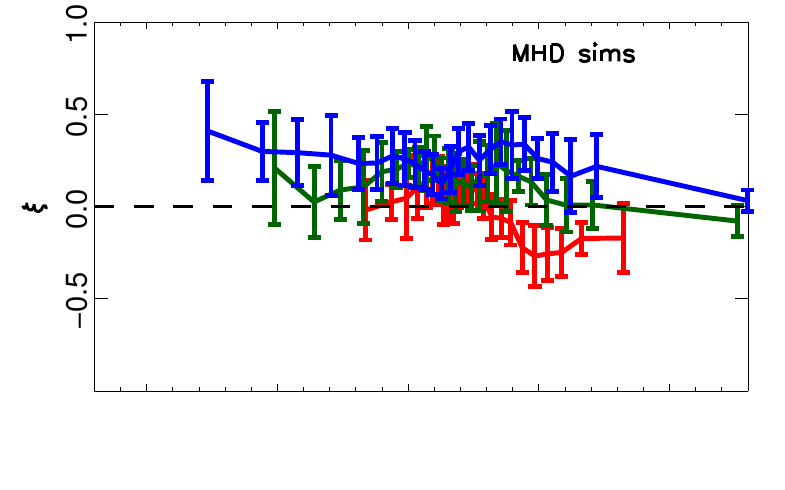}
}
\vspace{-1.3cm}
\centerline{
\includegraphics[width=0.5\textwidth,angle=0,origin=c]{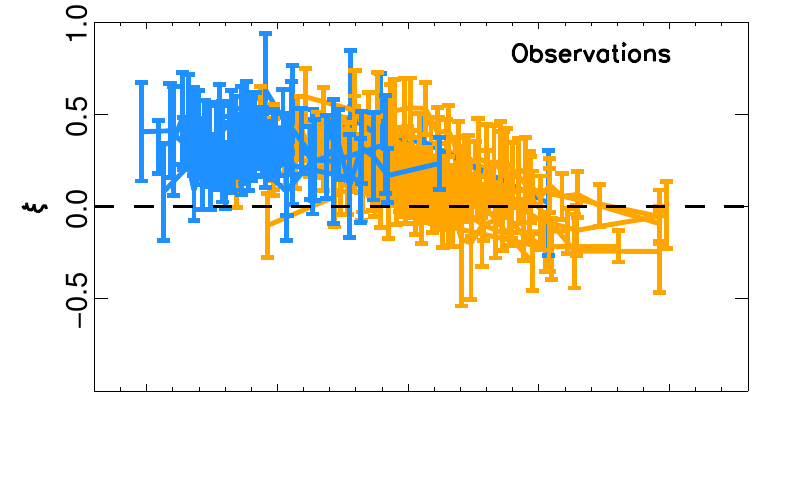}
}
\vspace{-1.3cm}
\centerline{
\includegraphics[width=0.5\textwidth,angle=0,origin=c]{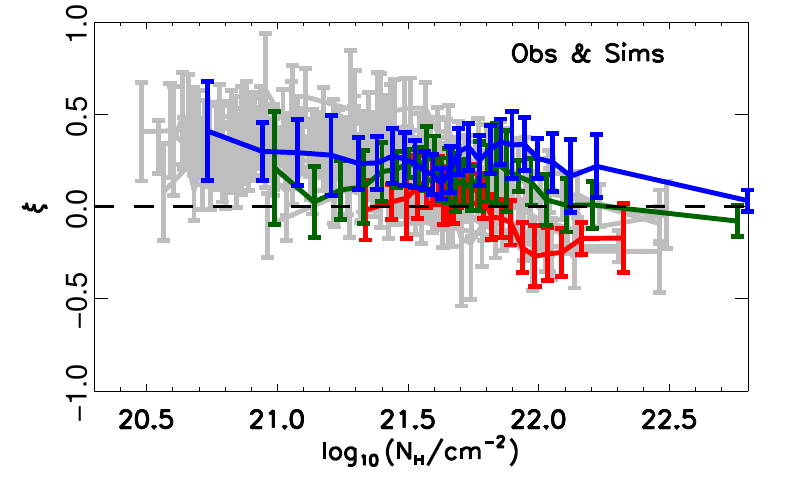}
}
\vspace{-0.3cm}
\caption{Histogram shape parameter $\zeta$ (Eqs.~\ref{eq:zeta} and \ref{eq:errzeta}) calculated for the different \nh\ bins in each region. 
\emph{Top:} relative orientation in synthetic observations of simulations with super-Alfv\'{e}nic (blue), Alfv\'{e}nic (green), and sub-Alfv\'{e}nic (red) turbulence, as detailed in \cite{soler2013}. 
\emph{Middle:} relative orientation in the regions selected from the \Planck\ all-sky observations, from Fig.~\ref{fig:HROzeta}. The blue data points correspond to the lowest \nh\ regions (CrA and the test regions in Fig.~\ref{fig:HROtest}, ChamSouth and ChamEast) and the orange correspond to the rest of the clouds. 
\emph{Bottom:} comparison between the trends in the synthetic observations (in colours) and the regions studied (grey). The observed smooth transition from \preferentially parallel ($\zeta > 0$) to perpendicular ($\zeta < 0$) is similar to 
%that
\langed{the transition} in the simulations for which the turbulence is Alfv\'{e}nic or sub-Alfv\'{e}ic.}
\label{fig:HROsims}
\end{figure}

\subsection{Comparison with simulations of MHD turbulence}

As a complement to observations, MHD simulations can be used to directly probe the actual 3D orientation of the magnetic field $\vec{B}$ with respect to the density structures. The change in the relative orientation with \nh\ was previously studied in MHD simulations using the inertia matrix and the HRO analysis \citep{hennebelle2013a,soler2013}. \cite{soler2013} showed that in 3D, the change in the relative orientation is related to the degree of magnetization.
If the magnetic energy is above or comparable to the kinetic energy (turbulence that is sub-Alfv\'{e}nic or close to equipartition), the less dense structures tend to be aligned with the magnetic field and the orientation progressively changes from parallel to perpendicular with increasing density. In the super-Alfv\'{e}nic regime, where the magnetic energy is relatively low, there appears to be no change in relative orientation with increasing density, with $\vec{B}$ and density structures being \preferentially parallel.

\cite{soler2013} describe 2D synthetic observations of the MHD simulations.
The synthetic observations are produced by integrating the simulation cubes along a direction perpendicular to the mean magnetic field and assuming a homogeneous dust grain alignment efficiency $\epsilon=1.0$. The angular resolution of the simulation is obtained by assuming a distance $d=150$\,pc and convolving the projected map with a Gaussian beam of 10\arcmin\ full width at half maximum (FWHM).
The trends in the relative orientation with \nh\ seen in 3D are also seen using the these 2D synthetic observations.
Given 
%the fact 
that sub-Alfv\'{e}nic or close to Alfv\'{e}nic turbulence does not significantly disturb the well-ordered mean magnetic field, the orientation of $\vec{B}$ perpendicular to the iso-density contours is 
%well projected
\langed{projected well} for lines of sight that are not close to the mean magnetic field orientation. In contrast, the projected relative orientation produced by super-Alfv\'{e}nic turbulence does not necessarily reflect the relative orientation in 3D as a result of the unorganized field structure.

The direct comparison between the HROs of the regions in this study and of the synthetic observations is presented in Fig.~\ref{fig:HROsims}. 
The trends in the relative orientation parameter, $\zeta$, show that the simulation with super-Alfv\'{e}nic turbulence does not undergo a transition in relative orientation from parallel to perpendicular for $\log_{10}(\nhd/\mbox{cm}^{-2}) < 23$. 
In contrast, most of the observed clouds show a decrease 
%of
\langed{in} $\zeta$ with increasing \nh, close to the trends seen from the simulations with Alfv\'{e}nic or sub-Alfv\'{e}nic turbulence for $\log_{10}(\nhd/\mbox{cm}^{-2}) < 23$. 
Furthermore, $X_{\textsc{HRO}}$, the value of $\log_{10}(\nhd/\mbox{cm}^{-2})$ where $\zeta$ goes through zero, is 
near \thresh, which is consistent with the behaviour seen in the simulations with super-Alfv\'{e}nic or Alfv\'{e}nic turbulence. 
Given that the physical conditions in the simulations ($\sigma_{v}=2.0\, $km\,s$^{-1}$ and $n=500$\,cm$^{-3}$) are typical of those in the selected regions ($\sigma_{v_{\parallel}}$ is given in Table~\ref{table-magnparameters}), the similarities in the dependence of $\zeta$ on \nh\ suggest that the strength of the magnetic field in most of the regions analysed 
 would be about the same as the mean magnetic fields in the Alfv\'{e}nic and sub-Alfv\'{e}nic turbulence simulations, which are 3.5 and 11\,\microG, respectively. However, more precise estimates of the magnetic field strength coming directly from the HROs would require further sampling of the magnetization in the MHD simulations and detailed modelling of the effects of the \LOSh\ integration.

Indirectly, the presence of a \nh\ threshold in the switch in
%preferential
preferred relative orientation between $\vec{B}_{\perp}$ and the \nh\ structures hints that gravity plays a significant role.  By contrast, for the regions with low average \nh\ (i.e., CrA and the two test regions; 
%blue points in 
Fig.~\ref{fig:HROsims}), 
there is little change in $\zeta$ and certainly no switch in the 
%preferential
preferred relative orientation to perpendicular.

\subsection{Physics of the relative orientation}\label{physicsro}

The finding of dense structures \preferentially perpendicular to the magnetic field (and the small mass-to-flux ratios discussed in Appendix~\ref{mtfr}) 
%suggest
\langed{suggests} that the magnetic field in most of the observed regions is significant for the structure and dynamics. 
However, discerning the underlying geometry is not obvious.  As one guide, 
a frozen-in and strong interstellar magnetic field would naturally cause a self-gravitating, static cloud to become oblate, with its major axis perpendicular to the field lines, because gravitational collapse would be restricted to occurring along field lines \citep{mouschovias1976I,mouschovias1976II}. 
In the case of less dense structures that are not self-gravitating, the velocity shear can stretch matter and field lines in the same direction, thereby producing aligned structures, as discussed in \cite{hennebelle2013a} and \cite{planck2014-XXXII}.

If the MCs are isolated entities and the magnetic field is strong enough to set a 
%preferential
\langed{preferred} direction for the gravitational collapse, the condensations embedded in the cloud are not very likely to have higher column densities than their surroundings \citep{nakano1998}. This means that the formation of dense substructures, such as prestellar cores and stars, by gravitational collapse would be possible only if the matter decouples from the magnetic field. This is possible through the decoupling between neutral and ionized species \juan{\citep[ambipolar diffusion,][]{mouschovias1991,li2008}} or through removal of magnetic flux from clouds via turbulent reconnection \citep{lazarian1999,santoslima2012}.

Alternatively, if we regard MCs not as isolated entities but \langed{as} the result of an accumulation of gas by large-scale flows \citep{ballesteros1999,hartmann2001,koyama2002,audit2005,heitsch2006}, the material swept up by colliding flows may eventually form a self-gravitating cloud. If the magnetic field is strong the accumulation of material is favoured along the magnetic field lines, thus producing dense structures that are \preferentially perpendicular to the magnetic field. The inflow of material might eventually increase the gravitational energy in parts of the cloud, \langed{thereby} producing supercritical structures such as prestellar cores.

For supersonic turbulence in the ISM and MCs, density structures can be formed by gas compression in shocks. If the turbulence is strong with respect to the magnetic field (super-Alfv\'{e}nic), gas compression by shocks is approximately isotropic; because magnetic flux is frozen into matter, field lines are dragged along with the gas, forming structures that tend to be aligned with the field. If the turbulence is weak with respect to the magnetic field (sub-Alfv\'{e}nic), the fields produce a clear anisotropy in MHD turbulence \citep{sridhar1994,goldreich1995,matthaeus2008,banerjee2009} and compression by shocks 
%occurs preferentially
\langed{that is favored to occur} along the magnetic field lines, creating structures perpendicular to the field. 
The cold phase gas that constitutes the cloud 
%has
\langed{receives} no information about the original flow direction because the magnetic field redistributes the kinetic energy of the inflows \citep{heitsch2009,inoue2009,burkhart2014}. This seems to be the case in most of the observed regions, where the \preferentially perpendicular relative orientation between the magnetic field and the high 
column density structures is an indication of the anisotropy produced by the field. 

The threshold of $\log_{10}(\nhd/\mbox{cm}^{-2}) \approx \thresh$ above which 
the preferential orientation of $\vec{B}_{\perp}$ switches to being perpendicular to the \nh\ contours is intriguing.
Is there a universal threshold column density that is independent of the particular MC environment and relevant in the context of star formation?
In principle, this threshold might be related to the column density of filaments at which substructure forms, as reported in an analysis of \Herschel\ observations \citep{arzoumanian2013}, but the \Planck\  polarization observations leading to $B_{\perp}$ do not fully resolve such filamentary structures. In principle, this threshold might also be related to the column density at which the magnetic field starts scaling with density, according to the Zeeman effect observations  of $B_{\parallel}$  (Figure~7 in \citealp{crutcher2012}). However, establishing such relationships requires further studies with MHD simulations to identify what densities and scales influence the change 
%of
\langed{in} relative orientation between $B_{\perp}$ and \nh\ structures and to model the potential imprint in $B_{\parallel}$ observations and in $B_{\perp}$ observations to be carried out at higher resolution.

\subsection{Effect of dust grain alignment}

Throughout this study we assume that the polarized emission observed by \Planck\ at 353\,GHz is representative of the projected morphology of the magnetic field in each region; i.e., we assume a constant dust grain alignment efficiency ($\epsilon$) \langed{that is} independent of the local environment. Indeed, 
observations and MHD simulations under this assumption \citep{planck2014-XIX, planck2014-XX} indicate that depolarization effects at large and intermediate scales in MCs might arise from the random component of the magnetic field along the line of sight.  On the other hand\langed{,} the sharp drop in the polarization fraction at $\nhd > 10^{22}$\,cm$^{-2}$ \citep[reported in][]{planck2014-XIX}, when seen at small scales, might be interpreted in terms of a decrease of $\epsilon$ with increasing column density \citep{matthews2001II,whittet2008}.

A leading theory for the process of dust grain alignment involves radiative torques by the incident radiation \citep{lazarian2007,hoang2009,A15}. A critical parameter for this mechanism is the ratio between the dust grain size and the radiation wavelength. As the dust column density increases, only the longer wavelength radiation penetrates the cloud and the alignment decreases. Grains within a cloud (without embedded sources) should have lower $\epsilon$ than those at the periphery of the same cloud. 
There is evidence for this from near-infrared interstellar polarization and submillimetre polarization along lines of sight through starless cores \citep{jones2015}, albeit 
%at
\langed{on} smaller scales and higher column densities than considered here.  If $\epsilon$ inside the cloud is very low, the observed polarized intensity would arise from the dust in the outer layers, tracing the magnetic field in the ``skin'' of the cloud. Then the observed orientation of $\vec{B}_{\perp}$ is not necessarily correlated with the column density structure, which is seen in total intensity, or with the magnetic field deep in the cloud. 

\cite{soler2013} presented the results of HRO analysis on a series of synthetic observations produced using models of how $\epsilon$ might decrease with increasing density. They showed that with a steep decrease there is no visible correlation between the inferred magnetic field orientation and the high-\nh\ structure, corresponding to nearly flat HROs. 

%In any case, the 
\langed{The} HRO analysis of MCs carried out here reveals 
%the presence of 
a correlation between the polarization orientation and the column density structure.  This suggests that the dust polarized emission 
%is sampling
\langed{samples} the magnetic field structure 
homogeneously on the scales being probed at the resolution of the \Planck\ observations or\langed{,} 
alternatively\langed{,} that the field deep within high-\nh\ structures 
%shares
\langed{has} the same orientation 
%as that probed
\langed{of the field} in the skin.

% --------------------------------------------------------------------------------------------------------------------------------------------------------------------
% CONCLUSIONS CONCLUSIONS CONCLUSIONS CONCLUSIONS CONCLUSIONS CONCLUSIONS CONCLUSIONS %---------------------------------------------------------------------------------------------------------------------------------------------------------------------

\section{Conclusions}\label{section:conclusions}

We have presented a study of the relative orientation of the magnetic field projected on the plane of the sky ($\vec{B}_{\perp}$), as inferred from the \Planck\ dust polarized thermal emission, with respect to structures detected in gas column density (\nh).
The relative orientation study 
%has
\langed{was} performed by using the histogram of relative orientations (HRO), 
a novel statistical tool 
%to characterize
\langed{for characterizing} extended polarization maps. With the unprecedented statistics of polarization observations in extended maps obtained by \Planck, we analyze the HRO in regions with different column densities within ten nearby molecular clouds (MCs) and two test fields.

In most of the regions analysed we find that the relative orientation between $\vec{B}_{\perp}$ and \nh\ structures changes systematically with \nh\, from \langed{being} parallel in the lowest column density areas to perpendicular in the highest column density areas. The switch occurs at $\log_{10}(\nhd/\mbox{cm}^{-2}) \approx \thresh$.  This change in relative orientation is particularly significant given that projection tends to produce more parallel \pseudovectors\ in 2D (the domain of observations) than exist in 3D.

The HROs in these MCs reveal that most of the high \nh\ structures in each cloud are 
%oriented preferentially
\langed{mostly oriented} perpendicular to the magnetic field, suggesting that they may have formed by material accumulation and gravitational collapse along the magnetic field lines.  According to a similar study where the same method was applied to MHD simulations, this trend is only possible if the turbulence is Alfv\'{e}nic or sub-Alfv\'{e}nic. This implies that the magnetic field is significant for the gas dynamics 
%at
\langed{on} the scales sampled by \Planck.
The estimated mean magnetic field strength is about 4 and 12\,\microG\ for the case of Alfv\'{e}nic and sub-Alfv\'{e}nic turbulence, respectively.

We also estimate the magnetic field strength in the MCs studied using the DCF and \hkd\ methods.  The estimates found seem consistent with the above values from the HRO analysis\langed{,} but given the assumptions and systematic effects involved\langed{,} we recommend that these rough estimates 
%should 
be treated with caution.  According to these estimates the \langed{analysed regions} appear to be magnetically sub-critical. This result is also consistent with the conclusions of the HRO analysis.  Specific tools, such as the DCF and \hkd\ methods, are best suited 
%for
\langed{to} the scales and physical conditions in which their underlying assumptions are valid.  The study of large polarization maps covering multiple scales calls for generic statistical tools, such as the HRO, 
%to characterize
\langed{for characterizing} their properties and 
%establish
\langed{establishing} a direct relation 
%with
\langed{to} the physical conditions included in MHD simulations. 

The study of the structure 
%at
\langed{on} smaller scales is beyond the scope of this work\langed{,} however, the presence of gravitationally bound structures within the MCs, such as prestellar cores and stars, suggests that the role of magnetic fields is changing 
%at
\langed{on} different scales. Even if the magnetic field is important in the accumulation of matter that leads to the formation of the cloud, effects such as matter decoupling from the magnetic field and the inflow of matter from the cloud environment lead to the formation of magnetically supercritical structures 
%at
\langed{on} smaller scales. Further studies will help to identify the dynamical processes that connect the MC structure with the process of star formation.

\begin{acknowledgements}
The development of \Planck\ has been supported by: ESA; CNES and CNRS/INSU-IN2P3-INP (France); ASI, CNR, and INAF (Italy); NASA and DoE (USA); STFC and UKSA (UK); CSIC, MICINN, JA, and RES (Spain); Tekes, AoF, and CSC (Finland); DLR and MPG (Germany); CSA (Canada); DTU Space (Denmark); SER/SSO (Switzerland); RCN (Norway); SFI (Ireland); FCT/MCTES (Portugal); and PRACE (EU). A description of the Planck Collaboration and a list of its members, including the technical or scientific activities in which they have been involved, can be found at \url{http://www.sciops.esa.int/index.php?project=planck&page=Planck_Collaboration}.
The research leading to these results has received funding from the European
Research Council under the European Union's Seventh Framework Programme
(FP7/2007-2013) / ERC grant agreement No.\ 267934.
\end{acknowledgements}

\bibliographystyle{aa}

{\raggedright
\bibliography{Planck_bib.bib,jdslib.bib,placeholder.bib}
%\bibliography{PIP_113_Soler_01JUL2015.bbl}
}

% -------------------------------------------------------------------------------------------------------------------------------------------------------------------------
% APPENDICES APPENDICES APPENDICES APPENDICES APPENDICES APPENDICES APPENDICES APPENDICES -------------------------------------------------------------------------------------------------------------------------------------------------------------------------

\appendix

% ===============================================================================================
\section{Selection of data}\label{appendix:selection}

The HRO analysis is applied to each MC using common criteria for selecting the areas in which the relative orientation is to be assessed.

\subsection{Gradient mask}\label{appendix:selectiong}

The dust optical depth, $\tau_{353}$, observed in each region, can be interpreted as
\begin{equation}
\tau^{\textsc{obs}}_{353} = \tau^{\textsc{mc}}_{353} +  \tau^{\textsc{bg}}_{353} + \delta_{\tau_{353}}\, ,
\end{equation}
where $\tau^{\textsc{mc}}_{353}$ is the optical depth of the MC, $\tau^{\textsc{bg}}_{353}$ is the optical depth of the diffuse regions behind and/or in front of the cloud (background/foreground), and $\delta_{\tau_{353}}$ 
%is 
the noise in the optical depth map with variance $\sigma^{2}_{\tau_{353}}=\overline{\delta^{2}_{\tau_{353}}}$.

The gradient of the optical depth can be then written as
\begin{equation}
\nabla\tau^{\textsc{obs}}_{353} = \nabla\tau_{353}^{\textsc{mc}} +  \nabla(\tau_{353}^{\textsc{bg}} + \delta_{\tau_{353}})\, .
\end{equation}
We quantify the contribution of the background/foreground and the noise, $\nabla(\tau_{353}^{\textsc{bg}} + \delta_{\tau_{353}})$, by evaluating $\nabla\tau_{353}$ in a reference field with lower submillimetre emission. Given that the dominant contribution to the background/foreground gradient would come from the gradient in emission from the Galactic plane, for each of the regions
%analysed we choose
\langed{analysed we chose} a reference field of the same size at the same Galactic latitude and with the lowest average \nh\ in the corresponding latitude band. We compute the average of the gradient norm in the reference field, $\left<|\nabla\tau^{\textsc{ref}}_{353}|\right>$, and 
%we 
use this value as a threshold for selecting the regions of the map where $\nabla\tau_{353}$ carries significant information about the structure of the cloud. We note that this threshold includes a contribution from the noise $\nabla\delta_{\tau_{353}}$. 
The HROs presented in this study correspond to regions in each field where $|\nabla\tau_{353}|>\left<|\nabla\tau^{\textsc{ref}}_{353}|\right>$.

\subsection{Polarization mask}\label{appendix:selectionp}

The total Stokes parameters $Q$ and $U$ measured in each region can be interpreted as
\begin{equation}
Q^{\textsc{obs}} = Q^{\textsc{mc}} + Q^{\textsc{bg}} + \delta_{\rm{Q}}\, , \,\, U^{\textsc{obs}} = U^{\textsc{mc}} + U^{\textsc{bg}} + \delta_{\rm{U}}\, ,
\end{equation}
where $Q^{\textsc{mc}}$ and $U^{\textsc{mc}}$ correspond to the polarized emission from the MC, $Q^{\textsc{bg}}$
%\peter{The suggested comma is not pertinent}
 and $U^{\textsc{bg}}$ correspond to the polarized emission from the diffuse background/foreground, and $\delta_{Q}$ and $\delta_{U}$ are the noise contributions to the observations, such that the variances $\sigma^{2}_{Q}=\overline{\delta^{2}_{Q}}$ and $\sigma^{2}_{U}=\overline{\delta^{2}_{U}}$.

As in the treatment of the gradient, we estimate the contributions of the background/foreground polarized emission and the noise using the rms of the Stokes parameters in the same reference field, 
$Q^{\textsc{ref}}_{\rm rms}$ and $U^{\textsc{ref}}_{\rm rms}$.
The HROs presented in this study correspond to pixels in each region where $|Q| > 2Q^{\textsc{ref}}_{\rm rms}$ or $|U| > 2U^{\textsc{ref}}_{\rm rms}$. The ``or'' conditional avoids biasing the selected values of polarization. This first selection criterion provides a similar sample 
%as
\langed{to} the alternative coordinate-independent criterion $\sqrt{Q^{2}+U^{2}}>2\sqrt{(Q^{\textsc{ref}}_{\rm rms})^{2}+(U^{\textsc{ref}}_{\rm rms})^{2}}$. 
This first criterion aims to distinguish between the polarized emission coming from the cloud and the polarized emission coming from the background/foreground estimated in the reference regions. 

Additionally, as a second criterion, our sample is restricted to polarization measurements where $|Q| > 3\sigma_{Q}$ or $|U| > 3\sigma_{U}$. 
This aims to select pixels where the uncertainty in the polarization angle is smaller than the size of the angle bins used for the constructions of the HRO \citep{serkowski1958,montier2014a}. 
In terms of the total polarized intensity, $P=\sqrt{Q^{2}+U^{2}}$, and following equations.~B.4 and B.5 in \cite{planck2014-XIX}, the second criterion corresponds to $P/\sigma_{P}>3$ and uncertainties in the polarized orientation angle $\sigma_{\psi}<10$\deg.
\\

% this following paragraph does not belong to the subsection so use \\ above
\noindent
The fractions of pixels considered in each region, after applying the selection criteria described above, are summarized in Table~\ref{table-graderror}. The largest masked portions of the regions correspond to the gradient mask, which selects mostly \langed{those} areas of column density above the mean column density of the background/foreground $\left<\nhd^{\textsc{bg}}\right>$. The polarization mask provides an independent criterion that is less restrictive.  The intersection of these two masks selects the fraction of pixels considered for the HRO analysis.

\begin{table}[tmb]  % table* is a two-column table.  Drop the * for one column.
\begingroup
\newdimen\tblskip \tblskip=5pt
\caption{Selection of data$^a$.}
\label{table-graderror}                            % Label goes here.
\nointerlineskip
\vskip -3mm
\footnotesize
\setbox\tablebox=\vbox{
   \newdimen\digitwidth 
   \setbox0=\hbox{\rm 0} 
   \digitwidth=\wd0 
   \catcode`*=\active 
   \def*{\kern\digitwidth}
   \newdimen\signwidth 
   \setbox0=\hbox{+} 
   \signwidth=\wd0 
   \catcode`!=\active 
   \def!{\kern\signwidth}
\halign{\hbox to 1.15in{#\leaderfil}\tabskip 2.2em&
\hfil#&\hfil#&\hfil#&\hfil#\tabskip 0pt\cr
\noalign{\doubleline}
\omit\hfil Region\hfil & \hfil$\left<\nhd^{\textsc{bg}}\right>$\hfil & $f_{\nabla}$ &  \hfil $f_{\rm pol}$\hfil &  \hfil$f_{\rm tot}$ \hfil \cr
\omit & \hfil[10$^{20}$\,cm$^{-2}$]\hfil & \hfil[\%]\hfil & \hfil[\%]\hfil & \hfil[\%]\hfil\cr
\noalign{\vskip 4pt\hrule\vskip 6pt}
%----------------------------------------------------------------------------------------------------------------
Taurus & \hfil\phantom{0}6.2\hfil & \hfil66\hfil & \hfil78\hfil & \hfil28\hfil \cr
%----------------------------------------------------------------------------------------------------------------
Ophiuchus & \hfil\phantom{0}5.6\hfil & \hfil65\hfil & \hfil82\hfil & \hfil31\hfil \cr
%----------------------------------------------------------------------------------------------------------------
Lupus & \hfil\phantom{0}9.6\hfil & \hfil65\hfil & \hfil67\hfil & \hfil24\hfil   \cr
%----------------------------------------------------------------------------------------------------------------
Chamaeleon-Musca & \hfil\phantom{0}6.3\hfil & \hfil49\hfil & \hfil83\hfil & \hfil31\hfil \cr
%----------------------------------------------------------------------------------------------------------------
Corona Australia (CrA) & \hfil\phantom{0}5.4\hfil & \hfil34\hfil & \hfil40\hfil & \hfil\phantom{0}4\hfil  \cr
%----------------------------------------------------------------------------------------------------------------
\noalign{\vskip 4pt\hrule\vskip 6pt}
%----------------------------------------------------------------------------------------------------------------
Aquila Rift & \hfil18.4\hfil & \hfil48\hfil & \hfil96\hfil & \hfil32\hfil \cr
%----------------------------------------------------------------------------------------------------------------
Perseus & \hfil\phantom{0}5.3\hfil & \hfil60\hfil & \hfil59\hfil & \hfil16\hfil \cr
%----------------------------------------------------------------------------------------------------------------
\noalign{\vskip 4pt\hrule\vskip 6pt}
IC\,5146 & \hfil26.4\hfil & \hfil38\hfil & \hfil93\hfil & \hfil29\hfil  \cr
%---------------------------------------------------------------------------------------------------------------
Cepheus &  \hfil\phantom{0}7.5\hfil & \hfil66\hfil & \hfil80\hfil & \hfil36\hfil \cr
%----------------------------------------------------------------------------------------------------------------
Orion &  \hfil\phantom{0}6.1\hfil & \hfil63\hfil & \hfil67\hfil & \hfil24\hfil \cr
%----------------------------------------------------------------------------------------------------------------
\noalign{\vskip 4pt\hrule\vskip 6pt}
%----------------------------------------------------------------------------------------------------------------
ChamEast &  \hfil\phantom{0}6.3\hfil & \hfil33\hfil & \hfil38\hfil & \hfil10\hfil \cr
%----------------------------------------------------------------------------------------------------------------
ChamSouth &  \hfil\phantom{0}4.9\hfil & \hfil36\hfil & \hfil42\hfil & \hfil13\hfil \cr
%----------------------------------------------------------------------------------------------------------------
\noalign{\vskip 3pt\hrule\vskip 4pt}}}
\endPlancktable                    % ends one-column \halign
%\endPlancktablewide                 % ends two-column \halign
\tablenote a Mean column density of the background/foreground for each region $\left<\nhd^{\textsc{bg}}\right>$ estimated from a reference field at the same Galactic latitude; 
%fraction
\langed{percentage} $f_{\nabla}$ of all pixels where $|\nabla\tau_{353}|>\left<|\nabla\tau^{\textsc{ref}}_{353}|\right>$; 
%fraction 
\langed{percentage} $f_{\rm pol}$ of all pixels where $|Q| > 2Q^{\textsc{ref}}_{\rm rms}$ or $|U| > 2U^{\textsc{ref}}_{\rm rms}$ (first polarization criterion) and where $|Q| > 3\sigma_{Q}$ or $|U| > 3\sigma_{U}$ (second polarization criterion); 
and 
%fraction
\langed{percentage} $f_{\rm tot}$ of all pixels used for the HRO analysis.\par
\endgroup
\end{table}  

% ===============================================================================================

\section{Construction of the histogram of relative orientations and related uncertainties}\label{appendix:hro}

The HROs were calculated for 25 column density bins having equal numbers of selected pixels (10 bins in two regions with fewer pixels, CrA and IC 5146).  For each of these HROs we use 12 angle bins of width 15\,\deg\ (see Sect.~\ref{introhro}).

\subsection{Calculation of $\nabla\tau_{353}$ and the uncertainty of its orientation}\label{appendix:gradtauunc}

The optical depth gradient ($\nabla \tau_{353}$) is calculated by convolving the $\tau_{353}$ map with a Gaussian derivative kernel \citep{soler2013}, such that $\nabla \tau_{353}$ corresponds to
\begin{equation}\label{equation:grad}
\nabla \tau_{353} = (G_{x} \otimes \tau_{353})\,\vec{\hat{i}} + (G_{y} \otimes \tau_{353})\,\vec{\hat{j}} = g_{x}\,\vec{\hat{i}} + g_{y}\,\vec{\hat{j}}\, ,
\end{equation}
where $G_{x}$ and $G_{y}$ are the kernels calculated using the $x$- and $y$-derivatives of a symmetric 
%2-dimensional
\langed{two-dimensional} Gaussian function. 
The orientation of the iso-$\tau_{353}$ contour is calculated from the components of the gradient vector,
\begin{equation}\label{equation:gradtheta}
\theta = \arctan\left(-g_{x}\, , \, g_{y}\right)\, .
\end{equation}

Because the calculation of the gradient through convolution is a linear operation, the associated uncertainties can be calculated using the same operation,so that
\begin{equation}\label{equation:vargx}
\delta_{g_{x}} = G_{x} \otimes \delta_{\tau_{353}}\, ; \,\, \delta_{g_{y}} = G_{y} \otimes \delta_{\tau_{353}}\, ,
\end{equation}
from which we obtain the $\sigma^{2}_{g_{x}}=\overline{\delta^{2}_{g_{x}}}$ and $\sigma^{2}_{g_{y}}=\overline{\delta^{2}_{g_{y}}}$.

The standard deviation of the angle $\theta$ can be written as
\begin{equation}\label{equation:vartheta}
\sigma_{\theta} = \sqrt{\left(\frac{\partial\theta}{\partial g_{x}}\right)^{2}\sigma_{g_{x}}^{2} + \left(\frac{\partial\theta}{\partial g_{y}}\right)^{2}\sigma_{g_{y}}^{2}}\, ,
\end{equation}
which corresponds to
\begin{equation}\label{equation:varthetafull}
\sigma_{\theta} = \frac{1}{g_{x}^{2}+g_{y}^{2}}\,\sqrt{g_{y}^{2}\sigma_{g_{x}}^{2} + g_{x}^{2}\sigma_{g_{y}}^{2}}\, .
\end{equation}
In the application discussed here, the standard deviations in the $\tau_{353}$ map within the selected areas are much less than a few 
%percent, and
\langed{percentage points,} so their effect on the estimate of the orientation of the gradient is negligible.

% -------------------------------------------------------------------------------------------------------------------------------------------------------------
\subsection{Uncertainties affecting the characterization of relative orientations within MCs}\label{appendix:relorientunc}

\subsubsection{Uncertainties in the construction of the histogram}\label{HROuncertainties}

\begin{figure}[ht!]
\centerline{
\includegraphics[width=0.45\textwidth,angle=0,origin=c]{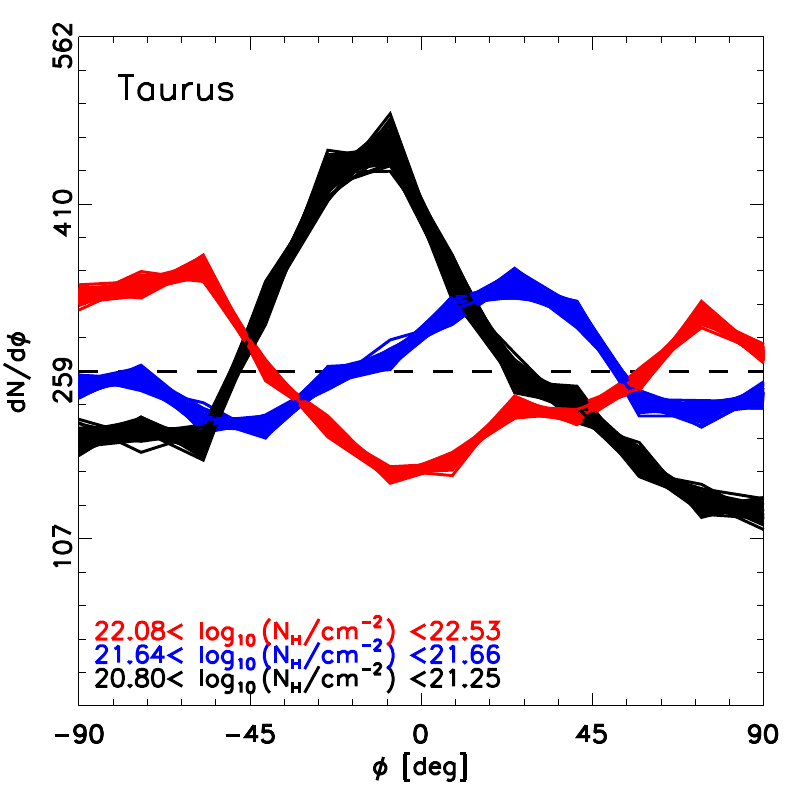}
}
\vspace{-0.2cm}
\caption{HROs in the Taurus region 
%corresponding
\langed{that correspond} to the indicated \nh\ bins. The plotted values are obtained using the original $\tau_{353}$ map at 10\arcmin\ resolution and maps of the Stokes parameters $Q_{r}$ and $U_{r}$, 
%corresponding
\langed{which correspond} to 1000 random noise realizations. Each realization is generated using a Gaussian probability density function centred on the measured values $Q$ and $U$ with variances $\sigma^{2}_{Q}$ and $\sigma^{2}_{U}$.
}
\label{fig:errorsMChist}
\end{figure}

To estimate the uncertainty associated with the noise in Stokes $Q$ and $U$, we produce 1000 noise realizations, $Q_{r}$ and $U_{r}$, using Monte Carlo sampling. We assume that the errors are normally distributed and are centred on the measured values $Q$ and $U$ with dispersions $\sigma_{\rm{Q}}$ and $\sigma_{\rm{U}}$ \citep{planck2013-p02,planck2013-p03,planck2013-p02b,planck2013-p03f}. Given that $\sigma_{\rm{QU}}$ is 
%small compared to
\langed{smaller than} $\sigma^2_{\rm{Q}}$ and $\sigma^2_{\rm{U}}$, it is justified to generate $Q_{r}$ and $U_{r}$ independently of each other. 
We then introduce $Q_{r}$ and $U_{r}$ in the analysis pipeline and compute the HRO using the corresponding $\tau_{353}$ map in each region. The results, presented in Fig.~\ref{fig:errorsMChist} for the Taurus region, show that the noise in $Q$ and $U$ does not critically affect the shape of the HROs or the trend in $\zeta$.  
Together\langed{,} the low noise in the maps of $\tau_{353}$ and the selection criteria for the polarization measurements ensure that the HRO is well determined.

Another source of uncertainty in the HRO resides in the histogram binning process. The variance in the $k$th histogram bin is given by
\begin{equation}\label{equation:histvar}
\sigma^{2}_{k} = h_{k}\left(1-\frac{h_{k}}{\htot}\right)\, ,
\end{equation}
where $h_{k}$ is the number of samples in the $k$th bin\langed{,} and $\htot$ is the total number of samples.

Of the 
%above 
two independent sources of uncertainty \langed{above,} we find that the largest contribution comes from the binning process\langed{, so that} 
%and therefore 
these are the ones shown as the shaded uncertainty ranges in all figures of HROs, for example for Taurus in Fig.~\ref{fig:HRO1}.  Because of the large number of samples in each histogram bin, the uncertainties in the HRO do not significantly affect the results of this study.

\subsubsection{Uncertainties in the histogram shape parameter $\zeta$}

\begin{figure}[ht!]
\centerline{
\includegraphics[width=0.5\textwidth,angle=0,origin=c]{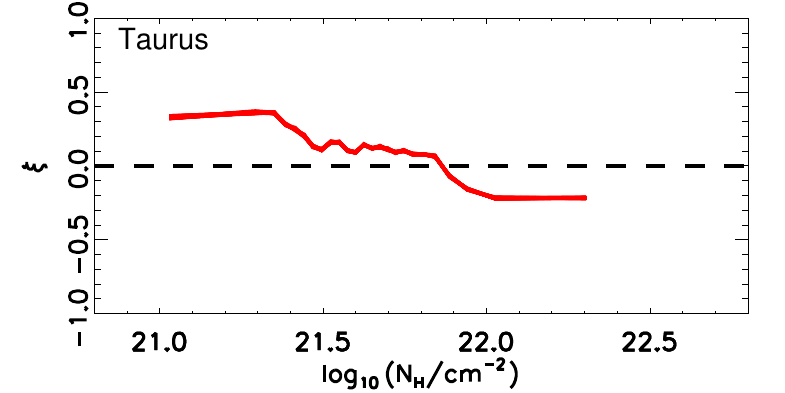}
}
\vspace{-0.0cm}
\caption{Histogram shape parameter, $\zeta$, as a function of $\log_{10}(\nhd/\mbox{cm}^{-2})$ in the Taurus region. The values are obtained using the $\tau_{353}$ map at 10\arcmin\ resolution and maps of the Stokes parameters $Q_{r}$ and $U_{r}$ 
%corresponding
\langed{that correspond} to 1000 
%random noise
\langed{random-noise} realizations. Each realization is generated using a Gaussian probability density function centred on the measured values $Q$ and $U$with variances $\sigma^{2}_{Q}$ and $\sigma^{2}_{U}$.  By joining the values at each \nh\ bin\langed{,} we find a trend very close to the black line in Fig.~\ref{fig:HROzeta}, with little dispersion from the noise; much larger are the uncertainties in evaluating $\zeta$ at each \nh\ bin, as given in  Fig.~\ref{fig:HROzeta} but not shown here.
}
\label{fig:errorsMCzeta}
\end{figure}

As in the case of the HRO, the uncertainty in $\zeta$, as defined in Eq.~\eqref{eq:zeta}, can be quantified using the random realizations introduced in the previous section. Figure~\ref{fig:errorsMCzeta} shows the dependence of $\zeta$ on $\log_{10}$\nh\ obtained using $Q_{r}$ and $U_{r}$ for the Taurus region. The small variations 
%about
\langed{around} the trend line 
indicate that the uncertainties in $Q$ and $U$ do not significantly affect the trends discussed in this study, as expected from the behaviour of the histograms presented in Fig.~\ref{fig:errorsMChist}. The main source of uncertainty in the estimation of $\zeta$ is related to the histogram binning, characterized by the error bars calculated using Eq.~\eqref{eq:errzeta}.  As seen in Fig.~\ref{fig:HROzeta} and reproduced in Fig.~\ref{fig:MChroUniform}, these are much larger than the dispersion of the values of $\zeta$ in Fig.~\ref{fig:errorsMCzeta}.

\section{Statistical significance of the HRO signal}\label{appendix:statistics}

\subsection{A product of chance?}\label{appendix:statisticsc}

To investigate various potential sources of the signal found in the HROs we use the $\tau_{353}$ map at 10\arcmin\ resolution in each region in combination with $Q$ and $U$ maps that are produced with different recipes, each with 1000 realizations.

\begin{figure}[ht!]
\centerline{
\includegraphics[width=0.5\textwidth,angle=0,origin=c]{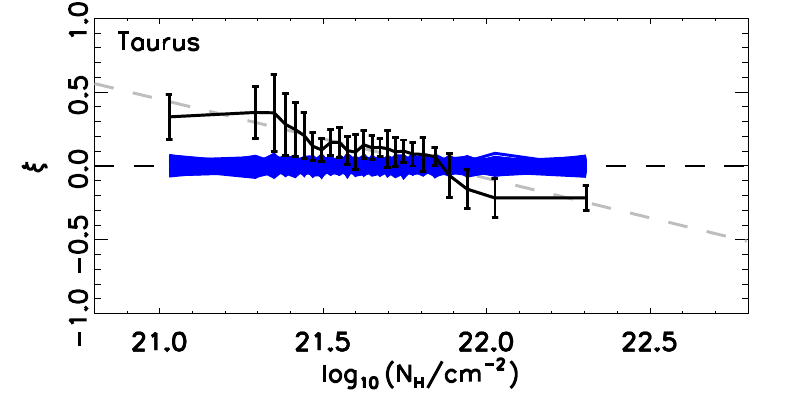}
}
\caption{Histogram shape parameter, $\zeta$, as a function of $\log_{10}(\nhd/\mbox{cm}^{-2})$ in the Taurus region. 
Blue lines join the $\zeta$ values obtained using the $\tau_{353}$ map at 10\arcmin\ resolution and each of 1000 random realizations of $Q$ and $U$ maps corresponding to a uniform distribution of $\psi$.
Results in black and grey are from the analysis of the \planck\ data, as reported in Fig.~\ref{fig:HROzeta}. 
}
\label{fig:MChroUniform}
\end{figure}

%%%%%%%% totally random field
To eliminate random fields, we use $Q$ and $U$ maps produced with a random realization of $\psi$ with a uniform distribution and 
%using
\langed{with} unit-length polarization \pseudovectors. The results of this numerical experiment are shown in Fig.~\ref{fig:MChroUniform} for the Taurus region.  For each of the 1000 realizations\langed{,} we join the values at each \nh\ bin to show the trend lines in order to compare with 
%that
\langed{the lines} in Fig.~\ref{fig:HROzeta}.

\begin{figure}[ht!]
\centerline{
\includegraphics[width=0.5\textwidth,angle=0,origin=c]{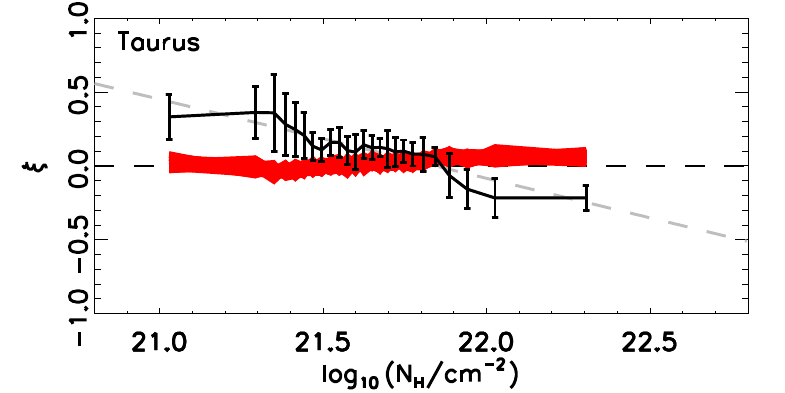}
}
\caption{
Like Fig.~\ref{fig:MChroUniform} but for values of $\zeta$ (red) 
obtained using 1000 random realizations of $Q$ and $U$ maps corresponding to a Gaussian distribution of $\psi$ centred on $\psi_{0}=0$\deg\ and with standard deviation $\varsigma_{\psi}=45$\deg.}
\label{fig:MChroGauss}
\end{figure}

%%%%%%% large scale field with random fluctuations
To eliminate the large-scale magnetic field as the source we use $Q$ and $U$ maps produced with random realizations of $\psi$ with a Gaussian distribution and unit-length polarization \pseudovectors. The polarization angle distribution is centred at $\psi_{0}=0$\deg\ with a standard deviation $\varsigma_{\psi}=45$\deg. The results of this numerical experiment are shown in Fig.~\ref{fig:MChroGauss} for the Taurus region.

\begin{figure}[ht!]
%\centerline{\includegraphics[width=0.5\textwidth,angle=0,origin=c]{DX11/TaurusDX11TopPolTaufwhm10_HROzetaJackKnifePS.eps}}
\centerline{\includegraphics[width=0.5\textwidth,angle=0,origin=c]{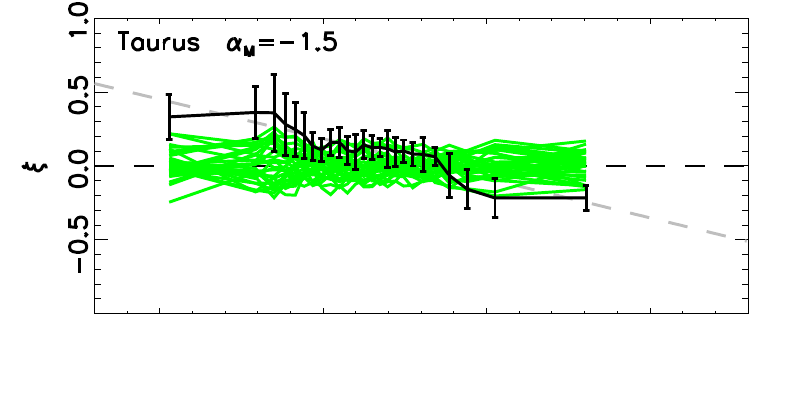}}
\vspace{-1.1cm}
%\centerline{\includegraphics[width=0.5\textwidth,angle=0,origin=c]{DX11/TaurusDX11TopPolTaufwhm10_HROzetaJackKnifePSn25.eps}}
\centerline{\includegraphics[width=0.5\textwidth,angle=0,origin=c]{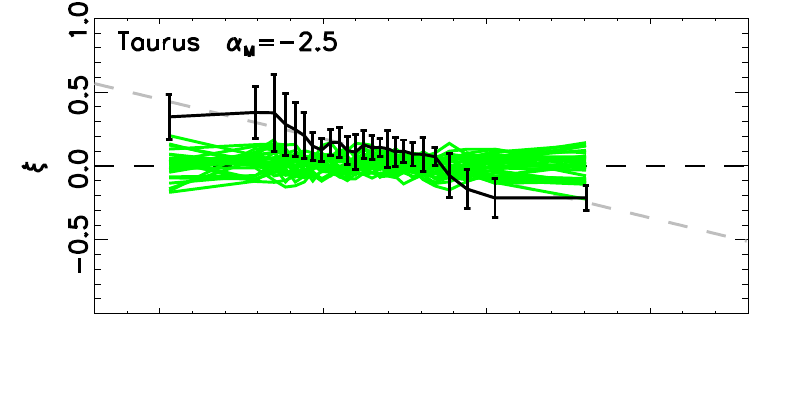}}
\vspace{-1.1cm}
%\centerline{\includegraphics[width=0.5\textwidth,angle=0,origin=c]{DX11/TaurusDX11TopPolTaufwhm10_HROzetaJackKnifePSn35.eps}}
\centerline{\includegraphics[width=0.5\textwidth,angle=0,origin=c]{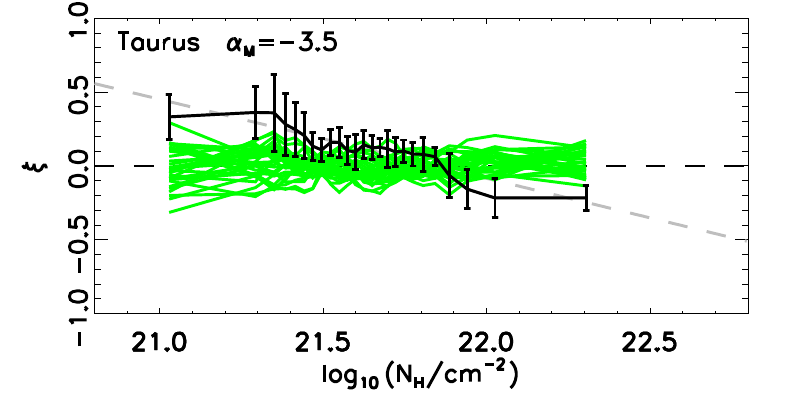}}
\caption{
Like Fig.~\ref{fig:MChroUniform} but for values of $\zeta$ (green) 
obtained using 1000 random realizations of $Q$ and $U$ maps corresponding to a power spectrum $P(k)\propto k^{\,\alpha_{\textsc{m}}}$, with $\alpha_{\textsc{m}}=-1.5$, $-2.5$, and $-3.5$. 
}
\label{fig:MChroPS}
\end{figure}

%%%%%%%% power spectrum
To eliminate random spatial correlations\langed{,} we use $Q$ and $U$ maps produced from random realizations with a power spectrum $P(k)\propto k^{\,\alpha_{\textsc{m}}}$. For the spectral indices \langed{we} adopted, $\alpha_{\textsc{m}}=-1.5$, $-2.5$, and $-3.5$, correlations are introduced in the orientation of the polarization \pseudovectors\ that are independent of the structure of matter. This test evaluates the statistical significance of the random sampling of spatial correlations in the magnetic field implied when we calculate the HROs in a finite region of the sky. The results of this numerical experiment are shown in Fig.~\ref{fig:MChroPS} for the Taurus region.

The results of these numerical experiments show that the trends with \nh\ found in the HROs and $\zeta$ do not arise by chance from these potential sources. 

\subsection{A product of random correlations in the polarization maps?}\label{appendix:statisticsr}

The random realizations of the magnetic field presented above are useful for characterizing the behaviour of $\zeta$.  However, they are bound to produce very different statistics compared to those in the real observations. In reality the orientations of $\vec{B}_{\perp}$ have some non-trivial correlation structure in the map.  In principle\langed{,} the difference in the HROs might be due to the different correlations of the observed $\vec{B}_{\perp}$ and of the random realizations of $\vec{B}_{\perp}$.

To evaluate whether the signal present in the HROs arises from an actual physical relationship between $\vec{B}_{\perp}$ and $\tau_{353}$, we introduce randomness by shifting the Stokes $Q$ and $U$ maps with respect to the $\tau_{353}$ map and then calculate the corresponding HROs and $\zeta$ as a function of \lognh.  The intrinsic statistical properties of the two maps are unchanged because the two maps are unchanged, only shifted.  

\begin{figure}
\centerline{
\includegraphics[width=0.5\textwidth,angle=0,origin=c]{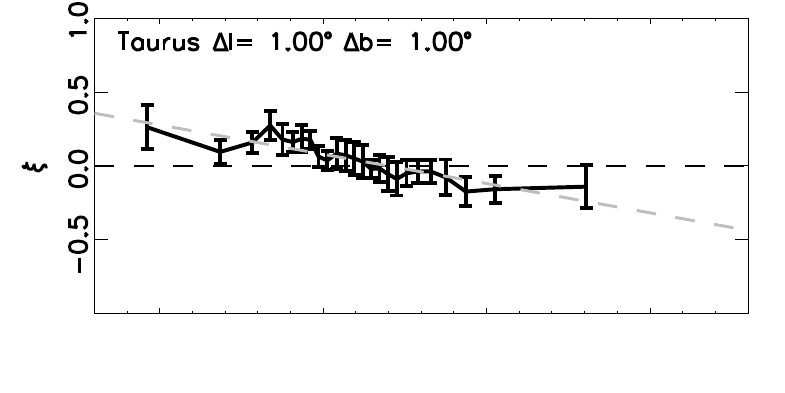}
}
\vspace{-1.1cm}
\centerline{
\includegraphics[width=0.5\textwidth,angle=0,origin=c]{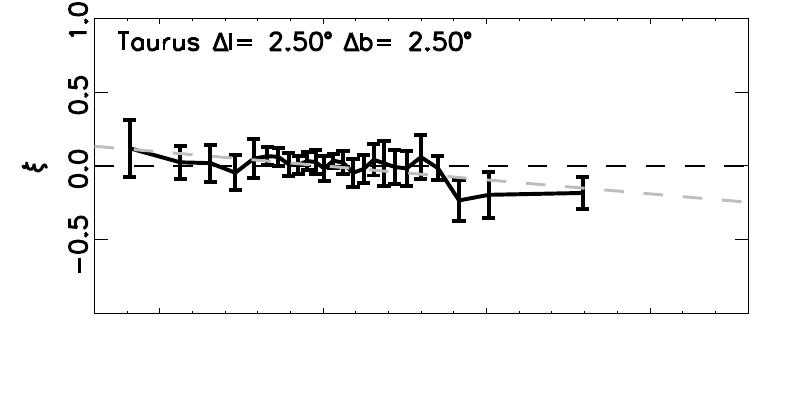}
}
\vspace{-1.1cm}
\centerline{
\includegraphics[width=0.5\textwidth,angle=0,origin=c]{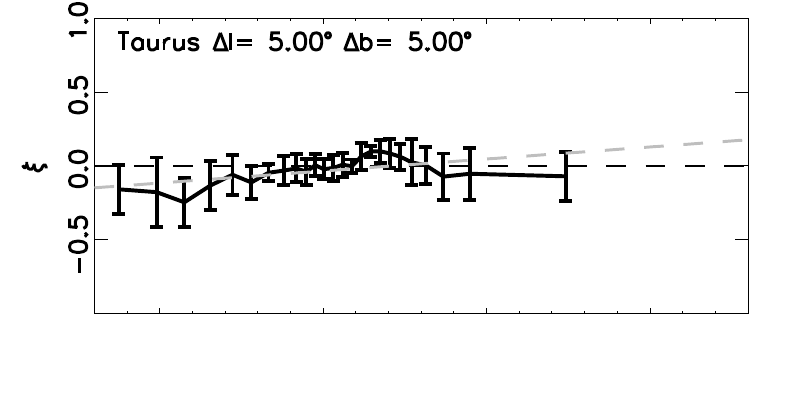}
}
\vspace{-1.1cm}
\centerline{
\includegraphics[width=0.5\textwidth,angle=0,origin=c]{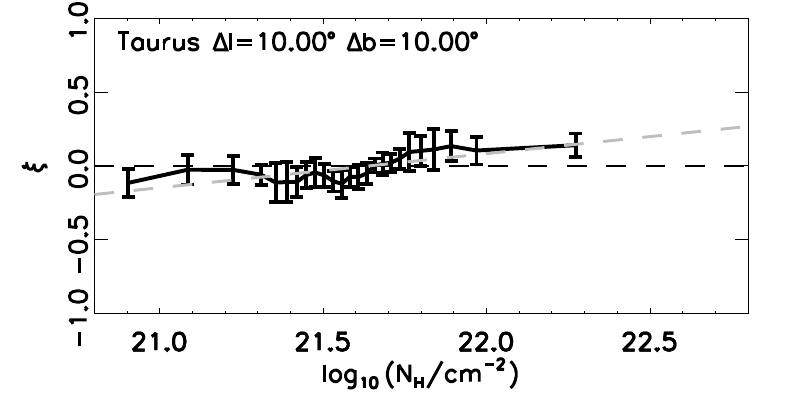}
}
\caption{Histogram shape parameter, $\zeta$, as a function of $\log_{10}(\nhd/\mbox{cm}^{-2})$ calculated from the $\tau_{353}$ map at 10\arcmin\ resolution in the Taurus region and the Stokes $Q$ and $U$ maps shifted  in Galactic longitude and latitude by the values indicated, for $\Delta l$ and $\Delta b$\langed{,} respectively.}
\label{fig:hroCorr}
\end{figure}

If the trend 
%of
\langed{in} $\zeta$ as a function of \lognh\ were arising randomly, these trends would be unchanged even for significant shifts.
Instead the results of this experiment, illustrated in Fig.~\ref{fig:hroCorr} for the Taurus region, show that the trends tend to disappear with increasing values of the size of the shift. For shifts of about 1\degr, the correlation between the magnetic field and the matter is still present, as expected from the results presented in Fig.~\ref{fig:TaurusMultiScale}, but for larger shifts the correlation is lost and the trend does not survive.  
Over the many MCs studied, the nature of the trends at large shifts appears to be random.

\subsection{Projection effects}\label{appendix:statisticsp}

We evaluate the statistical significance of the relative orientation between $\vec{B}_{\perp}$ and $\nabla\tau_{353}$ by considering the distribution of relative orientations between two vectors in 3D space compared to the distribution of relative orientation between their projections in 2D.

The projection of a vector $\vec{v}$ onto a plane normal to the unit vector $\vec{\hat{n}}$ can be written as
\begin{equation}\label{eq:uprojection}
\vec{u} = \vec{v} - (\vec{v}\cdot\vec{\hat{n}})\,\vec{\hat{n}}\, .
\end{equation}

Consider two unit vectors in 3D, $\vec{v}_{1}$ and $\vec{v}_{2}$, separated by an angle $\alpha$, such that
\begin{equation}\label{eq:angle3D}
\cos\alpha =  \vec{v}_1\cdot\vec{v}_2 \, .
\end{equation}
The angle $\beta$ between the projections of these two vectors onto a plane normal to $\vec{\hat{n}}$, $\vec{u}_{1}$, and $\vec{u}_{2}$, can be written as
\begin{center}
\begin{eqnarray}\label{eq:angle2D}
\cos\beta &=&  \frac{\vec{u}_1\cdot\vec{u}_2}{|\vec{u}_{1}||\vec{u}_{2}|} \nonumber \\ 
 &=& \frac{\left(\, \vec{v}_{1}\cdot\vec{v}_{2} - (\vec{v}_{1}\cdot\vec{\hat{n}})(\vec{v}_{2}\cdot\vec{\hat{n}})\, \right)}{\left(\, \vec{v}_{1}\cdot\vec{v}_{1}-(\vec{v}_{1}\cdot\vec{\hat{n}})^{2}\, \right)^{1/2}\, \left(\, \vec{v}_{2}\cdot\vec{v}_{2}-(\vec{v}_{2}\cdot\vec{\hat{n}})^{2}\, \right)^{1/2}}\, .
\end{eqnarray}
\end{center}
Given a particular distribution of angles between the vectors $\vec{v}_1$ and $\vec{v}_2$, this expression, which is solved numerically, is useful for evaluating the resulting distribution of angles between the projected vectors $\vec{u}_1$ and $\vec{u}_2$.
Without \langed{any} loss of generality\langed{,} we can assume that $\vec{v}_1$ is oriented along the axis of a spherical coordinate system\langed{,} such that $\vec{v}_{1}=\hat{\vec{k}}$\langed{,} and that $\vec{v}_2$ is oriented at polar angle $\theta$ and azimuth $\phi$, such that $\vec{v}_{2} = \cos\phi\sin\theta\,\vec{\hat{i}} + \sin\phi\sin\theta\,\vec{\hat{j}} + \cos\theta\,\hat{\vec{k}}$. Then, we can simulate a distribution of $\cos\alpha$ by simulating a distribution of $\cos\theta$ and generating $\phi$ with a uniform distribution.

\begin{figure}[ht!]
\vspace{-0.3cm}
\centerline{
\includegraphics[width=0.5\textwidth,angle=0,origin=c]{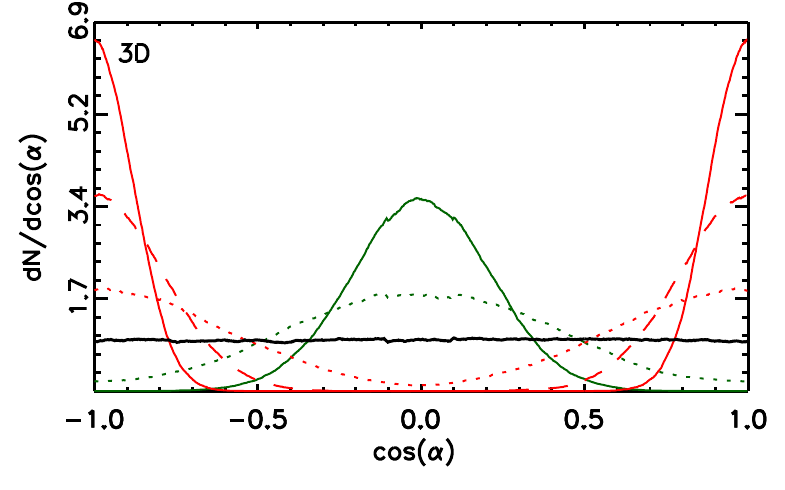}
}
\vspace{-0.3cm}
\centerline{
\includegraphics[width=0.5\textwidth,angle=0,origin=c]{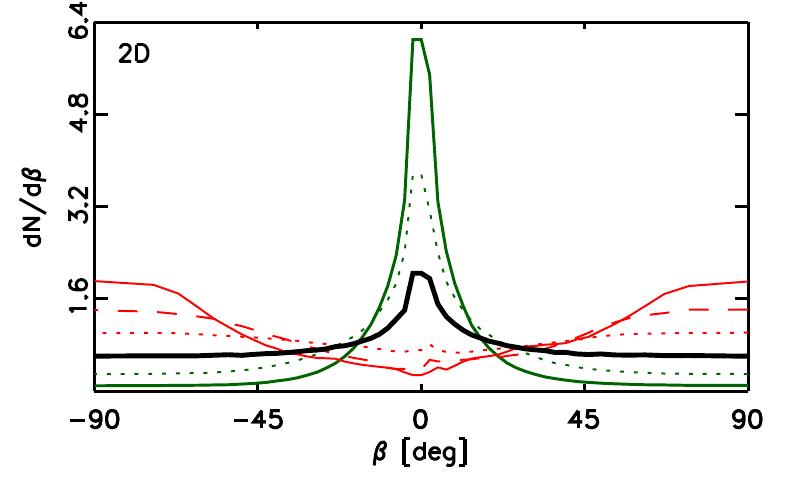}
}
\vspace{-0.3cm}
\caption{Normalized distributions of relative orientations of random vectors in 3D (top) and of their projections in 2D (bottom). The curves correspond to \preferentially perpendicular vectors (red), uniform distribution of relative orientations (black), and \preferentially parallel vectors (green). The solid, dashed, and dotted lines correspond to dispersions of the relative orientation angle $\varsigma_{\alpha}=$ 18\deg, 36\deg, and 72\deg, respectively.
}
\label{fig:orientationMC}
\end{figure}

We thus generate a set of vectors $\vec{v}_2$ that follow a particular distribution of $\cos\alpha$. We choose three examples: a uniform distribution of relative orientation between $\vec{v}_2$ and $\vec{v}_1$, $\vec{v}_2$ vectors that are 
%preferentially
\langed{mostly} parallel to $\vec{v}_{1}$, and $\vec{v}_2$ vectors that are 
%preferentially
\langed{mostly} perpendicular to $\vec{v}_{1}$. The last two are Gaussian distributions centred at $\cos\alpha = 0$ (for 
%preferentially
\langed{mostly} parallel) or $\cos\alpha=\pm1$ (for 
%preferentially
\langed{mostly} perpendicular, given the periodicity of $\alpha$) with a dispersion $\varsigma_{\alpha}$. These distributions are shown in the top panel of Fig.~\ref{fig:orientationMC} for three values of $\varsigma_{\alpha}$. 

Using Eq.~\eqref{eq:angle2D} we calculate the distribution of the angles between the projected vectors and the results are shown in the bottom panel of Fig.~\ref{fig:orientationMC}.   Projection generally tends to make vector pairs look more parallel in 2D than they are in 3D, regardless of the orientation in 3D, in agreement with the results of the Gaussian models presented in \cite{planck2014-XXXII}.
Nevertheless\langed{,} the distribution of relative orientations in 2D is qualitatively similar (though not identical) to the distribution in 3D.  
In particular, the signature of perpendicular orientation is not erased.
We conclude that the observation in 2D of $\vec{B}_{\perp}$ close to perpendicular to the column density structures is a direct indicator of the perpendicular configuration of $\vec{B}$ with respect to the density structures in 3D. 
In contrast, the parallel alignment of $\vec{B}_{\perp}$ with the structure, or no signs of preferential orientation, while suggestive of parallel orientation in 3D, does not unambiguously rule out 
the presence of some close-to-perpendicular orientations within the distribution in 3D.

% -----------this is now an appendix, formerly in main text as section 5----------------------------------------------

\section{Alternative estimates of magnetic field strength}\label{section:bestimates}

Given the historic importance of the DCF method 
\citep{davis1951a,chandrasekhar1953} and the related \hkd\ method \citep{hildebrand2009}\langed{,} we have used them to estimate the magnetic field strength in each region.  We also evaluated the mass-to-flux ratio, which is important in investigating the stability against gravitational collapse.  We now provide a critical  assessment of the applicability of these results.

\subsection{Davis-Chandrasekhar-Fermi method}

The DCF method estimates the strength of $\vec{B}_{\perp}$ in a region using the dispersion of the polarization angles $\varsigma_{\psi}$.  Assuming that the magnetic field is frozen into the gas and that the dispersion of the $\vec{B}_{\perp}$ orientation angles (or equivalently $\varsigma_{\psi}$) is due to transverse incompressible Alfv\'{e}n waves, then
\begin{equation}\label{eq:cfmethod}
B^{\textsc{DCF}}_{\perp} = \sqrt{4\,\pi\,\rho}\frac{\sigma_{v_{\parallel}}}{\varsigma_{\psi}}\, ,
\end{equation}
where $\sigma_{v_{\parallel}}$ is the dispersion of the radial velocity of the gas (Appendix~\ref{bcalc}) 
and $\rho$ %is 
the gas mass density.

We calculate $\varsigma_{\psi}$ directly from Stokes $Q$ and $U$ using
\begin{equation}\label{eq:psidispersion}
\varsigma_{\psi} =\sqrt{\left< (\Delta\psi)^{2} \right>}
\end{equation}
and
\begin{equation}\label{eq:moredispersion}
\Delta\psi = \frac{1}{2}\arctan\left(Q\left<U\right>-\left<Q\right>U\,, \,Q\left<Q\right>+\left<U\right>U \right)\, ,
\end{equation}
where  $\left<\dots\right>$  denotes an average over the selected pixels in each map \citep{planck2014-XIX}.

\begin{table*}[tmb]  % table* is a two-column table.  Drop the * for one column.
\begingroup
\newdimen\tblskip \tblskip=5pt
\caption{Magnetic properties of the selected regions.$^a$}
\label{table-magnparameters}                            % Label goes here.
\nointerlineskip
\vskip -3mm
\footnotesize
\setbox\tablebox=\vbox{
   \newdimen\digitwidth 
   \setbox0=\hbox{\rm 0} 
   \digitwidth=\wd0 
   \catcode`*=\active 
   \def*{\kern\digitwidth}
   \newdimen\signwidth 
   \setbox0=\hbox{+} 
   \signwidth=\wd0 
   \catcode`!=\active 
   \def!{\kern\signwidth}
\halign{\hbox to 1.15in{#\leaderfil}\tabskip 2.2em&
\hfil#&\hfil#&\hfil#&\hfil#&\hfil#&\hfil#&\hfil#&\hfil#\tabskip 0pt\cr
\noalign{\doubleline}
\omit\hfil Region\hfil&\hfil $\sigma_{v_{\parallel}}$ \hfil&\hfil $\varsigma_{\psi}$ \hfil&\multispan2\hfil $b$\hfil & \hfil $B^{\textsc{DCF}}_{\perp}$  \hfil & \hfil $B^{\textsc{\HIL}}_{\perp}$  \hfil &\hfil$\lambda^{\textsc{DCF}}_{\rm obs} $\hfil & \hfil$\lambda^{\textsc{\HIL}}_{\rm obs} $\hfil \cr
\omit\hfil \hfil&\hfil [km\,s$^{-1}$] \hfil&\hfil [deg] \hfil&\hfil [deg] \hfil& [rad]\hfil & \hfil[\microG]\hfil & \hfil[\microG]\hfil & & \cr
\noalign{\vskip 4pt\hrule\vskip 6pt}
%----------------------------------------------------------------------------------------------------------------
%\multirow{2}{*}{this} 
%Taurus & 24.0$\pm$0.1 &    2.5$\pm$0.0 & 46.8$\pm$1.1 &   12.9$\pm$   0.3 &   62.6$\pm$   0.7 &  0.4$\pm$   0.3 &  0.1$\pm$   0.1  \cr
Taurus &   1.2$\pm$0.5 &  43$\pm$0.1 &  \hfil23$\pm$0.05\hfil &  \hfil0.39$\pm$0.01\hfil &13$\pm$5 &  32$\pm$13 & 0.4$\pm$0.4 & 0.2$\pm$0.1 \cr
%----------------------------------------------------------------------------------------------------------------
%Ophiuchus & 24.3$\pm$   0.1 &    2.5$\pm$   0.1 &   36.1$\pm$   0.7 &    22.$\pm$   0.7 &    44.$\pm$   1.2 &  0.2$\pm$   0.2 &  0.1$\pm$   0.1 \cr
Ophiuchus & 0.9$\pm$0.4 &  29$\pm$0.3 & \hfil20$\pm$0.04\hfil &  \hfil0.36$\pm$0.01\hfil & 13$\pm$6 &  25$\pm$11 & 0.4$\pm$0.4 & 0.2$\pm$0.2 \cr
%----------------------------------------------------------------------------------------------------------------
%Lupus & 40.8$\pm$   0.2 &    3.2$\pm$   0.1 &   35.2$\pm$   0.6 &   28.8$\pm$   0.8 &   30.3$\pm$   0.8 &  0.2$\pm$   0.1 &  0.1$\pm$   0.1 \cr
Lupus &  1.5$\pm$0.6 & 46$\pm$0.7 & \hfil30$\pm$0.06\hfil & \hfil0.52$\pm$0.01\hfil  &  14$\pm$5 &  29$\pm$11 & 0.3$\pm$0.2 & 0.2$\pm$0.1  \cr
%----------------------------------------------------------------------------------------------------------------
%Cham.-Musca & 15.3$\pm$   0.1 &    1.7$\pm$   0.1 &   24.6$\pm$   0.2 &   21.3$\pm$   0.7 &   47.4$\pm$   1.5 &  0.1$\pm$   0.1 &  0.1$\pm$   0.0  \cr
Chamaeleon-Musca &  1.0$\pm$0.4 &  36$\pm$0.3 & \hfil23$\pm$0.05\hfil & \hfil0.40$\pm$0.01\hfil &  12$\pm$5 &  27$\pm$11 & 0.4$\pm$0.3 & 0.2$\pm$0.2 \cr
%----------------------------------------------------------------------------------------------------------------
%CrA & 26.6$\pm$   0.1 &    1.8$\pm$   0.0 &   44.2$\pm$   1.1 &   13.1$\pm$   0.3 &   29.1$\pm$   0.2 &  0.0$\pm$   0.1 &  0.0$\pm$   0.0  \cr
Corona Australis (CrA) &  0.6$\pm$0.2 &  59$\pm$0.1 & \hfil30$\pm$0.07\hfil & \hfil0.52$\pm$0.01\hfil &   5$\pm$2 &  12$\pm$\phantom{0}5 & 0.9$\pm$0.9 & 0.3$\pm$0.3 \cr
%----------------------------------------------------------------------------------------------------------------
\noalign{\vskip 4pt\hrule\vskip 6pt}
%----------------------------------------------------------------------------------------------------------------
%Aquila Rift & 36.1$\pm$13.5 & & 33.6$\pm$0.4 & 23.0$\pm$ 1.0 & 18.6$\pm$ 7.0 & 3.5 & 4.4 \cr
Aquila Rift &  1.9$\pm$0.6 &  43$\pm$0.5 & \hfil23$\pm$0.09\hfil & \hfil0.40$\pm$0.01\hfil &  20$\pm$6 &  50$\pm$15 & 0.3$\pm$0.2 & 0.1$\pm$0.1 \cr
%----------------------------------------------------------------------------------------------------------------
%Perseus & 33.0$\pm$13.2 & & 44.7$\pm$1.0 & 26.3$\pm$1.1 & 39.3$\pm$15.7 & 1.2 & 0.8 \cr
Perseus &  1.5$\pm$0.6 &  38$\pm$0.3 & \hfil29$\pm$0.11\hfil & \hfil0.50$\pm$0.01\hfil &  17$\pm$7 &  30$\pm$11 & 0.3$\pm$0.3 & 0.2$\pm$0.2 \cr
%----------------------------------------------------------------------------------------------------------------
\noalign{\vskip 4pt\hrule\vskip 6pt}
%IC5146 & 61.2$\pm$20.4 & & 27.9$\pm$  0.3 & 42.3$\pm$2.0 & 21.8$\pm$ 7.3 & 0.8 & 1.5 \cr
IC\,5146 &  1.7$\pm$0.6 &  69$\pm$0.1 & \hfil49$\pm$0.11\hfil & \hfil0.85$\pm$0.01\hfil &  11$\pm$4 &  18$\pm$\phantom{0}6 & 0.5$\pm$0.3 & 0.3$\pm$0.2 \cr
%---------------------------------------------------------------------------------------------------------------
%Cepheus & 31.8$\pm$12.5 & & 34.4$\pm$0.3 & 49.6$\pm$2.3 & 54.4$\pm$21.6 & 0.7 & 0.7 \cr
Cepheus &  1.6$\pm$0.6 &  43$\pm$0.2 & \hfil20$\pm$0.04\hfil & \hfil0.35$\pm$0.01\hfil &  16$\pm$6 &  47$\pm$18 & 0.3$\pm$0.1 & 0.1$\pm$0.0 \cr
%----------------------------------------------------------------------------------------------------------------
%Orion & 5.2$\pm$ 3.6 & & 38.0$\pm$0.7 & 22.5$\pm$ 0.9 & 221.8$\pm$151.8 & 1.9 & 0.2 \cr
Orion &  1.7$\pm$0.6 &  36$\pm$0.1 & \hfil26$\pm$0.06\hfil & \hfil0.45$\pm$0.01\hfil &  20$\pm$7 &  38$\pm$14 & 0.3$\pm$0.3 & 0.2$\pm$0.2 \cr
%----------------------------------------------------------------------------------------------------------------
\noalign{\vskip 3pt\hrule\vskip 4pt}}}
%\endPlancktable                    % ends one-column \halign
\endPlancktablewide                 % ends two-column \halign
\tablenote a 
Tabulated values are: the velocity dispersion, $\sigma_{v_{\parallel}}$; 
the dispersion of the polarization orientation angle, $\varsigma_{\psi}$; 
the turbulent contribution to the angular dispersion, $b$, calculated from a linear fit to Eq~\eqref{eq:strucfuncfit}; 
the magnetic field strengths, $B^{\textsc{DCF}}_{\perp}$ and $B^{\textsc{\HIL}}_{\perp}$, calculated with the DCF method, of Eq.~\eqref{eq:cfmethod}, and the \hkd\ method, of Eq.~\eqref{eq:cfmethodHild}, with $b$ values obtained from a fit to Eq.~\eqref{eq:strucfuncfit} in the range $50$\arcmin$\le \ell \le 200$\arcmin; 
and their corresponding observed mass-to-flux ratios, $\lambda^{\textsc{DCF}}_{\rm obs}$ and $\lambda^{\textsc{\HIL}}_{\rm obs}$. 
The reported uncertainties are from the appropriate propagation of errors.\par
\endgroup
\end{table*}

%\subsection{\citeauthor{hildebrand2009} method}
\subsection{Davis-Chandrasekhar-Fermi plus structure function method}

As described by \cite{hildebrand2009}, the \hkd\ method 
characterizes the magnetic field dispersion about local structured fields by considering the difference in angle,  $\Delta\psi(\boldsymbol{\ell}) = \psi(\vec{x}) - \psi(\vec{x+\boldsymbol{\ell}})$, between pairs of $\vec{B}_\perp$ vectors separated by displacements $\boldsymbol{\ell}$ in the plane of the sky. Assuming that the angle differences are statistically isotropic (i.e., they depend only on $\ell=|\boldsymbol{\ell}|$ and not on the orientation of $\boldsymbol{\ell}$), they can binned by distance, $\ell$.  From the $N(\ell)$ pairs of $\vec{B}_\perp$ vectors for that bin, the square of the second-order structure function is
\begin{equation}\label{eq:strucfunc}
S^{2}_{2}(\ell) = \left< [\Delta\psi(\vec{x},\ell)]^{2} \right>_{x} =  \left<\frac{1}{N(\ell)}\sum\limits_{i=1}^{N(\ell)} (\Delta\psi_{x,i})^{2} \right>_{x}\, ,
\end{equation}
\juan{as introduced by \cite{kobulnicky1994} and \cite{falceta2008}}.
In terms of the Stokes parameters, each term in the sum can be written
\begin{equation}\label{eq:dispfunc}
\Delta\psi_{x,i} = \frac{1}{2}\arctan\left(Q_{i}U_{x}-Q_{x}U_{i}\,, \,Q_{i}Q_{x}+U_{i}U_{x}\right)\, ,
\end{equation}
where the subscripts $x$ and $i$ represent the central and displaced positions, respectively.

\cite{hildebrand2009} 
%assumed
\langed{assume} that $\vec{B}(\vec{x})$ is composed of a large-scale structured field, $\vec{B}_{0}(\vec{x})$, and a random component, $\vec{B}_{{\rm r}}(\vec{x})$, which are statistically independent of each other. Assuming that $\vec{B}_{0}(\vec{x})$ is a smoothly varying quantity, its contribution to $S^{2}_{2}(\ell)$ should increase in proportion to $\ell^{2}$ for distances that are much smaller than the scales at which $\vec{B}_{0}$ itself fluctuates. Additionally, assuming that turbulence occurs 
%at
\langed{on} scales that are small enough to completely decorrelate $\vec{B}_{{\rm r}}$ for the range of scales probed by the displacements $\ell$, \cite{hildebrand2009} 
derived the approximation
\begin{equation}\label{eq:strucfuncfit}
S^{2}_{2}(\ell) = b^{2} + m^{2}\ell^{2}\, ,
\end{equation}
where the two terms on the right-hand side give the contributions from the random and large-scale magnetic fields, respectively. These assumptions are not necessarily valid for the range of scales in the \planck\ data.

\cite{hildebrand2009} used Eq.~\eqref{eq:strucfuncfit} to estimate $b^{2}$ as the intercept 
and then related $b$ to the ratio of the random to the large-scale magnetic field strength, both projected onto the plane of the sky, through
\begin{equation}\label{eq:bratios}
\frac{\left<B^{2}_{{\rm r},\perp}\right>^{1/2}}{B_{0,\perp}} = \frac{b}{\sqrt{2-b^{2}}}\, ,
\end{equation}
where $\left<B^{2}_{{\rm r},\perp}\right>^{1/2}$ stands for the root mean square (rms) variations about the large-scale magnetic field, $B_{0,\perp}$.
The same assumptions that result in Eq.~\eqref{eq:cfmethod} -- 
i.e., considering only incompressible and isotropic turbulence, magnetic fields frozen into the gas, and dispersion of the $\vec{B}_{\perp}$ orientation originating in transverse incompressible Alfv\'{e}n waves -- lead to
\begin{equation}\label{cf1953}
\frac{\left<B^{2}_{{\rm r,\perp}}\right>^{1/2}}{B_{0,\perp}} = \frac{\sigma_{v_{\parallel}}}{V_{\textsc A,\perp}}\, ,
\end{equation}
where 
\begin{equation}\label{AlfvenSpeed}
V_{\textsc A,\perp} = \frac{B_{0,\perp}}{\sqrt{4\pi\rho}}
\end{equation}
is the speed of the transverse incompressible Alfv\'{e}n waves.
Combining Eqs.~\eqref{eq:bratios}\,--\,\eqref{AlfvenSpeed} results in
\begin{equation}\label{eq:cfmethodHild}
B^{\textsc{\HIL}}_{\perp} \equiv B^{\textsc{\HIL}}_{0,\perp} = \sqrt{4\,\pi\,\rho}\,\frac{\sigma_{v_{\parallel}}\,\sqrt{2-b^{2}}}{b}\, .
\end{equation}

%%%%%%%% CALCULATION

\begin{figure*}%[ht!]
\centerline{
\includegraphics[width=0.34\textwidth,angle=0,origin=c]{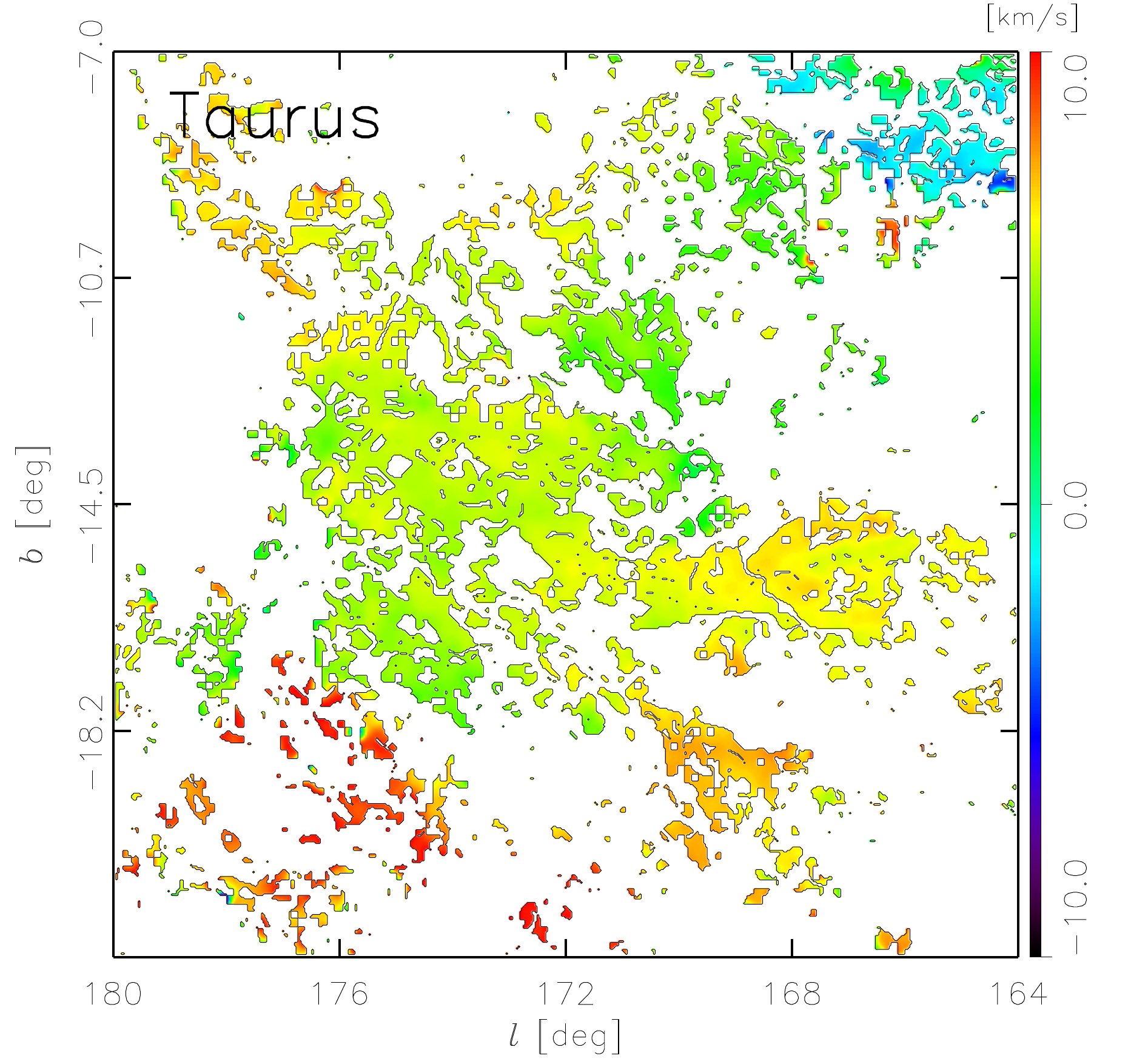}
\hspace{-0.35cm}
\includegraphics[width=0.34\textwidth,angle=0,origin=c]{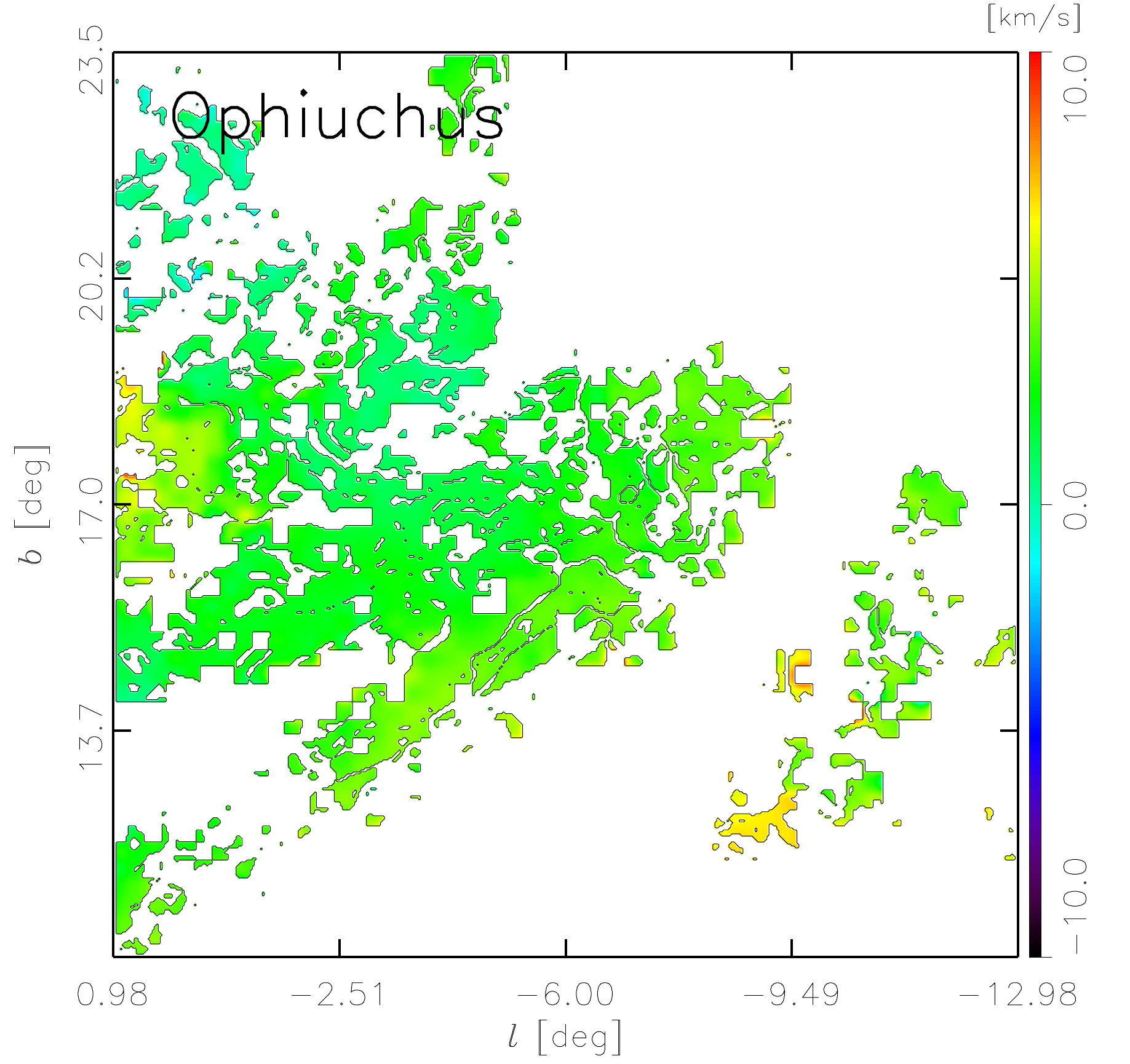}
\hspace{-0.35cm}
\includegraphics[width=0.34\textwidth,angle=0,origin=c]{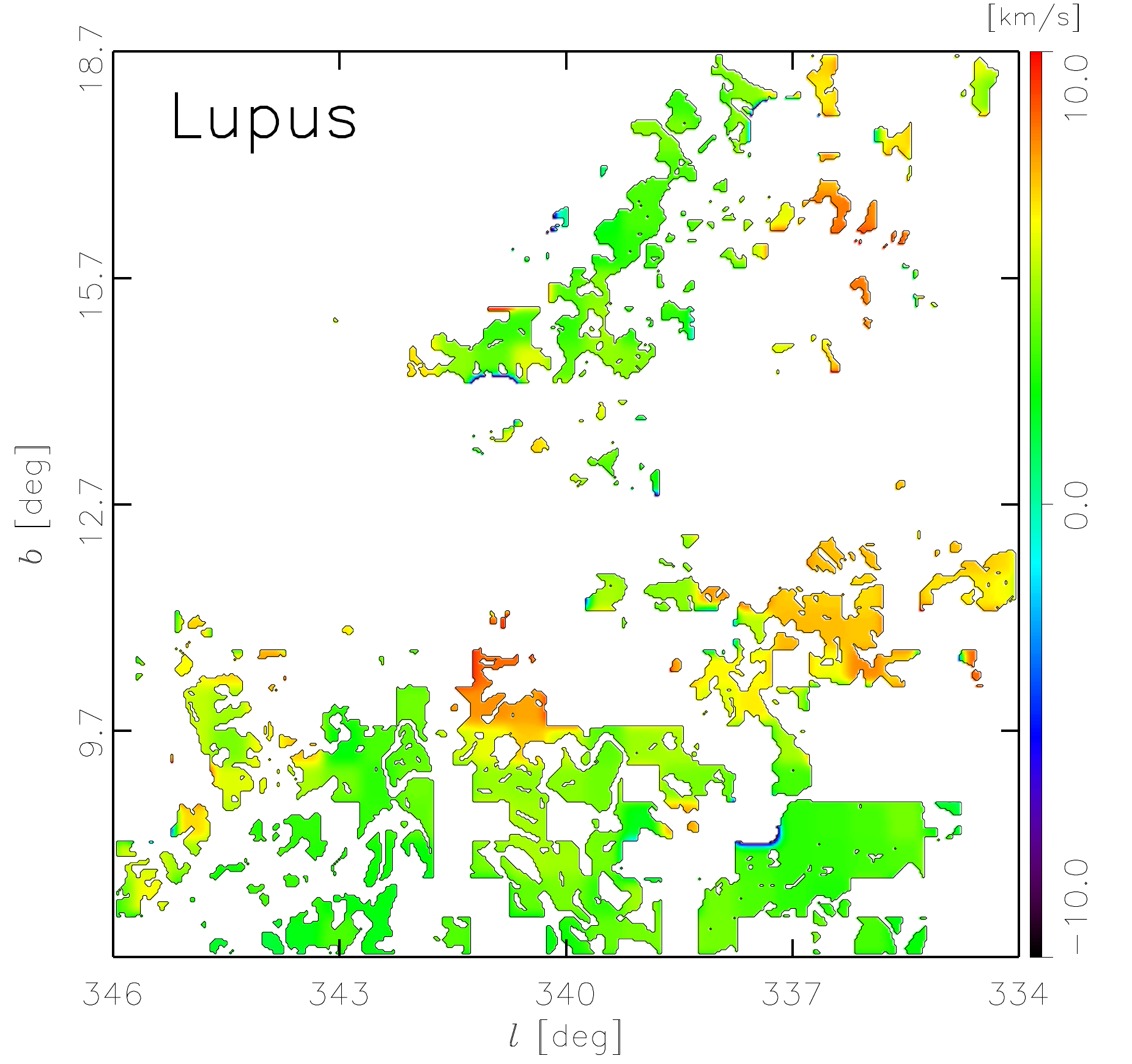}
}
\vspace{0.01cm}
\centerline{
\includegraphics[width=0.34\textwidth,angle=0,origin=c]{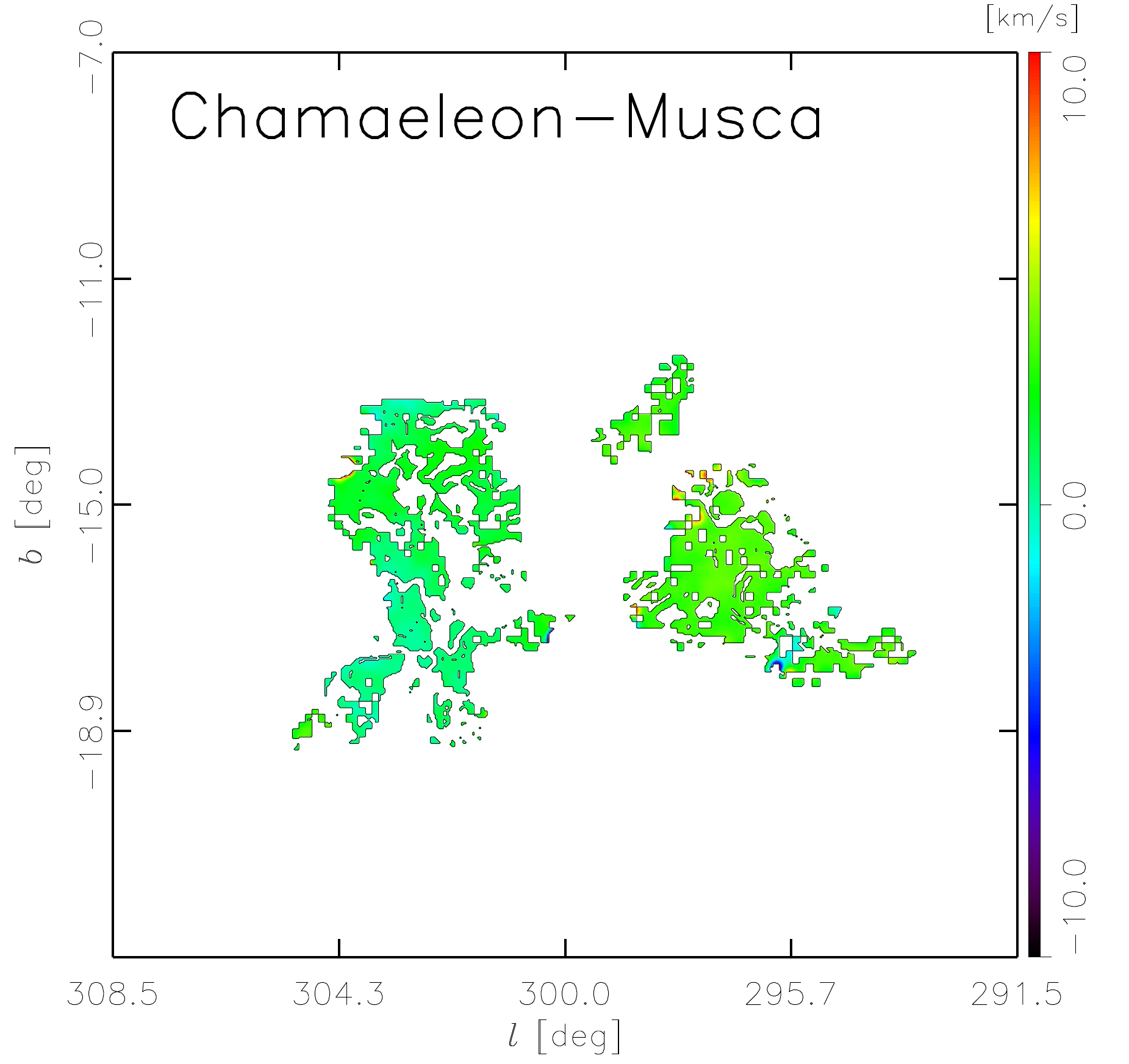}
\hspace{-0.2cm}
\includegraphics[width=0.34\textwidth,angle=0,origin=c]{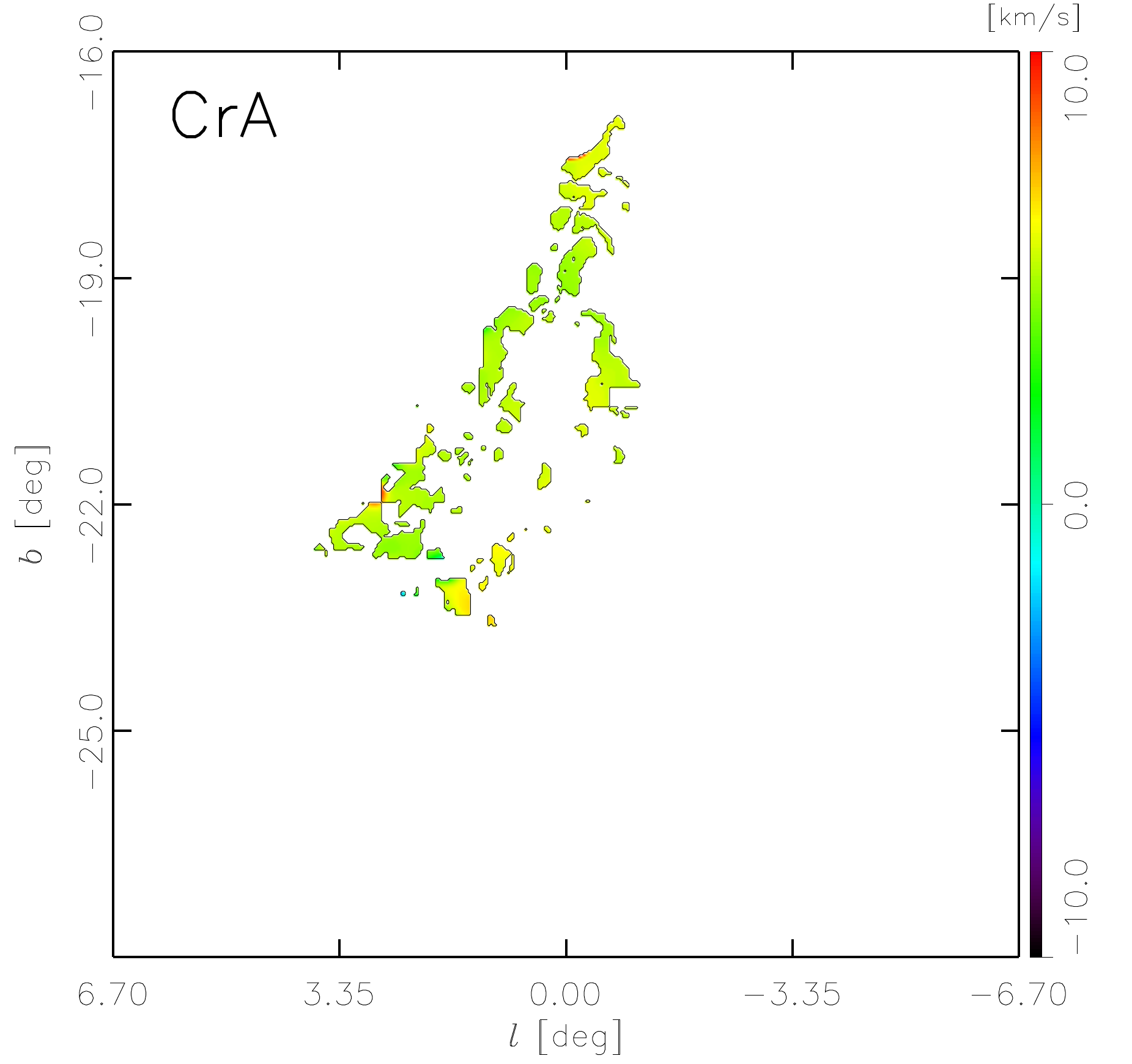}
}
\vspace{0.01cm}
\centerline{
\includegraphics[width=0.34\textwidth,angle=0,origin=c]{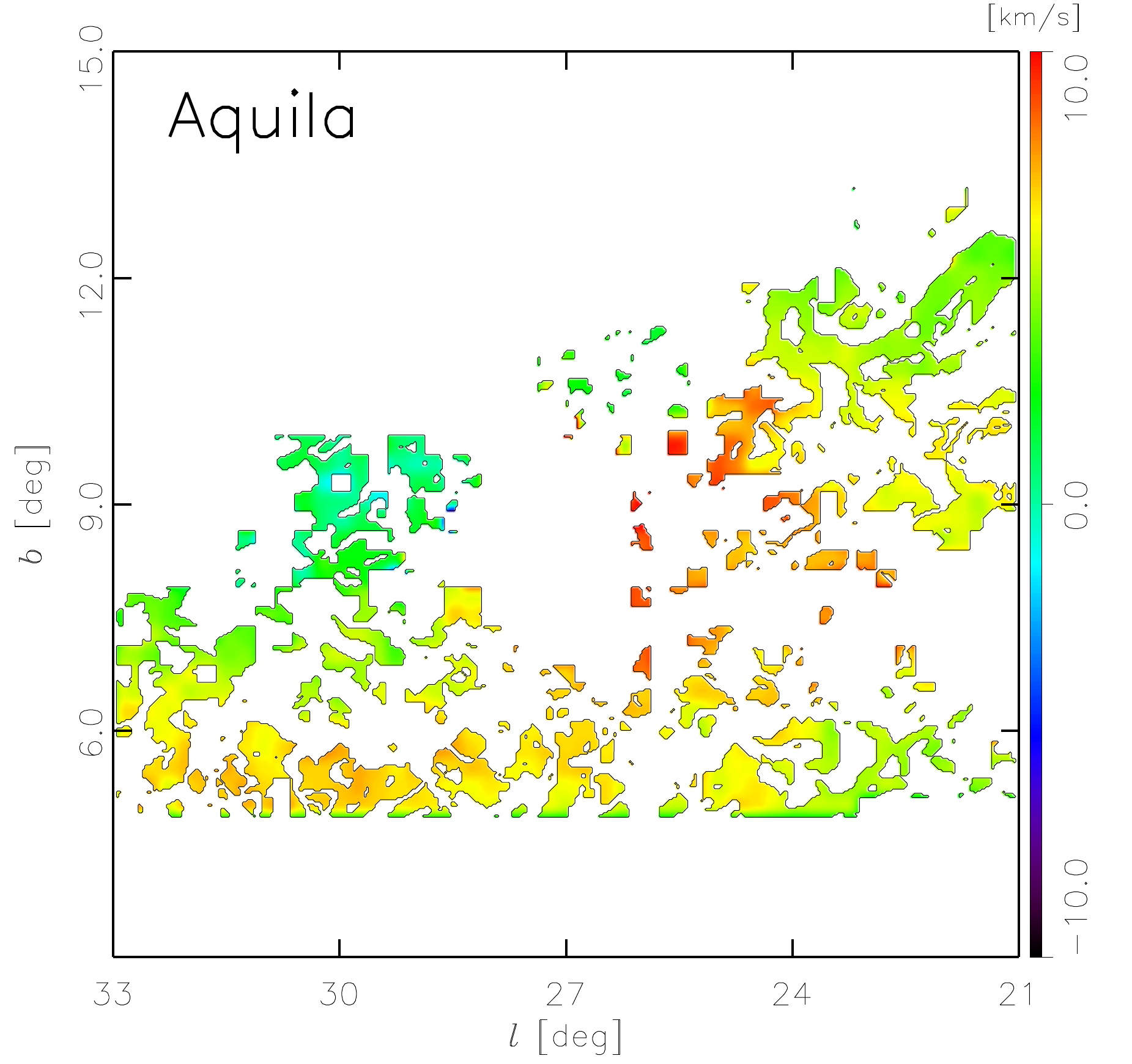}
\hspace{-0.15cm}
\includegraphics[width=0.34\textwidth,angle=0,origin=c]{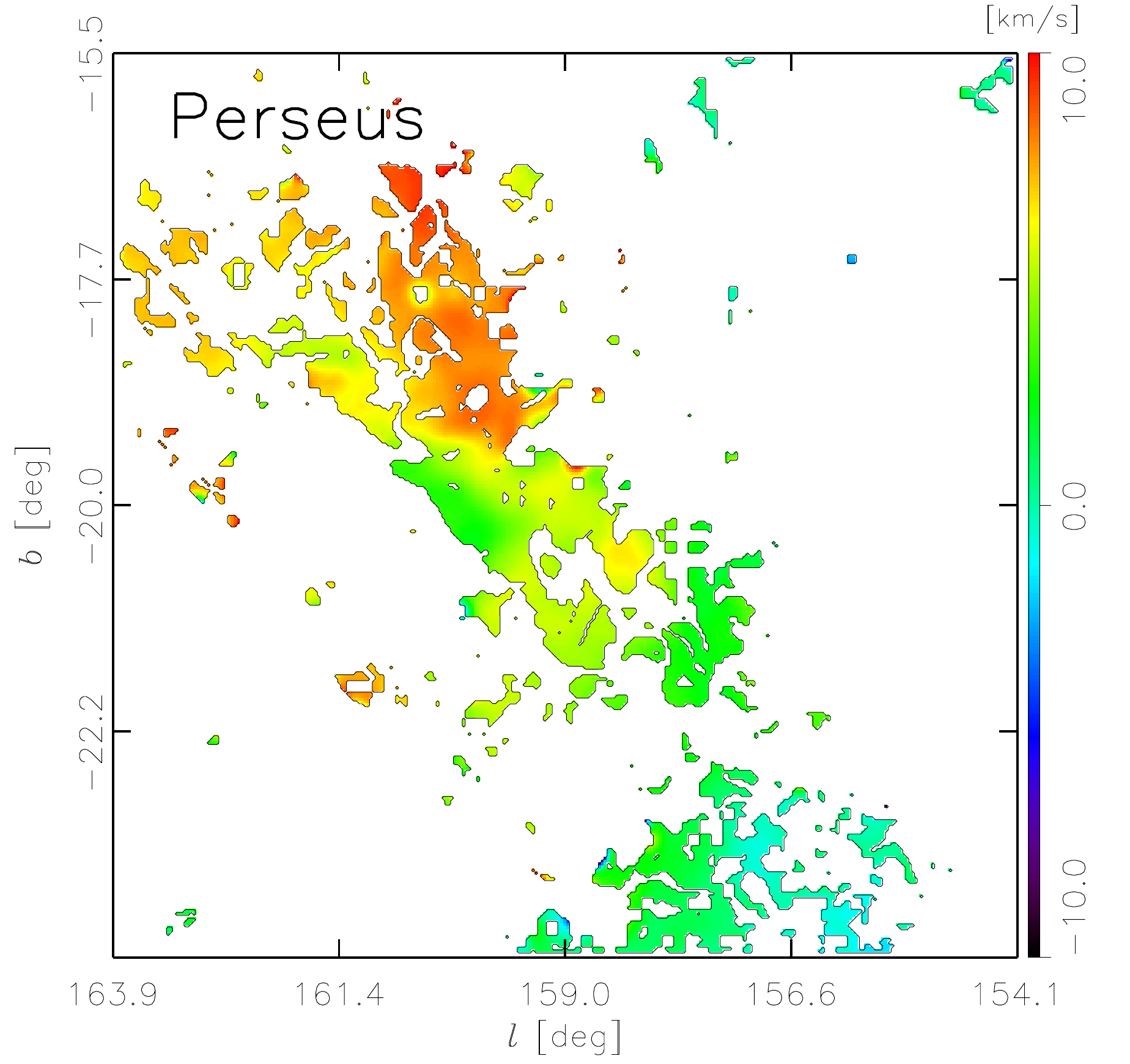}
}
\vspace{0.01cm}
\centerline{
\includegraphics[width=0.34\textwidth,angle=0,origin=c]{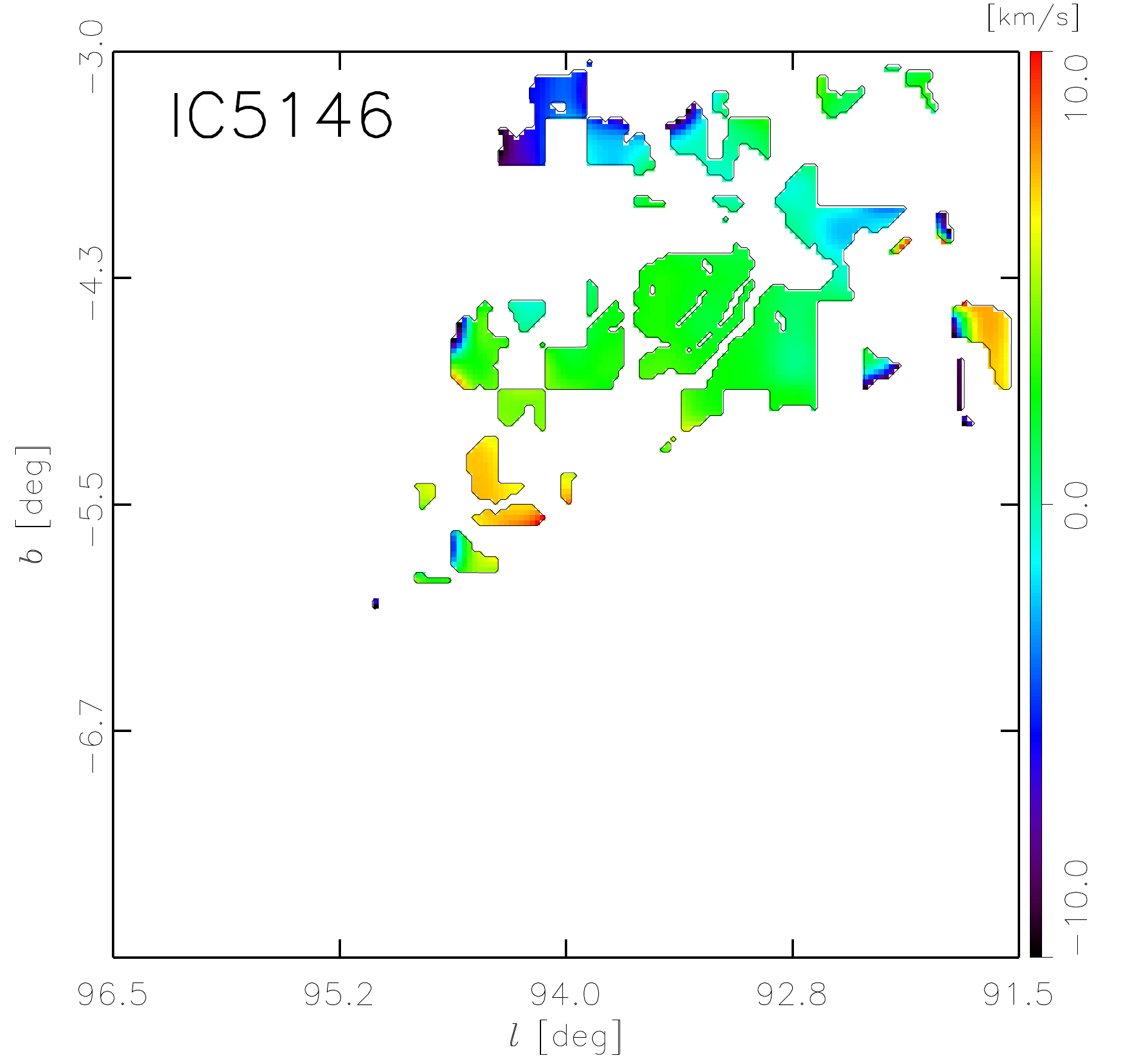}
\hspace{-0.35cm}
\includegraphics[width=0.34\textwidth,angle=0,origin=c]{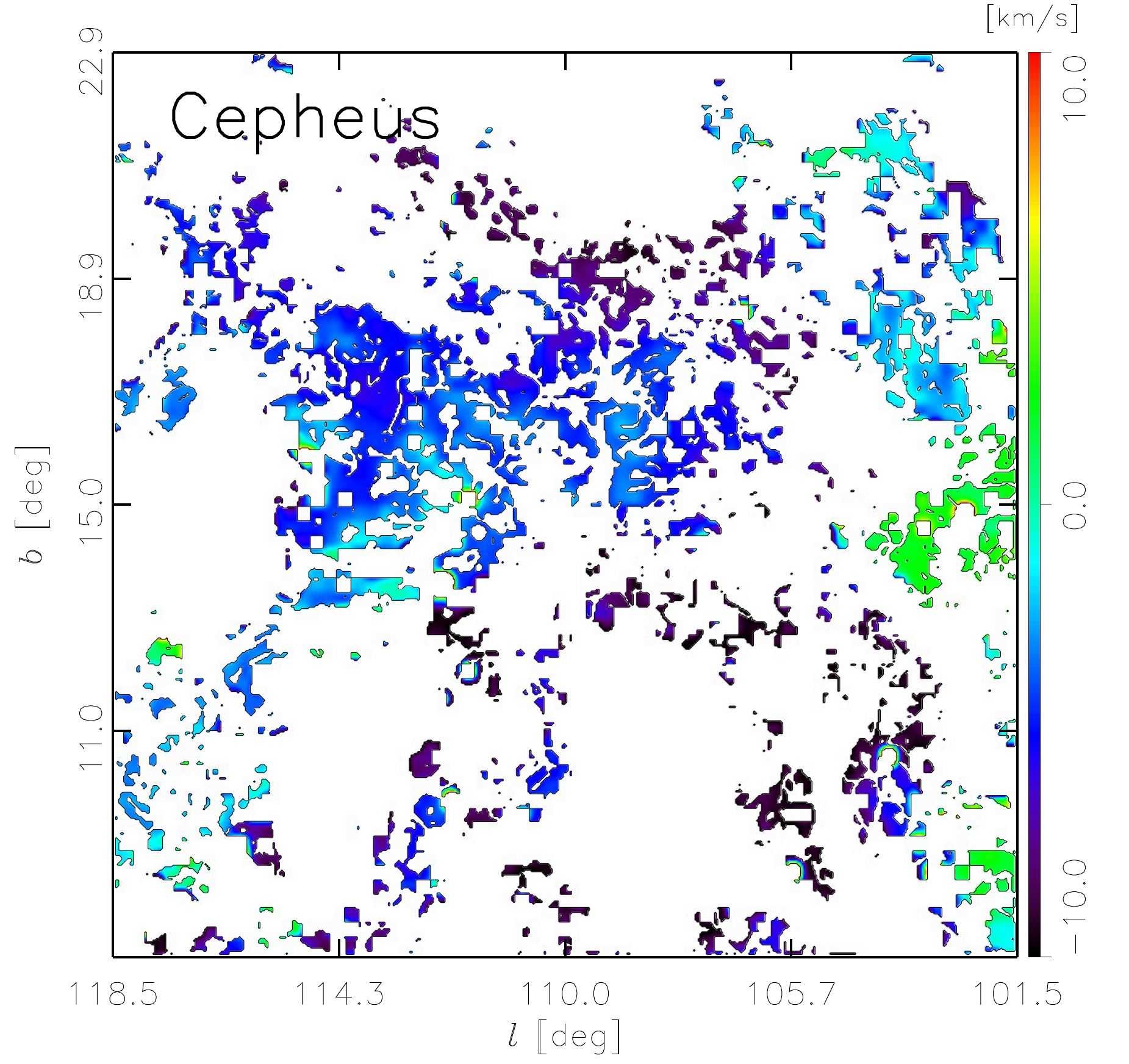}
\hspace{-0.35cm}
\includegraphics[width=0.34\textwidth,angle=0,origin=c]{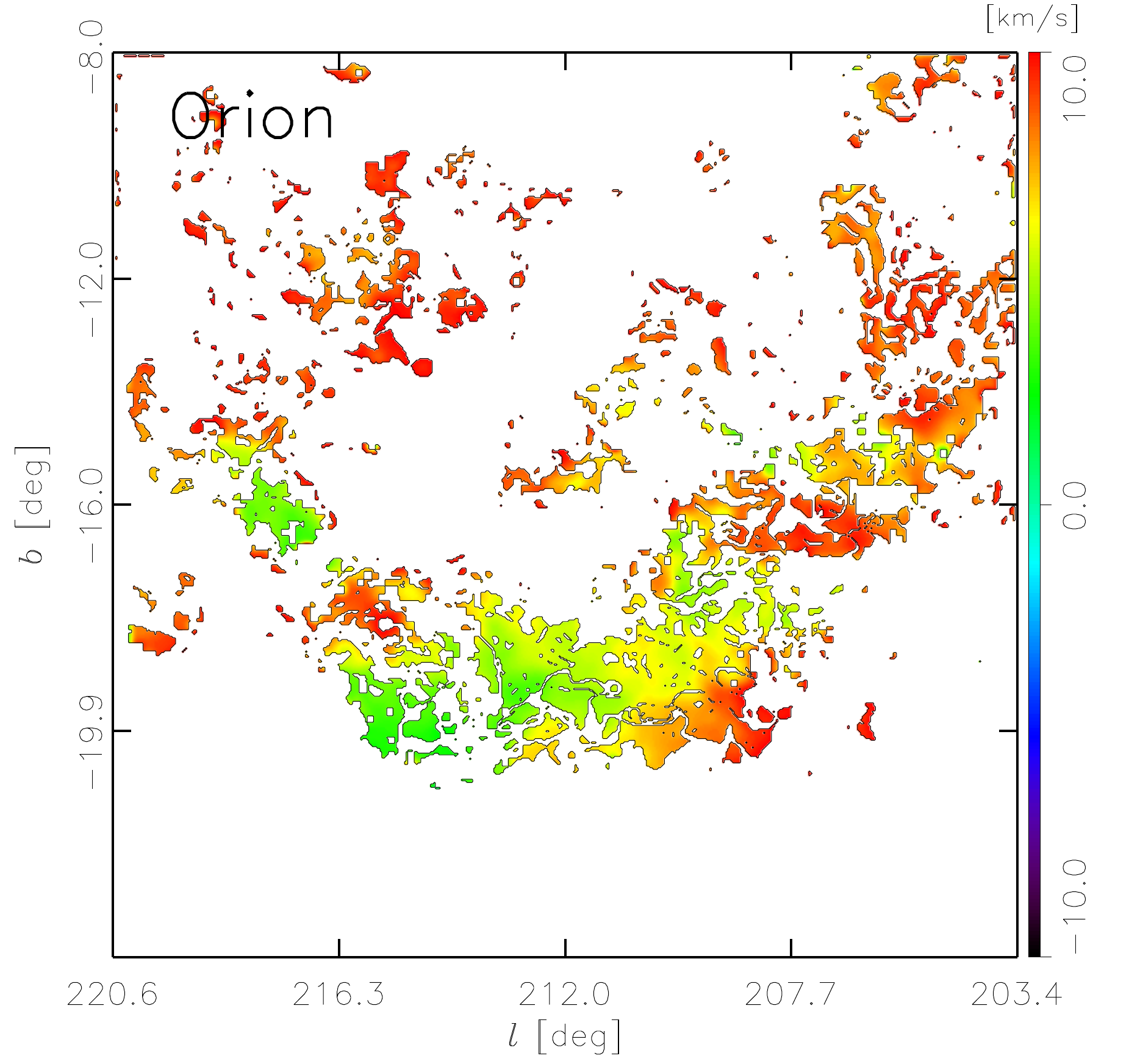}
}
\vspace{-0.15cm}
\caption{Line-of-sight centroid velocity $v_{\parallel}$ maps inferred from CO emission \citep{dame2001}. 
Areas shown have sufficient S/N in the polarization observations (second polarization criterion in Appendix~\ref{appendix:selectionp}) for the DCF and \hkd\ analyses and lie in the range $-10<v_{\parallel}/\mbox{(km\,s$^{-1}$)}<10$.
}
\label{fig:VelRegions1}
\end{figure*}

%%%%% for the HKD only %%%%%%%%%%%%%%%%%%%%%%%%%%%%%%%

\begin{figure*}%[ht!]
\centerline{
\includegraphics[width=0.50\textwidth,angle=0,origin=c]{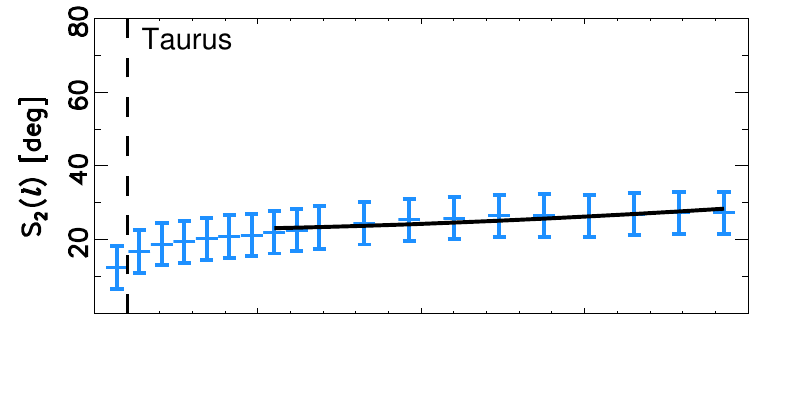}
\hspace{-0.35cm}
\includegraphics[width=0.50\textwidth,angle=0,origin=c]{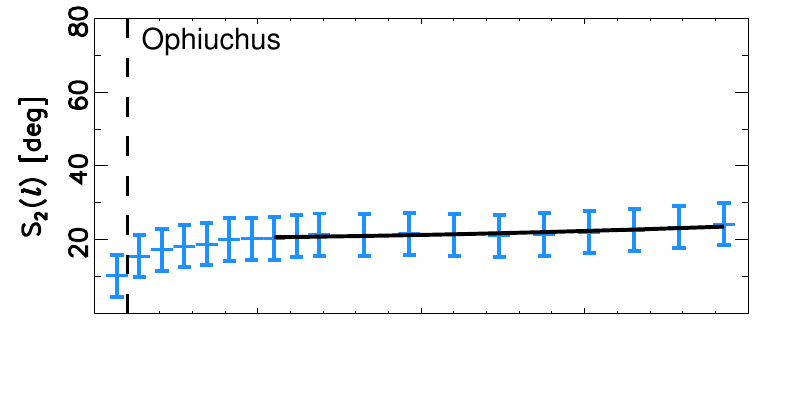}
}
\vspace{-1.1cm}
\centerline{
\includegraphics[width=0.50\textwidth,angle=0,origin=c]{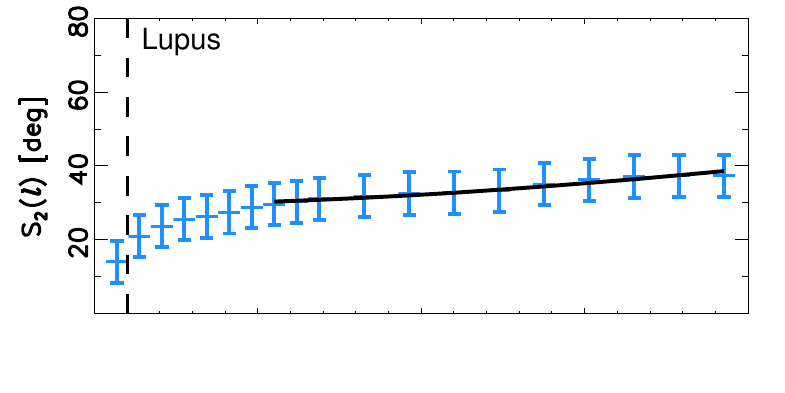}
\hspace{-0.35cm}
\includegraphics[width=0.50\textwidth,angle=0,origin=c]{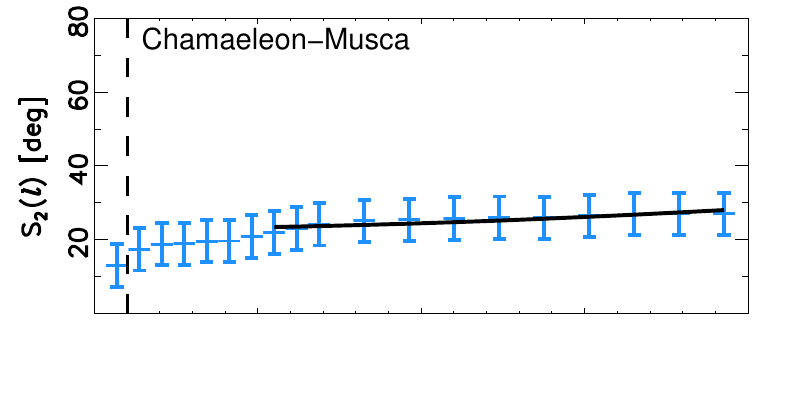}
}
\vspace{-1.1cm}
\centerline{
\includegraphics[width=0.50\textwidth,angle=0,origin=c]{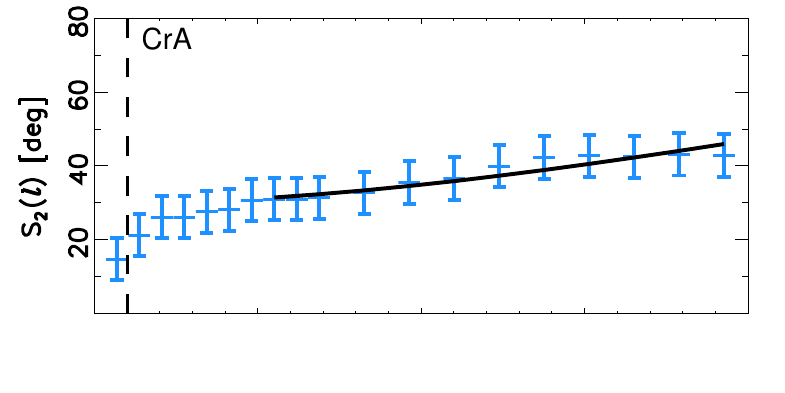}
\hspace{-0.35cm}
\includegraphics[width=0.50\textwidth,angle=0,origin=c]{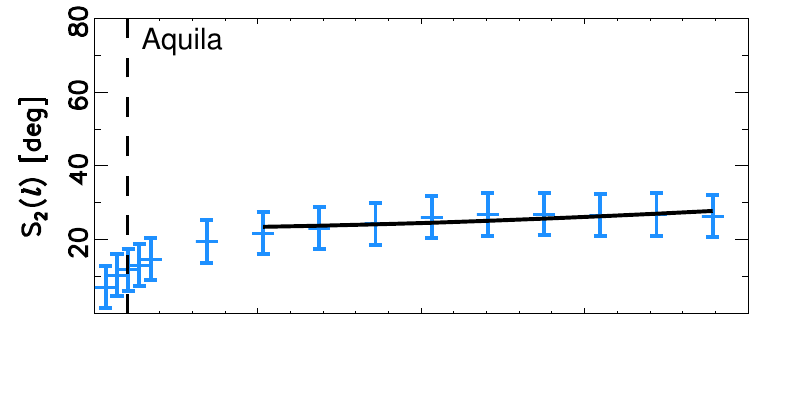}
}
\vspace{-1.1cm}
\centerline{
\includegraphics[width=0.50\textwidth,angle=0,origin=c]{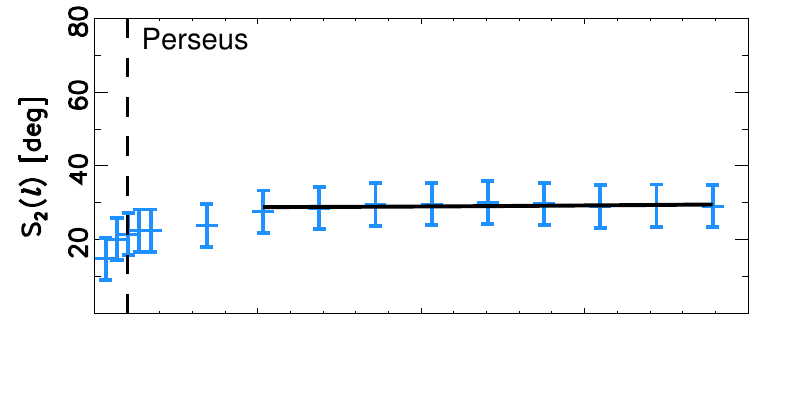}
\hspace{-0.35cm}
\includegraphics[width=0.50\textwidth,angle=0,origin=c]{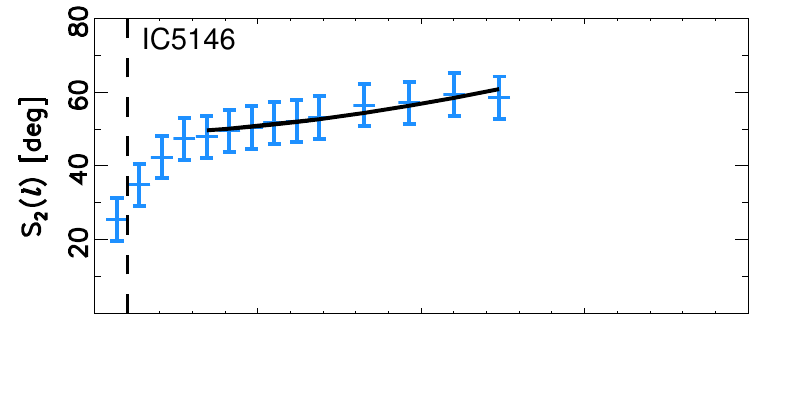}
}
\vspace{-1.1cm}
\centerline{
\includegraphics[width=0.50\textwidth,angle=0,origin=c]{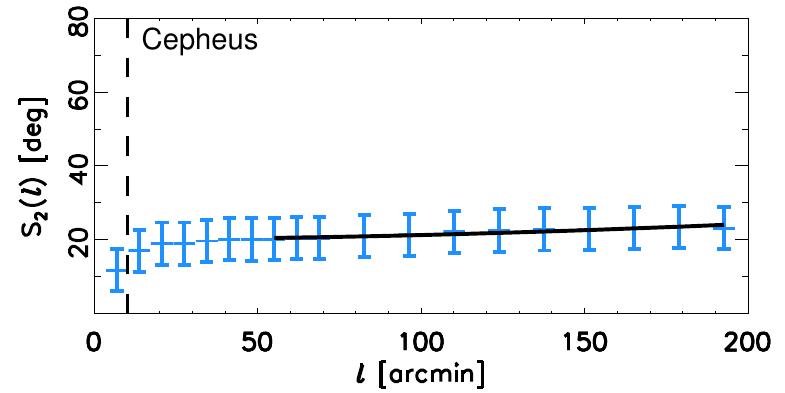}
\hspace{-0.35cm}
\includegraphics[width=0.50\textwidth,angle=0,origin=c]{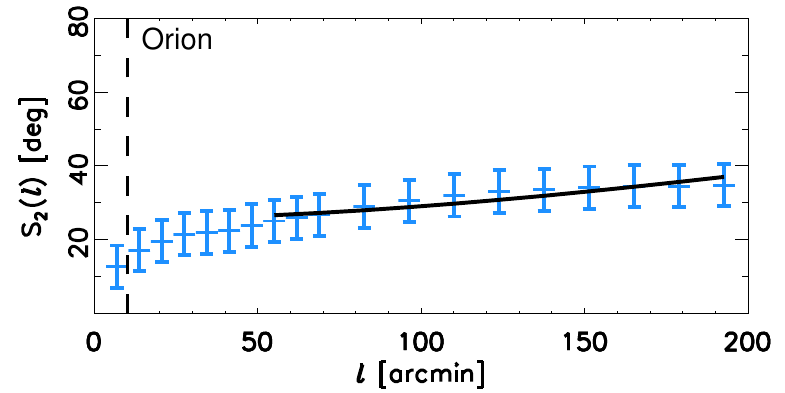}
}
\caption{Structure function $S_{2}(\ell)$ calculated from the $Q$ and $U$ maps of the selected regions following Eq.~\eqref{eq:strucfunc}. The black lines indicate the fits given by Eq.~\eqref{eq:strucfuncfit}.  The vertical dashed line marks the common 10\arcmin\ resolution of the data used in the analysis; the data are correlated for 
%small
\langed{low} values of $\ell$ causing the drop in $S_{2}(\ell)$. 
}
\label{fig:StrucFunc}
\end{figure*}

%%%%% end: for the HKD only %%%%%%%%%%%%%%%%%%%%%%%%%%%%%%%

\subsection{Calculation}\label{bcalc}

The estimates of velocity dispersion $\sigma_{v_{\parallel}}$ and mass density $\rho$ are 
%in common for
\langed{the same in} the two methods (DCF and \hkd).

We obtain $\sigma_{v_{\parallel}}$ from the most complete CO emission-line survey of the Milky Way currently available, that of \cite{dame2001}. This data set consists of 488\,000 spectra that beam-sample (1/8\deg) a set of MCs and the Galactic plane over a strip from 4\deg\ to 10\deg\ wide in latitude.
We find that material in the MCs has centroid velocities in the range $-10<v_{\parallel}/\mbox{(km\,s$^{-1}$)}<10$ in the first moment map.  Furthermore, for the calculations below\langed{,} we need to select pixels with sufficient S/N in the polarization observations and so use the second criterion described in Appendix~\ref{appendix:selectionp}.  
This combination of cuts selects the areas illustrated in Fig.~\ref{fig:VelRegions1}.
Over these areas we calculate the average of the velocity dispersion from the second moment map and use this value as $\sigma_{v_{\parallel}}$.  
This is tabulated along with other properties in Table~\ref{table-magnparameters}.

% MASS %%%%%%%%%%%%%%%%%%%%%%%%%%%%%%
The mass density, $\rho = \mu\,n\,m_{\rm p}$, is the product of the proton mass, $m_{\rm p}$, the mean molecular weight per hydrogen molecule, $\mu=2.8$, and the mean number density, $n$.  We require a value of $n$, which for the discussions here we approximate to be $100\,$cm$^{-3}$.  This is a typical value for MCs \citep{draine2011} and in rough agreement with the column densities and cloud sizes presented in Table~\ref{table-fields}. In practice, $n$ varies from one cloud to the next\langed{,} and its estimation involves the assumption of a particular cloud geometry, resulting in additional uncertainties that are not considered in this study.

%%%%% for the DCF only %%%%%%%%%%%%%%%%%%%%%%%%%%%%%%%

For the DCF method, the
dispersion of the orientation angle $\varsigma_{\psi}$ corresponds to the centred second moment of the $\psi$ distribution evaluated in pixels within the selected region. Following the analysis presented in \cite{planck2014-XIX} for $\mathcal{S}(\ell)$, the variance on $\varsigma_{\psi}$ can be expressed as
\begin{equation}\label{eq:varsigmadebiassigma}
\sigma^{2}_{\varsigma_{\psi}} = \frac{1}{N^{2}\varsigma^{2}_{\psi}}\left[\left(\sum\limits_{i=1}^{N}\varsigma_{i}\right)^{2}\sigma^{2}_{\psi} + \sum\limits_{i=1}^{N}\varsigma^{2}_{i}\sigma^{2}_{\psi_{i}} \right]\, ,
\end{equation}
where $\psi_{i}$ and $\sigma_{\psi_i}$ are the orientation angle and uncertainty for each polarization \langed{pseudo-vector}, 
$N$ is the number of $\psi$ measurements,
$\varsigma_{i} = \psi_{i}-\left<\psi\right>$, and 
$\left<\psi\right>$ is the average orientation angle in each region. 
Like other quadratic functions, $\sigma^{2}_{\varsigma_{\psi}}$ is biased positively when noise is present, leading to an overestimation of this quantity. However, given that we limit our analysis to polarization measurements with high S/N and that the uncertainties in $\sigma_{v_{\parallel}}$ are considerably larger, the bias correction does not have a significant effect on our estimates. We apply Eq.~\eqref{eq:cfmethod} to estimate $B^{\textsc{DCF}}_{\perp}$ and propagate the errors to obtain the values listed in Table~\ref{table-magnparameters}.

For the \hkd\ method, 
to estimate the parameter $b$\langed{,} we first calculate $S_{2}^2(\ell)$ for each of the regions.
The resolution of the data used is 10\arcmin.  
We evaluate $S_{2}^2(\ell)$ in steps of 3\parcm44 for lags 0\arcm\,$<\ell<34$\parcm4 and in steps of 34\parcm4 for lags 40\arcm\,$<\ell<200$\arcm. For each considered lag $\ell$, $\Delta\psi_{x,i}$ is evaluated pixel by pixel, considering all the pixels located in an annulus with radius $\ell$. The term $\Delta\psi_{x,i}$ is only considered in the calculation of $S_{2}^2(\ell)$ if there are at least three pixels in the annulus.
As anticipated, the model from Eq.~\eqref{eq:strucfuncfit} does not agree with the data on all sampled scales, 
%and 
so the range of scales considered is limited to $\ell$ above the resolution of the data and between 50\arcmin\ and 200\arcmin, where the behaviour of $S^{2}_{2}(\ell)$ is approximately linear in $\ell^{2}$. We then make a linear fit of $S^{2}_{2}(\ell)$ using $\ell^2$ as the independent variable and estimate the value of $b$ and its uncertainty, $\sigma_{b}$.  
The results are plotted in Fig.~\ref{fig:StrucFunc}.

Using the calculated $b$ values,\footnote{For Eq.~\eqref{eq:cfmethodHild} $b$ is required to be in radians, and so those are also listed in Table~\ref{table-magnparameters}.} 
we apply Eq.~\eqref{eq:cfmethodHild} to estimate $B^{\textsc{\HIL}}_{\perp}$\langed{,} and the results are listed in Table~\ref{table-magnparameters}. As expected, the \hkd\ method produces estimates of $B_{\perp}$ that are about twice as large as those obtained with the DCF method. However, it is worth noting that because of the discrepancy between the model and the observations\langed{,} the value of $b$ depends on the selected $\ell$-range and so propagates into a different value of $B^{\textsc{\HIL}}_{\perp}$.

\subsection{Mass-to-flux ratios}\label{mtfr}

The critical value for the mass that can be supported against gravity by a magnetic flux $\Phi$ can be estimated to first order for a uniform disk from $(M/\Phi)_{\rm crit} \equiv 1/(2\pi G^{1/2})$ \citep{nakano1978}. 
%Note that
\langed{The} precise value of the right-hand side changes for different geometries \citep[e.g.,][]{spitzer1968,mckee1993}. 
Stability can be assessed using
\begin{equation}\label{magcritical}
\lambda = \frac{(M/\Phi)}{(M/\Phi)_{\rm crit}} = 7.6\times10^{-21}\,\frac{(\nhb/\mbox{cm}^{-2})}{(\bb/\mu\mbox{G})}\, ,
\end{equation}
where $\nhb$ and $\bb$ are the H$_{2}$ column density and magnetic field strength along a magnetic flux tube 
\citep{crutcher2004}. 
A cloud is supercritical \langed{and} prone to collapse under its own gravity, when $\lambda>1$; otherwise, when $\lambda<1$, the cloud is sub-critical, magnetically supported against gravitational collapse.

What is observable is $\lambda_{\rm obs}$, in which $\nhb/\bb$ is replaced by $N_{{\rm H}_{2}}/B_{\perp}$.  
We evaluate $\lambda_{\rm obs}$ by combining the value of $\left<N_{{\rm H}_2}\right>$ computed from the integrated CO line emission and the conversion factor $X_{\textsc{CO}}=(1.8\pm0.3)\times10^{20}$\,cm$^{-2}\,$K$^{-1}$\,km$^{-1}$\,s \citep{dame2001} and $B_{\perp}$ estimated with the DCF and \hkd\ methods. 
The $X_{\textsc{CO}}$ factor may show cloud-to-cloud or regional variations \citep{draine2011}, 
but we consider that these are not significant in comparison to the uncertainties involved in the estimation of $B_{\perp}$. 
The calculated values of $\lambda_{\rm obs}$ are listed in Table~\ref{table-magnparameters}.  They are consistent with being less than unity.

From this we might obtain $\lambda$ by judicious deprojection, since
\begin{equation}\label{relative}
\lambda =  (\nhb  / N_{{\rm H}_{2}}) \, (B_{\perp}/\bb) \times \lambda_{\rm obs}  \equiv f_{\rm dp} \, \lambda_{\rm obs}\,  .
\end{equation}
\langed{Here,} $B_{\perp}$ is always less than $\bb$, pushing $f_{\rm dp}$ below unity.  
The situation for column density 
%is dependent
\langed{depends} on the geometry of the structure relative \langed{to} the magnetic field. For a structure with the magnetic field along the short axis,  $\nhb <  N_{{\rm H}_{2}}$, again lowering $f_{\rm dp}$. 

Statistically, the mean mass-to-flux ratio can be related to the observed value by assuming a particular geometry of the cloud, an ellipsoid with equatorial radius $a$ and 
%centre to pole
\langed{centre-to-pole} distance $c$, and a magnetic field oriented along the polar axis of the ellipsoid. For an oblate spheroid, flattened perpendicular to the orientation of the magnetic field, $f_{\rm dp} = 1/3$, yielding 
\begin{equation}\label{average}
\overline{M/\Phi} = \int^{\pi/2}_{0} \frac{M\cos\beta}{\Phi/\sin\beta}\,\sin\beta\,d\beta \, = \frac{1}{3}(M/\Phi)_{\rm obs}  ,
\end{equation}
where $\beta$ is the inclination angle with respect to the \LOS, and the $\sin\beta$ dependence in the flux comes from $B = B_{\perp}\sin\beta$, \citep{crutcher2004}. For a sphere there is no $\cos\beta$ dependence and $f_{\rm dp} = 1/2$. For a prolate spheroid elongated along the orientation of the field, the mass is multiplied by $\sin\beta$ instead of $\cos\beta$, resulting in $f_{\rm dp} = 3/4$.
Investigating which geometry, if any, is most relevant to the actual MCs and the magnetized structures detected within them is obviously important before firm conclusions can be drawn regarding gravitational instability.

% Discussion Discussion Discussion 

\subsection{Discussion}\label{sect:cfdiscussion}

Our estimates of the magnetic field strengths in the MCs analysed and the mass-to-flux ratios, presented in Table~\ref{table-magnparameters}, stem from the classic calculation presented by \cite{chandrasekhar1953} and an updated interpretation presented by \cite{hildebrand2009}. Both of these methods assume very specific conditions\langed{,} 
%and 
so (as we discuss below) this limits the conclusions that can be drawn from these estimates of the magnetic field strength. 
The deduced mass-to-flux ratios suggest that the clouds are potentially magnetically sub-critical, but we again need to be cautious 
%regarding
\langed{about drawing} conclusions from this application of the DCF and \hkd\ analyses alone.

In the case of the DCF method, the values of $B^{\textsc{DCF}}_{\perp}$ are obtained by assuming that the structure of the magnetic field is the product only of incompressible Alfv\'{e}n waves, where the displacements are perpendicular to the direction of propagation. This is not the case for turbulence in MCs where the random component of the magnetic field can have any orientation. The dispersion measured about mean fields, assumed to be straight in DCF, may be much larger than should be attributed to MHD waves or turbulence, leading to an overestimation of $\varsigma_{\psi}$ and to low values of $B_{\perp}$.
Moreover true interstellar turbulence in MCs involves not only incompressible Alfv\'en waves, but also compressible magneto-sonic waves, which do not satisfy Eq.~\ref{eq:cfmethod}. Furthermore, depending on the scales examined, the magnetic field may have structures due to effects such as differential rotation, gravitational collapse, or expanding H$\textsc{ii}$ regions.

In the case of the \hkd\ method, the values of $B^{\textsc{\HIL}}_{\perp}$ are obtained by assuming a very specific model of the magnetic field\juan{, which is in principle just a first-order approximation}.
First, this model assumes that the effect of the large-scale structured magnetic field, $\vec{B}_{0}$, is to cause the square of the second-order structure function, $S^{2}_{2}(\ell)$, to increase as $\ell^{2}$. This corresponds to a very specific correlation function for $\vec{B}_{0}$.  Second, this model assumes that the dispersion of the random component of the field, $\vec{B}_{\rm r}$, is scale-independent, which is not realistic for the range of scales probed by \Planck\ \citep{elmegreen2004,hennebelle2012}.

In addition, the magnetic field orientation deduced from the polarization angle in a particular direction is not generally that of the field at a single point along the \LOS.  The observed polarization is the average of various field \pseudovectors\ weighted by local dust emission along the \LOS. The net effect of the integration of multiple uncorrelated components along the \LOS\ is an observed dispersion of the polarization angle that is smaller than the true 3D dispersion of the magnetic field orientation, thus leading to an overestimation of $B_{\perp}$. \cite{myers1991} presented an analysis of this effect in terms of the number of correlation lengths of the magnetic field along the \LOS\ through a cloud, which they calculated empirically. 

\juan{\cite{houde2009} presented an extension of the \HIL\ method that includes the effect of signal integration along the \LOS\ and across the area subtended by the telescope beam.
The extended method, also implemented in \cite{houde2011,houde2013}, is based on the identification of the magnetized turbulence correlation length ($\delta$) by means of the structure function of the polarization angles.
In the case of the \Planck\ 353\,GHz observations, the angular resolution is not sufficient to identify $\delta$ and the corrections would have to rely on rough estimates of this value and the depth of integration ($\Delta$).
Following equation\,29 in \cite{houde2009}, rough estimates $\delta\approx0.2$\,pc and $\Delta\approx10$\,pc result in a few correlation lengths across the beam, corresponding to correction factors around 0.4. Coincidentally, such correction factors lead to values of $B^{\textsc{\HIL}}_{\perp}$ close to those of $B^{\textsc{DCF}}_{\perp}$ in Table\,\ref{table-magnparameters}.
But note that this correction relies on specific assumptions on the nature of the turbulence correlation function, it does not circumvent the necessity to observe the dust polarized emission with higher angular resolution to fully characterize the magnetic field. 
%From spatial patterns of optical polarization, where the beam scale is essentially the angular size of the star, \cite{myers1991} estimated that the number of correlation lengths along the \LOS\ is of the order of a few in many clouds.
%However, it is important to note that as in the case of the \hkd\ method, the corrections presented in \cite{houde2009} rely on specific assumptions on the nature of the turbulence correlation function, which are not generally met at the scales considered in this study.
%%%This implies that the corrections of the \LOS\ integration effect do not circumvent the necessity to observe the dust polarized emission with higher angular resolution. 
}

MHD simulations provide a potentially useful guide to what modifications might be introduced into the DCF formula to allow for inhomogeneity and line-of-sight averaging. Using synthetic observations of MHD simulations, \cite{ostriker2001} showed that correcting  Eq.~\eqref{eq:cfmethod} by a factor $\mathcal{C}\approx0.5$ provides a good approximation to the actual magnetic field strength in cases where $\varsigma_{\psi} < 25$\degr. 
However, the effect of nonlinear amplitudes is uncertain \citep{zweibel1996}\langed{,} and the method fails for values of $\varsigma_{\psi} > 25$\degr, which is the case for all of the regions in this study (Table~\ref{table-magnparameters}). 
\cite{falceta2008} propose a method that is potentially valid for any value of $\varsigma_{\psi}$, based on a fit to the $B^{\textsc{DCF}}_{\perp}$ values obtained from maps at different resolutions, again concluding that the field should be lower than estimated with Eq.~\eqref{eq:cfmethod}. 
However, this approach was not tested in MHD simulations that include gravity, which is the critical process that we aim to evaluate by using the DCF method.

Another strong assumption is that the behaviour of the velocity and the magnetic field are 
%well represented
\langed{represented well} by the observed quantities $\sigma_{v_{\parallel}}$ and $\varsigma_{\psi}$ for a particular set of scales, which might not necessarily be the case. Even if the power spectra of $\vec{v}$ and $\vec{B}$ are comparable in 3D, the integration along the \LOS\ is different for the two quantities.  The dispersion $\sigma_{v_{\parallel}}$ is based on the emission-line profile $v_{\parallel}$ tracing a gas species, and while the emission is directly integrated along the \LOS, the line profile is possibly affected by radiative transfer and excitation effects.
The tracer of the magnetic field is the optically-thin polarized submillimetre emission of dust\langed{,} and both the polarization and $\vec{B}_{\perp}$ are projected and integrated along the \LOS\ as \pseudovectors. 

Finally, in the estimates of $B^{\textsc{DCF}}_{\perp}$ and $B^{\textsc{\HIL}}_{\perp}$, a common mean density $\rho$ has been adopted, while in practice $\rho$ is different from cloud to cloud. Direct estimation of the values of $\rho$ from the measured column densities $\left<\nhd \right>$ and $\left<N_{{\rm H}_{2}}\right>$ relies heavily on the geometrical modelling of the cloud and the filling factors of each species, introducing uncertainties that 
%impact
\langed{affect} the calculated values of $B_{\perp}$.

Given the \LOSh\ integration and the fact that $\varsigma_{\psi} > 25$\degr\ uniformly, the calculated values could be considered as upper limits to the actual $B_{\perp}$. However, other shortcomings, such as the assumptions about the correlation structure and the uncertainties in the determination of the density, do not necessarily bias the estimate of the magnetic field strength towards high values.  In conclusion, the values presented in Table~\ref{table-magnparameters} should be viewed only as a reference and only applied with caution, given the many assumptions in both methods at the scales considered.

\end{document}